\documentclass[preprint,floats,aps,12pt,superscriptaddress,nofootinbib,floatfix]{revtex4}
\usepackage[utf8]{inputenc}
\usepackage[sort&compress]{natbib} 
\usepackage{times} 
\usepackage{amssymb,amsmath}
\usepackage{prelim2e}\usepackage[none,bottom]{draftcopy}
\draftcopyName{preprint / }{1.2} %ADAPT TEXT
\ifx\pdfoutput\undefined
\usepackage[usenames,dvips]{color}                        %Timesschriftart benutzen
\else
\usepackage[usenames,dvipsnames]{color}
\usepackage{epstopdf}
\usepackage[pdftex]{graphicx}
\usepackage[pdftex]{hyperref}
\hypersetup{a4paper,
pdfproducer={lateX},
pdfview=FitV,       % FitH
pdfstartview=FitB,
linkcolor=blue,     % links to same page
citecolor=blue,     % citations
pagecolor=blue,     % links to same page
urlcolor=red,      % links to URLs
breaklinks=true,    % links may be split onto 2 lines
colorlinks=true,
citebordercolor=0 0 0,  % color for \cite
filebordercolor=0 0 0,
linkbordercolor=0 0 0,
menubordercolor=0 0 0,
pagebordercolor=0 0 0,
urlbordercolor=0 0 0,
pdfhighlight=/I,
pdfborder=0 0 0,   % no box around links
backref=false,
pagebackref=false,
bookmarks=true,
bookmarksopen=true,
bookmarksnumbered=true
}
\fi
\ifx\pdfoutput\undefined
\DeclareGraphicsExtensions{.eps}
\else
\DeclareGraphicsExtensions{.jpg, .pdf, .tif, .png}
\fi
\graphicspath{{.},{./}} %$ %ADAPT DIRECTORY
\usepackage[usenames,dvipsnames]{color}
\usepackage[normalem]{ulem}
\newcommand{\mylab}[1]{\label{#1}}
\renewcommand{\vec}[1]{\mathbf{#1}}
\newcommand{\vecg}[1]{\boldsymbol{#1}}
\newcommand{\tens}[1]{\mathbf{\underline{#1}}}

\begin{document}
%
%----------------------------------------------------------------%
\title{Suppression of coarsening and emergence of oscillatory behavior in a Cahn-Hilliard model with nonvariational coupling}

\author{Tobias Frohoff-H\"ulsmann}
\email{t\_froh01@uni-muenster.de}
\thanks{ORCID ID: 0000-0002-5589-9397}
\affiliation{Institut f\"ur Theoretische Physik, Westf\"alische Wilhelms-Universit\"at M\"unster, Wilhelm-Klemm-Str.\ 9, 48149 M\"unster, Germany}

\author{Jana Wrembel}
\affiliation{Institut f\"ur Theoretische Physik, Westf\"alische Wilhelms-Universit\"at M\"unster, Wilhelm-Klemm-Str.\ 9, 48149 M\"unster, Germany}

\author{Uwe Thiele}
\email{u.thiele@uni-muenster.de}
\homepage{http://www.uwethiele.de}
\thanks{ORCID ID: 0000-0001-7989-9271}
\affiliation{Institut f\"ur Theoretische Physik, Westf\"alische Wilhelms-Universit\"at M\"unster, Wilhelm-Klemm-Str.\ 9, 48149 M\"unster, Germany}
\affiliation{Center for Nonlinear Science (CeNoS), Westf{\"a}lische Wilhelms-Universit\"at M\"unster, Corrensstr.\ 2, 48149 M\"unster, Germany}
\affiliation{Center for Multiscale Theory and Computation (CMTC), Westf{\"a}lische Wilhelms-Universit\"at, Corrensstr.\ 40, 48149 M\"unster, Germany}
\begin{abstract}
We investigate a generic two-field Cahn-Hilliard model with variational and nonvariational coupling. It describes, for instance, passive and active ternary mixtures, respectively. Already a linear stability analysis of the homogeneous mixed state shows that activity not only allows for the usual large-scale stationary (Cahn-Hilliard) instability of the well known passive case but also for small-scale stationary (Turing) and large-scale oscillatory (Hopf) instabilities. In consequence of the Turing instability, activity may completely suppress the usual coarsening dynamics. In a fully nonlinear analysis we first briefly discuss the passive case before focusing on the active case. Bifurcation diagrams and selected direct time simulations are presented that allow us to establish that nonvariational coupling (i) can partially or completely suppress coarsening and (ii) may lead to the emergence of drifting and oscillatory states. Throughout, we emphasize the relevance of conservation laws and related symmetries for the encountered intricate bifurcation behavior.~\\
The published version of this preprint can be found under \newline
T. Frohoff-H\"ulsmann, J. Wrembel and U. Thiele.
Suppression of coarsening and emergence of oscillatory behavior in a Cahn--Hilliard model with nonvariational coupling.
\textit{Phys. Rev. E,} 103:042602, 2021.
DOI: 10.1103/PhysRevE.103.042602
\end{abstract}
%
%\begin{keyword} 
%Sliding drops \sep Heterogeneous substrates \sep Pinning and depinning
%\pacs{
%68.15.+e, % Thin films: Liquid thin films
%47.20.Ky  % Fluid dynamics: Nonlinearity (including bifurcation theory)
%47.55.Dz  % Drops and bubbles 
%68.08.-p  % Liquid-solid interfaces
%}
%\end{keyword} 
%
\maketitle
%
%\received{6.5.2002}
%
%----------------------------------------------------------------%
% referees: 
%
%%%%%%%%%%%%%%%%%%%%%%%%%%%%%%%%%%%%%%%%%%%%%%%%%%%%%%%%%%
%  INTRO
%%%%%%%%%%%%%%%%%%%%%%%%%%%%%%%%%%%%%%%%%%%%%%%%%%%%%%%%%%\section{The nonreciprocal Cahn-Hilliard model}
\section{Introduction} \mylab{sec:intro}
%%%%%%%%%%%%%%%%%%%%%%%%%%%%%%%%%%%%%%%%%%%%%%%%%%%%%%%%%%%%%%%%%%%%%%%%%%%%%%%
% 
Phase separation, also called demixing, unmixing or decomposition is a universal process occurring in many experimental systems where an initially homogeneous mixed state decomposes into different phases \cite{Lang1992,Jones2002,Onuki2002}. If quenched into a linearly unstable state, phase heterogeneities develop on a typical lengthscale determined by the quench. Over time, the developing structures continuously coarsen, i.e., their average size increases and their number decreases \cite{Lang1992}.
The simplest dynamical model for such processes is the Cahn-Hilliard (CH) equation, a nonlinear, dissipative model originally proposed to describe the dynamics of demixing of isotropic solid or fluid binary solutions \cite{CaHi1958jcp,Cahn1965jcp}. Extensions to decomposing mixtures of multiple components are also available \cite{Eyre1993sjam,HuOS1995m}. In the classification of Hohenberg and Halperin, the class of models is referred to as ``model-A'' \cite{HoHa1977rmp}. Already in the case of a binary mixture, the generic CH model captures many qualitative features of demixing and thus is widely applied from material science to soft matter. Variants and extensions are also increasingly used in biophysical contexts. Examples include descriptions of protein patterns near membranes of living cells \cite{RBGS2008jcp,JoBa2005prl}, of the motility-induced phase separation of active Brownian particles \cite{WTSA2014nc,CaTa2015arcmp,SBML2014prl,RaBZ2019epje}, and of the suppression of Ostwald ripening in active emulsions relevant for centrosome dynamics in biological cells \cite{ZwHJ2015pre,LeWu2018jpdap,WZJL2019rpp}.
% 
% coupled CH + reaction as membrane model \cite{JoBa2005pb}
% only coupled RD \cite{AlBa2010pb}, but also showing coarsening
% active membrane model 2 CH + 'recycling' \cite{ReBL2005pre,GoSR2009pre}

A common feature of most variants of CH models outside the biophysical context is that the described dynamics of a concentration or density field $\phi(x,t)$ conserves a mass-like quantity \textit{and} results in the decrease of an underlying energy $\mathcal{F[\phi]}$. Spatial derivatives only enter through a squared-gradient term representing the energetic cost of interfaces. These physical properties directly determine the form of the equation: a conservation law with a variational form. With other words, the CH model represents a mass-conserving gradient dynamics that describes the transition from an (unstable) initial state to a (stable or metastable) equilibrium state that minimizes $\mathcal{F}$. The final state is not necessarily the global energy minimum. If it is the global minimum, it corresponds to the thermodynamic equilibrium only in the thermodynamic limit, i.e., for diverging system size. For a discussion how this limit is approached with increasing system size see Ref.~\cite{TFEK2019njp}.

If the system boundaries do not sustain any throughflow, and no energy is fed into the system in other ways, e.g., by chemical reactions, we call the system ``passive''. This, together with the variational form implies that no sustained drift or time-periodic behavior can occur and, in particular, all linear modes are stationary.
However, there exist several settings where the system becomes ``driven'' or ``active''. One option is the addition of a lateral driving force in combination with a corresponding flux of material across the system boundaries. The resulting convective CH equation is studied, e.g., in \cite{EmBr1996pre,GoDN1998pd,WORD2003pd,TALT2020n}. In this case, the driving term breaks the parity symmetry of the CH equation, i.e., in a one-dimensional (1D) system the left-right symmetry. This case shall not concern us here.

Another option is to add ``activity'', normally, corresponding to additional terms that do not break the parity symmetry but are nevertheless nonvariational, i.e., they break the gradient dynamics structure of the equation. Often, such contributions result from a chemo-mechanical coupling, e.g., for self-propelled constituents, and indicate that the system acquires energy from outside that is then dissipated within. An example is an active CH type equation that describes phase separation processes in nonequilibrium systems. It models aspects of the so-called ``active phase separation'' in suspensions of self-propelled particles \cite{CaTa2015arcmp,SMBL2015jcp,BeRZ2018pre}, and is also relevant in the context of cell polarization and chemotactic aggregation \cite{BeZi2019po,RaZi2019pre}. Close to the corresponding critical point it can be systematically derived as leading (passive CH equation) and next-to-leading order (active extensions) dynamics \cite{BeRZ2018pre,RaBZ2019epje}. Despite its nonvariational character, generalized thermodynamic quantities can be defined such as nonequilibrium pressure and chemical potential which result in nonequilibrium coexistence conditions and an ``uncommon tangent (Maxwell) construction'' \cite{SSCK2018pre,WTSA2014nc}. Other active CH type equations do not allow for the definition of such generalized thermodynamic quantities. For systems of more than one dimension a term can be added that supports self-sustained circulating currents \cite{TjNC2018prx}. % the flux is composed of spatial gradient and rotational parts according to a Helmholtz decomposition \cite{TjNC2018prx}. 

In the context of applications, biophysical and other, often several degrees of freedom are involved, i.e., dynamic models describe the coupled evolution of several density- or concentration-like order parameter fields that each may follow a conserved or nonconserved dynamics. Again, models can be variational or nonvariational. In the former case, such models describe, e.g., phase separation in ternary \cite{Eyre1993sjam,HuOS1995m} and multicomponent \cite{SPNS1996pd,ToPG2015prb,MoCa1971am} mixtures including membranes (see model~I in \cite{JoBa2005prl}). Also thin-film models for layers of solutions and suspensions \cite{ThAP2016prf,Thie2018csa} belong to the same class of equations.
In the nonvariational case, typical examples are models for phase separation in ternary mixtures with chemical reactions \cite{ToYa2002jcp,OkOh2003pre}, membrane models that include chemical reactions \cite{JoBa2005prl,JoBa2005pb}, and thin-film models for active liquids \cite{TSJT2020pre}.
Such membrane models consist, e.g., of reaction-diffusion (RD) equations for three fields with one conservation law (see model~II in \cite{JoBa2005prl} and \cite{AlBa2010pb}), and of four fields with CH or RD dynamics with two conservation laws \cite{JoBa2005pb}. A five-field model with two conservation laws is considered in \cite{HaFr2018np} 
% ALSO CITE?: K. C. Huang, Y. Meir, and N. S. Wingreen, Proceedings of the National Academy of Sciences of the United States of America 100, 12724 (2003; J. Halatek and E. Frey, Cell reports 1, 741 (2012).
where also a simpler conceptual model is analyzed consisting of a two-field RD system with one conservation law. An active emulsion model describing, e.g., centrosome dynamics in biological cells, employs a reactive coupling of two CH equations keeping only one overall conservation law \cite{ZwHJ2015pre,WZJL2019rpp}.
%add introduction of nonreciprocal coupling
A characterizing property of multicomponent systems are the coupling terms. A so-called ``nonreciprocal coupling'' break the action-reaction symmetry (Newton's third law) and always renders the dynamics nonvariational. In biophysical applications such couplings are often based on effective interactions between two species that are meditated by a nonequilibrium environment \cite{IBHD2015prx}, but can also describe predator-prey interactions \cite{ChKo2014jrsi}. The statistical and thermodynamic properties of nonreciprocal systems are treated in \cite{IBHD2015prx,LoKl2020njp}.
A ``nonreciprocal CH model'' consisting of two CH equations with nonvariational coupling is investigated in Refs.~\cite{SaAG2020prx,YoBM2020pnas} as a description of interacting scalar active particles where both species are individually conserved. It shows demixing at small nonreciprocal coupling which transitions to oscillatory behavior at high activity, e.g., resulting in self-propelled globally ordered bands.
In another conceptual model, two CH equations, i.e., two conservation laws, are coupled in a way that breaks both conservation laws \textit{and} the variational structure \cite{SATB2014c}. It is found that the coupling can suppress the coarsening process typical for CH dynamics and may even result in oscillatory dynamics.  

%studied in the context of Min-protein dynamics 
%A work describing Min-protein dynamics constrained by two conserved quantities elaborate this relationship .
%
%Interpreting the conserved quantity as a local control parameter they can link its spatial distribution to the qualitative dynamical behavior of the system \cite{HaFr2018np,BrHF2018arxiv}.
%
%With that they contribute to a new understanding of a mechanism of pattern formation and the transition to chaos, too.
%
% A simple reaction-diffusion system with one conserved quantity however does not show the complex dynamical behavior described in the min-protein system.

A central feature of phase separation as modeled by the CH model is the already mentioned coarsening that results in a continuous increase of typical sizes of the developing phase-separated regions, i.e., drops (clusters), holes or labyrinthine structures \cite{Lang1992}. Coarsening proceeds through the two main modes of volume transfer (known as Ostwald ripening) and by translation (known as coalescence). The volume transfer mode moves material between structures without moving their centers, i.e., their sizes change. In contrast, the translation mode moves the structures without changing their sizes. More details on coarsening behavior in the CH equation and related thin-film equations are, e.g., given in \cite{Onuki2002,KoOt2002cmp,GORS2009ejam,Nepo2015crp,ACRT2005pre}.

Coarsening may be suppressed by heterogeneities in the (still variational) system, e.g., for drops on a substrate with wettability patterns \cite{TBBB2003epje} or phase separation in a spatially modulated temperature profile \cite{KrKr2004pre}. In diblock copolymer melts described by a single CH equation with long-range interactions (Oono-Shiwa model) the system is stabilized at a certain length scale, i.e., coarsening is partially suppressed \cite{PoTo2015jsm}. Such an arrest of coarsening was recently discussed for reaction-diffusion systems with weakly broken mass conservation \cite{BWHY2021prl}.
Coarsening can also be suppressed by driving or activity. Studies of the convective CH equation \cite{EmBr1996pre,GNDZ2001prl} show that an increase in the lateral driving force results in a transition towards chaotic wave patterns. This implies that there exist parameter regions where driving suppresses coarsening \cite{ZPNG2006sjam}. Aspects of the underlying bifurcation structure are presented in \cite{TALT2020n}. 

In most active one-field CH models employed to describe motility-induced phase separation, coarsening is not suppressed but closely resembles its counterpart in the standard passive model \cite{CaTa2015arcmp}. However, ``reverse Ostwald ripening'' for vapor bubbles and liquid clusters is described for a one-field active CH model in two dimensions with two types of nonvariational contributions: a nonequilibrium chemical potential and a nonequilibrium flux, itself related to a nonlocal chemical potential \cite{TjNC2018prx}. The latter's specific vectorial character allows for self-sustained circulating currents and is responsible for the suppression of coarsening that occurs if the system is at least two-dimensional.
Suppression of coarsening is also observed for active models involving coupled CH equations. Reference~\cite{SATB2014c} shows that suppression occurs already at weak nonvariational coupling between the two concentration fields. It is argued that each structured field acts as heterogeneity for the other one and the resulting pinning arrests coarsening. Linear stability analysis and direct time simulations show that besides the arrest of coarsening, the nonvariational coupling may also induce the structures to drift or oscillate. With other words the chosen coupling dramatically changes central features of the phase separation model. 
Similar phenomena are also observed in more complex models for reactive decomposition \cite{OkOh2003pre,JoBa2005pb,WZJL2019rpp}. 

Motivated by these rich phenomena in active phase-separating systems, we study a system of generic kinetic equations consisting of two coupled CH equations. The coupling maintains both conservation laws and consists of separated variational (reciprocal) and nonvariational (nonreciprocal) contributions. This allows us to analyze the qualitative transitions in the dynamics of two conserved quantities that occur when going from a variational to a nonvariational model. In this way, we can clearly relate occurring qualitative changes to the imposed changes in the variational character and avoid a potential interference with effects due to a changing conservation character.

In a nongeneric limiting case of our model, the nonvariational case is studied in Ref.~\cite{SaAG2020prx} and with some further simplification in Ref.~\cite{YoBM2020pnas} with a focus on the emergence of traveling states. Here, we systematically show that the nonvariationally coupled CH model exhibits a much richer selection of phenomena. Especially, our analysis allows for a deeper understanding of similarities and differences between Ref.~\cite{SaAG2020prx} and the study in Ref.~\cite{SATB2014c} where the coupling does not maintain the conservation properties and is \textit{purely} nonvariational.
As a result it shall be possible to identify features of related system-specific models in the literature as generic features resulting from conservation laws. In particular, we show that for such systems a nonvariational coupling can (i) partially or completely suppress coarsening
  and (ii) may lead to the emergence of drifting and oscillatory states. %Thereby, we will distinguish partial and (linear and nonlinear) complete suppression of coarsening.
  Further, we discuss why in the simplified models studied in Refs.~\cite{SaAG2020prx,YoBM2020pnas} coarsening can not be suppressed.

Our work is structured as follows. In Section~\ref{sec:model} we introduce the model and discuss our numerical approach. Subsequently, Section~\ref{sec:linear} provides a linear stability analysis of the uniform state in the variational and the nonvariational case. For the latter, we discuss the transition from a large-scale stationary (Cahn-Hilliard) to a small-scale stationary (Turing) instability and the occurrence of a large-scale oscillatory (Hopf) instability.
Section~\ref{sec:nonlinear-vari} briefly discusses coarsening dynamics and the corresponding bifurcation structure in the variational case. This provides a reference for the subsequent analysis of the nonvariational case: In Sections~\ref{sec:nonlinear-nonvari} and \ref{sec:timeperiodic} we investigate how an increase in the nonvariational coupling suppresses coarsening and results in the emergence of persistent drift and oscillatory behavior, respectively.
Section~\ref{sec:conc} concludes with a summary and outlook.  Note that data sets for all figures as well as examples of \textit{Matlab} codes for the employed numerical path continuation and \textit{python} codes for time simulations are provided on the open source platform \textit{zenodo} \cite{FrWT2021zenodo}.
% This shall enable everyone to reproduce and extend our results.
%
%%%%%%%%%%%%%%%%%%%%%%%%%%%%%%%%%%%%%%%%%%%%%%%%%%%%%%%%%%%%%%%%%%%%%%%%%%%%%%%
\section{Governing equations}\mylab{sec:model}
%%%%%%%%%%%%%%%%%%%%%%%%%%%%%%%%%%%%%%%%%%%%%%%%%%%%%%%%%%%%%%%%%%%%%%%%%%%%%%%
%
The classic Cahn-Hilliard model describes the dynamics of diffusive phase decomposition processes in various (solid-solid, liquid-liquid, liquid-gas) demixing processes of binary systems.
For a scalar order parameter field $\phi(\vec{x},t)$ the corresponding conserved gradient dynamics reads
\begin{equation}
\frac{\partial\phi}{\partial t}\,=\,\vec{\nabla}\cdot \left[Q(\phi)\, \vec{\nabla}\,
\frac{\delta \mathcal{F}[\phi]}{\delta \phi}\right]\, ,
\mylab{eq:CH1}
\end{equation}
where $Q(\phi)$ is a positive definite mobility function (or constant) and 
\begin{equation}
\mathcal{F}[\phi(\vec{x},t)] =\int_V \left[\frac{\kappa}{2}\, |\nabla \phi|^2  + f(\phi) \right] \mathrm{d}\vec{x}
\mylab{eq:helmhet2}
\end{equation}
is the underlying free energy: a square-gradient  interface contribution with interface stiffness $\kappa>0$ is combined with the simple bulk contribution
\begin{equation}
f(\phi)=\frac{a}{2}\phi^2 + \frac{b}{4}\phi^4.
\mylab{eq:helmhet2b}
\end{equation}
Here, $b>0$ and either $a>0$ (case of single minimum) or $a<0$ (double-well potential). Note that Eq.~(\ref{eq:CH1}) is parity and field-inversion symmetric, i.e., it does not change its form for $\vec{x}\to-\vec{x}$ and $\phi\to-\phi$, respectively.

The variation of the energy $\delta \mathcal{F}/\delta \phi$ corresponds to a chemical potential $\mu$ and Eq.~(\ref{eq:CH1}) can compactly be written as continuity equation $\partial_t\phi + \vec{\nabla}\cdot\vec{j}=0$ with the flux $\vec{j}=-Q\nabla\mu$. The energy monotonically decreases in time (see, e.g., \cite{Doi2013}), i.e., it is a passive system.

For $a<0$ there exists a $\phi$-range of unstable uniform states that develop into a fully phase-separated state. In the thermodynamic limit of an infinite system, the interface contribution in Eq.~(\ref{eq:helmhet2}) can be neglected and the two coexisting phases (obtained by a Maxwell construction) correspond to the minima of $f(\phi)$ as they have identical chemical potential and pressure. For a detailed discussion how this relates to bifurcation diagrams of steady states for finite-size systems see Ref.~\cite{TFEK2019njp}.

After revising the classic CH model, we next introduce the coupled system of two CH equations studied here. Without coupling, each of the two equations corresponds to Eq.~(\ref{eq:CH1}), though with different constants, and the simple coupling is chosen in such a way that it respects  the field inversion symmetry $(\phi_1, \phi_2) \to (-\phi_1, -\phi_2)$ of the equations. After restriction to one spatial dimension and nondimensionalization (see appendix~\ref{sec:app-nondim}) the kinetic equations are
\begin{alignat}{1}
\begin{aligned}
\qquad \frac{\partial\phi_1}{\partial t}  &= \frac{1}{\ell^2}\frac{\partial^2}{\partial x^2} \left[- \frac{1}{\ell^2}\frac{\partial^2\phi_1 }{\partial x^2} +f_1'(\phi_1) - \left(\rho + \alpha\right) \phi_2 \right]\\
\qquad \frac{\partial\phi_2}{\partial t}  &=  \frac{Q}{\ell^2}\frac{\partial^2}{\partial x^2} \left[-\frac{\kappa}{\ell^2}\frac{\partial^2 \phi_2}{\partial x^2} +f_2'(\phi_2) - \left(\rho - \alpha\right) \phi_1 \right]\, ,
\end{aligned}
\label{eq:nondim_bG_final}
\end{alignat}
with $f_1=  a \phi_1^2/2  + \phi_1^4/4$ and $f_2=(a+a_\Delta) \phi_2^2/2  + \phi_2^4/4$.
Both fields have a conserved dynamics, i.e., at all times
\begin{equation}\label{eq:mass}
 \int_{-1/2}^{1/2} \textrm{d}x \phi_1 = \bar{\phi}_{1} \quad\mathrm{and}\quad
 \int_{-1/2}^{1/2} \textrm{d}x \phi_2 = \bar{\phi}_{2} \, ,
\end{equation}
where the $\bar\phi_i$ are parameters set by the initial conditions. Note that the field inversion symmetry does not normally hold for the deviations $\phi_i-\bar\phi_i$ that are often the relevant quantities to consider. The other parameters are the nondimensional domain size $\ell$, mobility ratio $Q$, effective temperature $a$, temperature shift $a_\Delta$, and ratio of interface rigidities $\kappa$. The respective final terms in Eqs.~(\ref{eq:nondim_bG_final}) represent the coupling. It is linear and contains a variational part of strength $\rho$ and a nonvariational part of strength $\alpha$. Increasing or decreasing $\alpha$ from the passive reference case ($\alpha=0$) one can investigate the system behavior with increasing activity.
Eqs.~(\ref{eq:nondim_bG_final}) represent a generic model for passive and active ternary mixtures\footnote{Sometimes two-field models are also referred to as ``binary systems''. Then the naming focuses on the demixing of two molecular species and neglects the third option for the occupation of a volume element, namely, the absence of both molecule types. In our naming convention the part-per-volume concentrations of all species have to add up to unity. Therefore, a ternary system may consist of three molecular species that together fill the entire volume, or of two molecular species and 'vacancies'.}. 

In the passive case ($\alpha=0$), the governing equations (\ref{eq:nondim_bG_final}) are of simple gradient dynamics form $\partial_t\phi_i=\partial_x (Q_i \partial_x \delta\mathcal{F}/\delta\phi_i)$ with $i=1,2$. The energy
\begin{equation}
 \mathcal{F}(\phi_1,\phi_2) = \mathcal{F}_1(\phi_1) + \mathcal{F}_2(\phi_2) + \mathcal{F}_\text{c}(\phi_1,\phi_2)
\end{equation}
is the sum of the two energies $\mathcal{F}_{1}$ and $\mathcal{F}_{2}$ for the decoupled fields, that are of the form (\ref{eq:helmhet2}) and the coupling contribution $\mathcal{F}_\text{c}= -\int_V \rho \phi_1 \phi_2 \mathrm{d}x$. The coupling in the passive case is  purely energetic. Note that we do not consider dynamic coupling as encoded in a mobility matrix, e.g., we exclude cross-diffusion. For a discussion of some such systems see Ref.~\cite{Thie2018csa}. The chosen active coupling represents possibly the simplest way to break the variational form of the passive case while keeping both conservation laws intact. Other options are possible, for a tentative classification of nonvariational amendments of one-field systems see the introduction of Ref.~\cite{EGUW2019springer}.

We mainly investigate steady states and time-periodic states in a spatial domain with periodic boundaries employing numerical path continuation accompanied by selected time simulations.  Numerical path continuation -- we use the \textit{Matlab} package \textit{pde2path}~\cite{UeWR2014nmma,EGUW2019springer} -- allows us to track linearly stable and unstable steady and time-periodic states while varying a primary control parameter. Beginning with a known starting state at some parameter value, \textit{pde2path} applies tangent predictors and Newton correctors to converge to a state at neighboring parameter values. Especially pseudo-arclength continuation is a parametrization which allows for reversals in the direction of control parameter steps -- a feature essential to track solution branches through folds.

For steady states without mean flow, we can integrate Eqs.~\eqref{eq:nondim_bG_final} twice to obtain
 \begin{alignat}{1}
\begin{aligned}
\qquad 0  &= - \frac{1}{\ell^2}\frac{\partial^2\phi_1 }{\partial x^2} +f_1'(\phi_1) - \left(\rho + \alpha\right) \phi_2 -\mu_1\\
\qquad0  &=  -\frac{\kappa}{\ell^2}\frac{\partial^2 \phi_2}{\partial x^2} +f_2'(\phi_2) - \left(\rho - \alpha\right) \phi_1 - \mu_2 \,,
\end{aligned}
\label{eq:steady_stateEq}
\end{alignat}
where the integration constants $\mu_i$ are nonequilibrium ``chemical potentials''.
To impose the conservation of both fields, we expand Eqs.~\eqref{eq:steady_stateEq} by adding the constraints (\ref{eq:mass}) and using the $\mu_i$ as secondary control parameters.\footnote{Alternatively, directly using Eqs.~(\ref{eq:nondim_bG_final}) with $\partial_t\phi_i=0$ in the continuation, the role is taken by the strengths of additional ``virtual'' source terms, that are automatically kept at zero.} Furthermore, linear stability of steady states is determined and, hence, all kinds of local bifurcations are detected. This allows one to switch to other steady state branches. Branches of time-periodic states are also continued \cite{Ueck2019ccp}. %in Sec.~\ref{sec:timeperiodic} and the necessary extension of the continuation technique can be found in \cite{EGUW2019springer}.
To present the resulting bifurcation behavior, the norm
\begin{equation}\label{eq:norm}
||\delta \phi|| \equiv \sqrt{\int_{-1/2}^{1/2} \sum_{i=1,2}\left(\phi_i - \bar\phi_i\right)^2 {\rm d} x }
\end{equation}
is employed as solution measure.  %The effective temperature $a$ is used as primary control parameter except for part of Sec.~\ref{sec:nonlinear-vari}.

%\tobias{ternary system asymmetric?}
%\tobias{talk about literature of ternary systems?} \ttuwe{yes, would be useful to cite some papers that look at exactly such equations, even better if they provide some bifurcation diagrams.}
%
%If we make the system active no free energy can be defined which comes along that time-periodic behavior can be expected.
%

%%%%%%%%%%%%%%%%%%%%%%%%%%%%%%%%%%%%%%%%%%%%%%%%%%%%%%%%%%%%%%%%%%%%%%%%%%%%%%%
\section{Linear stability of homogeneous state}\mylab{sec:linear}
%%%%%%%%%%%%%%%%%%%%%%%%%%%%%%%%%%%%%%%%%%%%%%%%%%%%%%%%%%%%%%%%%%%%%%%%%%%%%%%
\subsection{Hopf, Turing and Cahn-Hilliard instability}
First, we analyze the linear stability of homogeneous steady states $\vecg{\phi}(x) \equiv (\phi_1(x),\phi_2(x)) = (\bar{\phi}_{1},\bar{\phi}_{2})=\bar{\vecg{\phi}}$ that, due to mass conservation, all solve Eqs.~\eqref{eq:nondim_bG_final}. For the perturbation we introduce  the harmonic ansatz
\begin{align}
 \vecg{\phi}(x,t)= \vecg{\bar{\phi}} + \delta \vecg{\widetilde{\phi}} e^{i k x + \lambda t}
\end{align}
into Eqs.~\eqref{eq:nondim_bG_final}, linearize in $\delta \ll 1$, and obtain the linear algebraic system
\begin{equation}\label{eq:lambda_original}
\lambda  \vecg{\widetilde{\phi}} = - \left(\frac{k}{\ell}\right)^2
\left( \begin{array}{c c}
\left(\frac{k}{\ell}\right)^2 + f_1'' & -\left(\rho + \alpha\right) \\
-Q\left(\rho - \alpha\right) & \, Q\left(\kappa \left(\frac{k}{\ell}\right)^2 + f_2''\right)
\end{array}\right) \vecg{\widetilde{\phi}} \equiv -\left(\frac{k}{\ell}\right)^2\, \tens{B} \,\vecg{\widetilde{\phi}}\,.
\end{equation}
Rewriting as
  \begin{equation}
    \label{eq:two-field-ce2}
    \tilde\lambda\vecg{\phi} = -\tens{B}\vecg{\phi}
  \end{equation}
  with
  $\tilde\lambda=\lambda/q^2$ and $q=k/\ell$,
the resulting dispersion relations are
\begin{align}\label{eq:lambda+-}
 \tilde\lambda_\pm = \frac{1}{2} \left[ -\mathrm{tr} \tens{B} \pm \sqrt{(\mathrm{tr} \tens{B})^2 - 4\, \det \tens{B}} \right] \\
  \text{with} \quad \mathrm{tr} \tens{B} = q^2 (1+Q\kappa) + f_1'' + Q f_2'' \quad \text{and} \nonumber\\
  \det \tens{B} = Q\left[q^2 + f_1''\right] \left[\kappa q^2 + f_2''\right] + Q \Delta\,. \nonumber
\end{align}
Here we defined the difference in coupling strengths $\Delta\equiv \alpha^2- \rho^2$.
The rescaled eigenvalues $\tilde\lambda_\pm$ are of exactly the same form as those obtained for two coupled RD equations, i.e., the classical Turing system \cite{Turi1952ptrslsbs}.
The original eigenvalues $\lambda$ are obtained by multiplying Eq.~\eqref{eq:lambda+-} with $k^2/\ell^2$ reflecting the conservation of both fields.

In the following we use $f''_1$ and $f''_2$ as primary and secondary control parameter, respectively. Analyzing Eq.~\eqref{eq:lambda+-} gives us conditions for three different primary instabilities: (i) large-scale oscillatory (Hopf) instability (ii) small-scale stationary (Turing) instability and (iii) large-scale stationary (Cahn-Hilliard) instability. In the Cross-Hohenberg classification they are termed (i) type II$_\mathrm{o}$, (ii) type I$_\mathrm{s}$, and (iii) type II$_\mathrm{s}$ instability \cite{CrHo1993rmp}.
Large-scale [small-scale] instabilities are also commonly termed long-wave [short-wave] instabilities. Note that only instability (iii) occurs in the decoupled CH equations. We will show that $|\alpha|>|\rho|$ is a necessary condition for instabilities (i) and (ii).

 \textbf{(i)}
%First we consider the large-scale oscillatory instability related to Hopf bifurcations.
First we consider the Hopf instability.
The onset of an oscillatory instability is characterized by $\lambda_{\pm,c}=\pm i \omega_\textrm{c}$, i.e., with Eq.~(\ref{eq:lambda+-}) this requires  
\begin{equation}
 \mathrm{tr}  \tens{B} = 0 \quad\Rightarrow\quad f''_1 = -(1+ Q \kappa)q^2 - Q f''_2 \, .
 \label{eq:hopfonset}
\end{equation}
Since $Q, \kappa>0$, the largest $f''_1$ always occurs at $q^2=q_{\text{c}}^2 =0$, i.e., here, the oscillatory instability is always large-scale (Hopf instability). Therefore, the Hopf threshold is
\begin{equation}
{f''_1}^H =- Q f''_2\,.
\mylab{eq:a11ho2}
\end{equation} 
However, both eigenvalues of the original linear system [Eq.~\eqref{eq:lambda_original}] remain real and zero at exactly $k=0$ due to the conservation properties, i.e., the critical frequency is 
\begin{align}
\omega_\textrm{c} &= q_\text{c}^2\, \tilde{\omega}_\textrm{c}= 0 ~\nonumber ~\\
  \text{with  }\, \,\tilde{\omega}_\textrm{c}&=\sqrt{\det \tens{B}\big|_{q=q_{\text{c}}}}%= \sqrt{Q}\sqrt{{f''_1}^H f''_2 +\Delta}
                                               = \sqrt{Q}\sqrt{- Q {f''_2}^2 +\Delta}\,.
\mylab{eq:a11hofreq}
\end{align}
In summary, the Hopf instability occurs if $\Delta>  Q {f''_2}^2$ at $f''_1={f''_1}^H $.
In particular, if the two coupled subsystems are identical ($f''_1=f''_2$) the Hopf threshold is at ${f''_1}^H =0$. This implies that for identical subsystems with purely nonvariational coupling ($\rho=0$), oscillatory behavior occurs at arbitrarily small nonvariational coupling $\alpha$. Appendix~\ref{sec:app:small} focuses on this special case. However, a stronger contrast between the two coupled systems implies that a larger coupling $|\alpha|$ is needed to obtain oscillatory behavior. 

\textbf{(ii)}  
Next we consider the Turing instability related to pattern formation.
It occurs if $\tilde\lambda_{+,\text{c}}=0$ at $q_\text{c}\neq0$ and requires
\begin{equation}
\det \tens{B}=0 \Rightarrow f''_1 = -\frac{\Delta}{f''_2+\kappa q^2} - q^2 \,.
\label{eq:alltu}
\end{equation}
The maximum of $f''_1(q^2)$ is obtained via $\mathrm{d}f''_1/\mathrm{d} q^2=0$ and yields the critical wavelength
\begin{equation}\label{eq:kcTuring}
q_\text{c}^2 =\frac{1}{\kappa}\left[\pm \sqrt{\kappa \Delta} -  f''_2\right]\,.
\end{equation}
For $\kappa>1/Q$ [$\kappa<1/Q$] the plus [minus] sign in Eq.~\eqref{eq:kcTuring} corresponds to a maximum, the minus [plus] sign to a minimum in the dispersion relation, the latter not being relevant for the onset of instability. That is, for a Turing instability we demand
\begin{equation}
  q_\text{c}^2 >0 \Rightarrow f''_2<
\pm\sqrt{\kappa \Delta}\text{ for } \kappa \gtrless 1/Q\,.
\end{equation}
In particular, it requires nonvariational coupling stronger than the variational one, i.e.~$|\alpha|>|\rho|$, otherwise the root becomes complex.
For comparison with other studies (see conclusion) it is important to note that for $Q\kappa=1$ and $\kappa=0$ no Turing instability is possible.
% \begin{equation}
%  k_\text{c}^2 >0 \Rightarrow \alpha^2> \rho^2 + \frac{{f''_2}^2}{\kappa}\,. 
% \end{equation}
Inserting $q_\text{c}$ in \eqref{eq:alltu} gives the Turing instability threshold
\begin{equation}
{f''_1}^T = \frac1\kappa \left[f''_2 \mp 2 \sqrt{\kappa \Delta} \right]\text{ for } \kappa \gtrless 1/Q \,.
\mylab{eq:a11tu2}
\end{equation}
relevant for the related pitchfork bifurcations.

\textbf{(iii)}  
Finally, we consider the Cahn-Hilliard (CH) instability, i.e., the only instability occurring for the classical CH equation.
It is characterized by $\tilde\lambda_{+,\text{c}}=0$ at $q_\text{c}=0$ and occurs at
\begin{equation}
{f''_1}^\textrm{CH} = -\frac{\Delta}{f''_2}\,.
\mylab{eq:a11bo}
\end{equation}
The related bifurcations are again pitchfork bifurcations. 
%for $f''_2\ge \sqrt{\kappa(\alpha^2-\rho^2)}$ where
%$k_\text{c}^2$ in Eq.~(\ref{eq:a11tu}) becomes negative.

\begin{figure}
\includegraphics[width=0.7\textwidth]{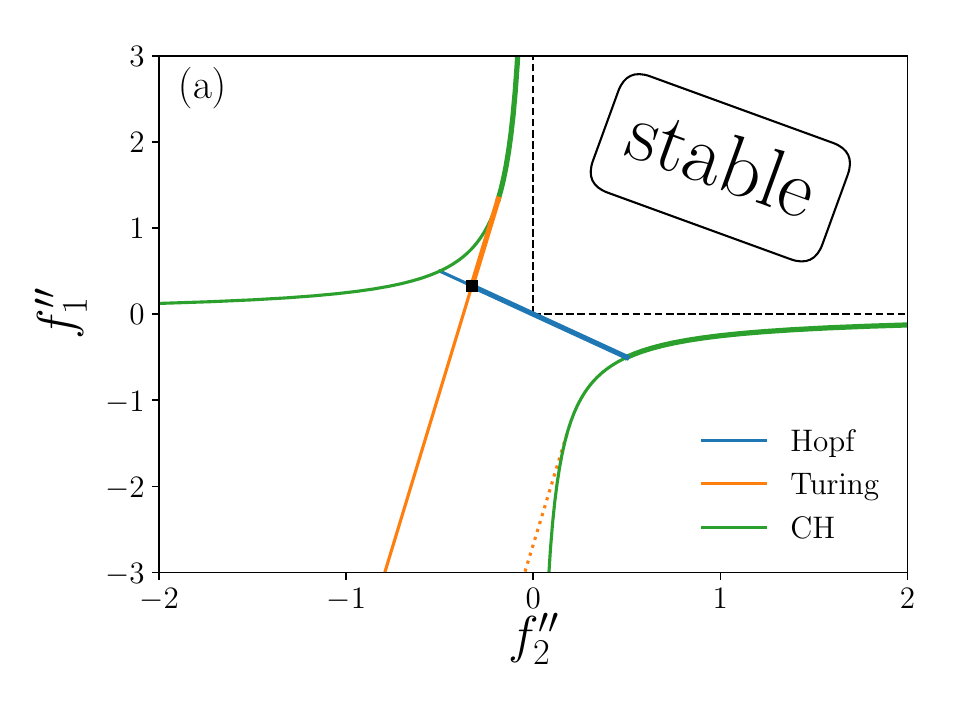}~\\[-3ex]
\includegraphics[width=0.45\textwidth]{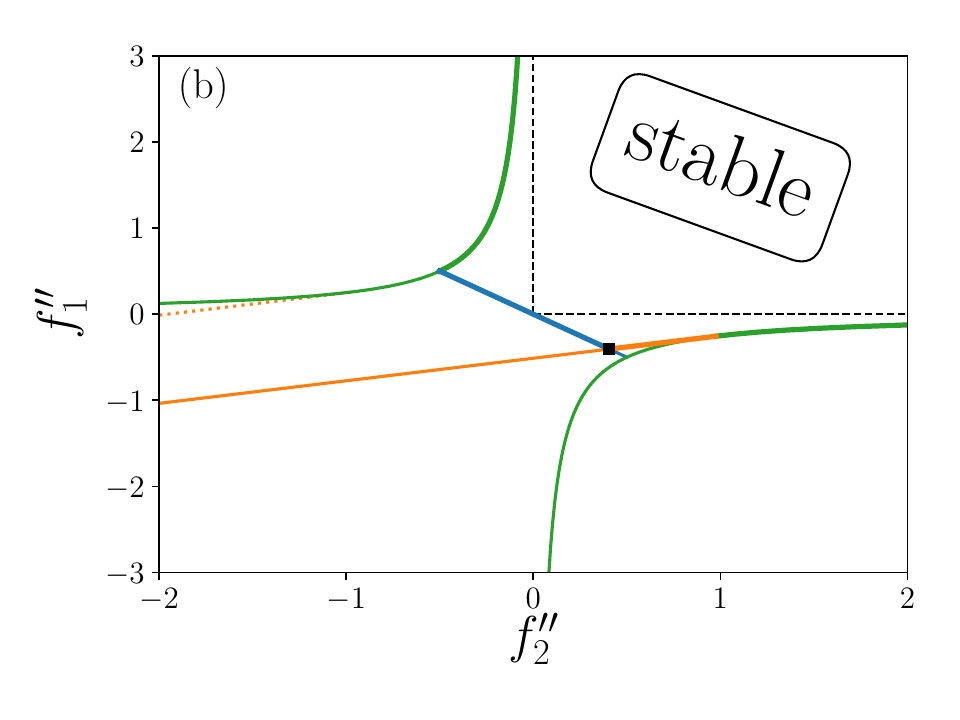}
\hfill
\includegraphics[width=0.45\textwidth]{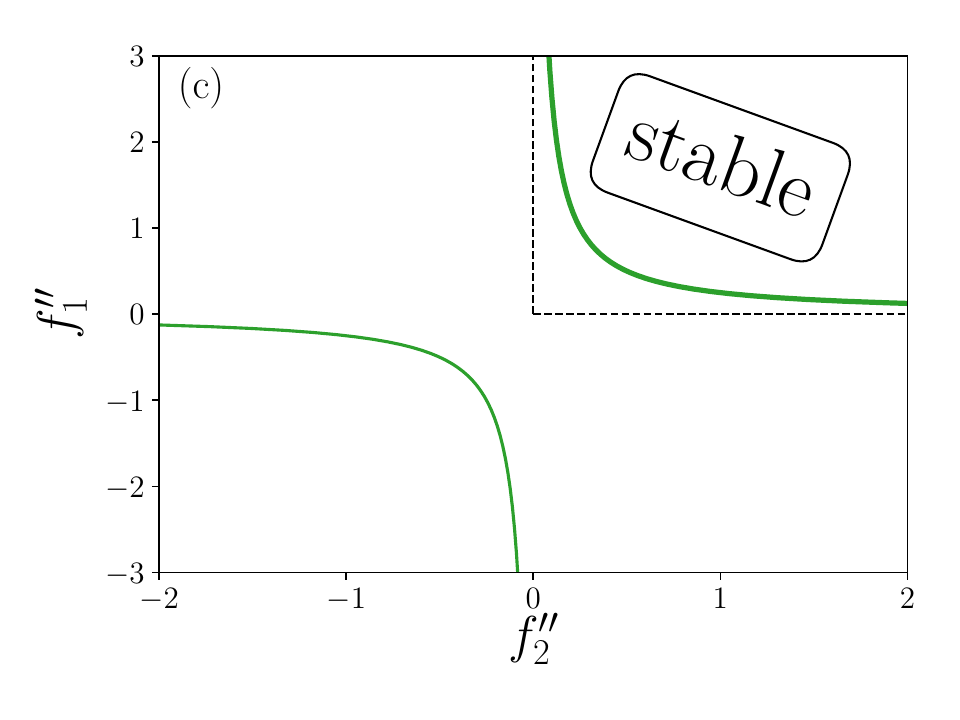}
\caption{\small \it Linear stability diagrams in the $(f''_2,f''_1)$-plane show thresholds of Hopf [Eq.~\eqref{eq:a11ho2}], Turing [Eq.~\eqref{eq:a11tu2}] and CH instabilities [Eq.~\eqref{eq:a11bo}] with blue, orange and green lines, respectively, for different values of $\kappa$ and $\Delta=\alpha^2-\rho^2$ at fixed $Q=1$. The boundary of the linearly stable region [upper right corner] is marked by heavy solid lines. The thin solid lines indicate where further instabilities set in beyond the dominating one. The dotted orange line indicates where the minimum of a dispersion relation of Turing type passes zero. The black dashed lines indicate the stability boundary of a decoupled system (or for identical coupling strengths, i.e.~ for $\Delta=0$). 
Panel (a) is for positive $\Delta=0.25$ and $\kappa=0.14<1/Q$, panel (b) is for $\kappa=3.82>1/Q$ and $\Delta=0.25>0$. The square symbol marks the codimension-2 point [Eq.~\eqref{eq:a22cd2}] where Hopf and Turing instabilities occur simultaneously. Panel (c) represents the case of $|\rho|>|\alpha|$ with $\Delta=-0.25$ and $\kappa=3.82$ where only the CH instability exists.
}
\label{fig:stab-diagram} 
\end{figure}

We see that the three parameters mobility ratio $Q$, rigidity ratio $\kappa$ and the difference in coupling strengths~$\Delta$ determine the three instability thresholds [cf.~Eqs.~\eqref{eq:a11ho2}, \eqref{eq:a11tu2}, \eqref{eq:a11bo}].
Figure \ref{fig:stab-diagram} provides a qualitative overview of the linear stability behavior in the $(f''_2,f''_1)$-plane. Hopf \eqref{eq:a11ho2}, Turing \eqref{eq:a11tu2} and CH \eqref{eq:a11bo} instability thresholds are given by blue, orange and green lines, respectively. The linearly stable region is in the upper right corner. Its boundary is marked by heavy colored lines that represent the onset of the different instabilities. For reference, dashed black lines indicate the instability thresholds for the CH instability of a decoupled system (also valid at $\Delta=0$).

Panel (a) and (b) show stability diagrams for positive $\Delta$, where Hopf, Turing, and CH instabilities occur while in panel (c) for $\Delta<0$ only CH instabilities exist. Further comparison reveals that the stable region widens [shrinks] for increasing [decreasing] $\Delta$. Hence, especially the purely variational coupling $\rho$ always acts destabilizing.
The CH instability thresholds (green lines) are hyperbolas [cf.~Eq.~\eqref{eq:a11bo}] which flip quadrants when $\Delta$ changes sign.
There are two Turing instability thresholds (orange lines) resulting from the two signs in Eq.~\eqref{eq:a11tu2}. For $\kappa<1/Q$ [panel (a)] the upper line corresponding to the plus sign refers to a maximum in the dispersion relation, thus, is relevant for the stability boundary (heavy orange line), whereas the lower line is related to a minimum (dotted orange line). In contrast for $\kappa>1/Q$ [panel (b)], the lower orange line matters.  
In both cases, the relevant Turing line crosses the Hopf line. The crossing point (black filled square) marks a codimension-2 point where both instabilities have their onset at the same value of the primary control parameter $f''_1$. This requires adjustment of a second control parameter, here 
\begin{equation}
{f''_2}^\mathrm{\,cd2} = 2 \frac{\pm \sqrt{\kappa \Delta}}{1+Q \kappa} \,=\, -\frac{{f''_1}^\mathrm{\,cd2}}{Q}\,.
\mylab{eq:a22cd2}
\end{equation}
The Turing lines terminate where they tangentially approach the CH lines at ${f''_2}^{T_{\text{end}}}= \pm \sqrt{\kappa \Delta}$ and the critical wavenumber reaches zero. The Hopf lines also end on the CH lines where $\tilde \omega_\text{c}$ becomes zero at ${f''_2}^{H_{\text{end}}}= \pm \sqrt{\frac{\Delta}{Q}}$. The two end points mark the transition from Turing and Hopf instability to CH instability, respectively. They do not correspond to a coexistence of different linear instabilities as does the codimension-2 point.
Especially, in the nongeneric case $\kappa=1/ Q$ one has
\begin{equation}
{f''_2}^\mathrm{\,cd2} ={f''_2}^{T_{\text{end}}}={f''_2}^{H_{\text{end}}}\,,
\end{equation}
and all three special points coincide. It is remarkable that in this case the Turing lines completely disappear since the eigenvalues become complex at the threshold implying that no Turing instability occurs (not shown).
% ~\\~\\~\\
% Hopf bifurcation [Turing bifurcation] dominant for
%   ${f''_1}^H>{f''_1}^T$ [${f''_1}^T>{f''_1}^H$]
% \tobias{hab ich nicht eingebaut, denn kann man das so einfach sagen? In panel(a) Turing bifurcation sets in for increasing $f''_1$ while Hopf instab. sets in for decreasing $f''_1$}
% ~\\~\\

Up to here, we have discussed the linear stability behavior of the model Eq.~\eqref{eq:nondim_bG_final} for arbitrary $f''_1$ and $f''_2$. In the following we focus on our specific case with $f''_1=a + 3 \bar{\phi}_1^2$, $f''_2=a+a_\Delta + 3 \bar{\phi}_2^2$ and $Q=1$. We discuss the resulting dispersion relations and stability boundaries for the passive (Sec.~\ref{sec:pass}) and active (Sec.~\ref{sec:act}) case. The effective temperature $a$ is employed as main control parameter corresponding to diagonal cuts through the stability diagrams in Fig.~\ref{fig:stab-diagram}. Some further details of our specific case are presented in appendix \ref{app:linear}.

% In the following, we discuss the dispersion relations [Eqs.~\eqref{eq:lambda+-}] and the resulting stability borders Eqs.~$\big($\ref{eq:azero},\ref{eq:azero-os}$\big)$ for passive ($\alpha=0$) and active ($\alpha \neq 0$) systems.
% %
%%%%%%%%%%%%%%%%%%%%%%%%%%%%%%%%%%%%%%%%%%%%%%%%%%%%%%%%%%%%%%%%%%%%%%%%%%%%%%%
\subsection{Passive system}\label{sec:pass}
%%%%%%%%%%%%%%%%%%%%%%%%%%%%%%%%%%%%%%%%%%%%%%%%%%%%%%%%%%%%%%%%%%%%%%%%%%%%%%%

\begin{figure}[tbh]
 \includegraphics[width=0.7\textwidth]{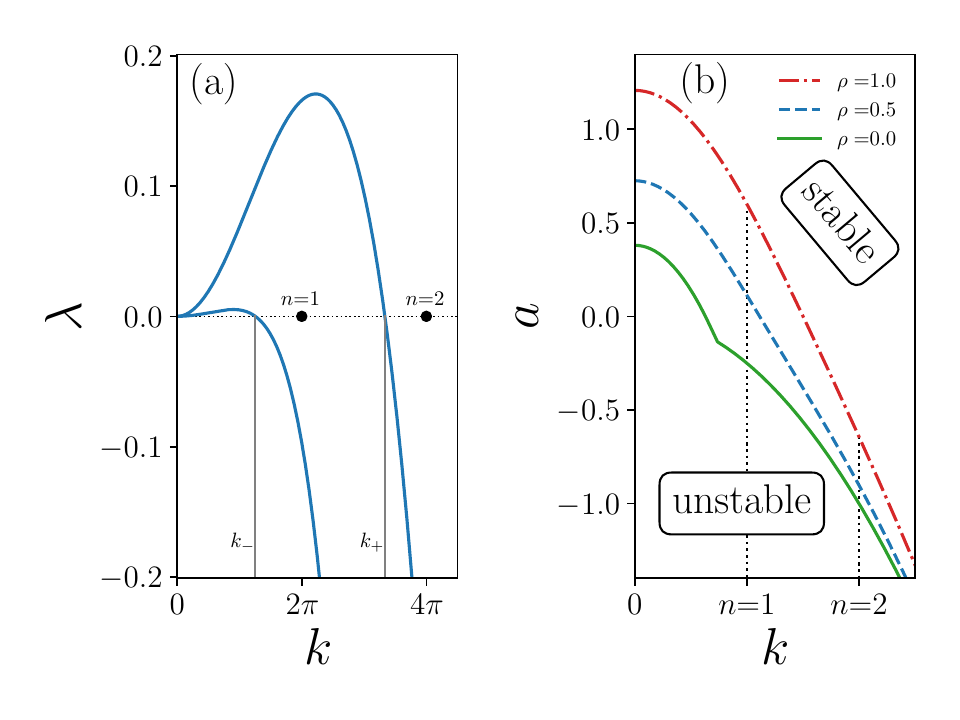}
 \caption{\small \it Linear stability behavior of homogeneous states for two passively coupled CH equations, i.e., in the variational case ($\alpha=0$). Panel (a) shows the dispersion relations $\lambda_\pm(k)$ [Eq.~\eqref{eq:lambda+-2}] at variational coupling strength $\rho = 0.5$ beyond the onset of the CH instability, i.e., here for $a=-0.55<a^{\text{CH}}$ [Eq.~\eqref{eq:a_critical}]. The respective critical wavenumbers $k_\pm$ [Eq.~\eqref{eq:k_pm}] are indicated by vertical gray lines. Panel (b) shows the stability borders $a_+(k)$ [Eq.~\eqref{eq:azero}] for three different coupling strengths $\rho=0,\, 0.5$ and $1.0$. In our scaling the selected wavenumbers are $k_n=2n\pi$. They are indicated by filled black circles in (a) and by vertical dotted lines in (b). The remaining parameters are $a_\Delta=-0.38\,,\,\, \kappa=3.82\,,\,\, \bar{\phi}_1=0.0\,,\,\, \bar{\phi}_2=0.0\,,\,\,\ell=4\pi$ and $Q=1$.}
\label{fig:dispersion} 
\end{figure}

In the variational case ($\alpha=0$ in Eqs.~\eqref{eq:nondim_bG_final}) the free energy is a Lyapunov functional, the discriminant in Eqs.~\eqref{eq:lambda+-} is always positive, all eigenvalues are real and instabilities are always stationary as for all gradient dynamics systems. A typical dispersion relation where both eigenvalues show bands of unstable wavenumbers is given in Fig.~\ref{fig:dispersion}~(a).

% %
% We set $\lambda_+(k)=0$ and find that $k=0$ is always a root (due to the conservation laws) and a second root $k_0$ occurs at 
% \begin{align}
% a_0(k_0)= & \frac{1}{2}\Bigg[-\left(\left(\frac{k_0}{L}\right)^2\left(1+\kappa\right)+a_\Delta+3 \left(\bar{\phi}_1^2+\bar{\phi}_2^2\right)\right)~\nonumber ~\\
%  & \left.+\sqrt{\left[\left(\frac{k_0}{L}\right)^2\left(1-\kappa\right)+3\left(\bar{\phi}_1^2- \bar{\phi}_2^2\right) - a_\Delta\right]^2+4 \rho ^2}\right]\, .\label{eq:azero}
% \end{align}
%
Stability borders $a_+(k)$ [see Eq.~\eqref{eq:azero}] for various variational coupling strengths $\rho$ are given in Fig.~\ref{fig:dispersion}~(b). They always show a single maximum at zero wavenumber, i.e.~the critical wavenumber is $k_\textrm{c}=0$. This shows that the variationally coupled system only exhibits CH instabilities as already concluded in the previous Section. An increase in the coupling strength acts destabilizing as it moves the instability onset $a^{\text{CH}}$ [Eq.~(\ref{eq:a_critical})] to higher temperatures and broadens the band of unstable wavenumbers.

The sign of $\rho$ does not influence the range and strength of instability, however, it influences the character of the resulting structures as it determines whether in-phase ($\rho>0$) or anti-phase ($\rho<0$) modulations of the two fields are favored.
Overall, in the case of passive coupling the CH instability of the one-field CH equation also characterizes the two-field case.
%Then, it can be expected that coarsening always prevails at large times.\tobiasbf{passt diese Aussage zu unseren Ergebissen zu nonlinear suppression of coarsening? Ich wuerde sie streichen}
%
%%%%%%%%%%%%%%%%%%%%%%%%%%%%%%%%%%%%%%%%%%%%%%%%%%%%%%%%%%%%%%%%%%%%%%%%%%%%%%%
\subsection{Active system}\label{sec:act}
%%%%%%%%%%%%%%%%%%%%%%%%%%%%%%%%%%%%%%%%%%%%%%%%%%%%%%%%%%%%%%%%%%%%%%%%%%%%%%%
%
\begin{figure}
\includegraphics[width=0.7\textwidth]{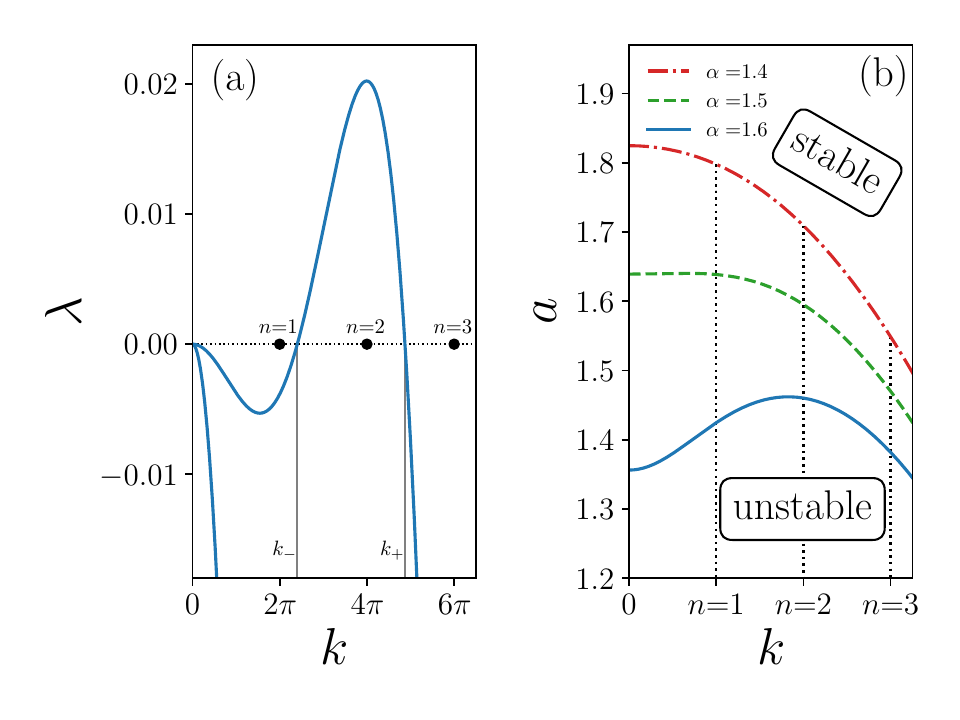}
\caption{\small \it Linear stability behavior of homogeneous states for two actively coupled CH equations, i.e., in the nonvariational case ($\alpha\neq0$). Panel (a) shows the dispersion relations $\lambda_\pm(k)$ [Eq.~\eqref{eq:lambda+-2}] at nonvariational coupling strength $\alpha=1.6$ beyond the onset of the Turing instability, i.e., here for $a=1.44<a^\text{T}$ [Eq.~\eqref{eq:a0+}]. The respective critical wavenumbers $k_\pm$ [Eq.~\eqref{eq:k_pm}] are indicated by vertical gray lines. Panel (b) shows stability borders $a_+(k)$ [Eq.~\eqref{eq:azero}] for three different coupling strengths $\alpha=1.4,\, 1.5$ and $1.6$. Parameters are $\rho=1.35\,,\,\,a_\Delta=-1.9\,,\,\,\kappa=0.14\,,\,\,\bar{\phi}_1=0\,,\,\,\bar{\phi}_2=0\,,\,\,\ell=4\pi$ and $Q=1$. The remaining symbols and lines are as in Fig.~\ref{fig:dispersion}.}
\label{fig:lin_supp_dispersion}
\end{figure}

In the nonvariational case, i.e., at $\alpha \neq 0$ no Lyapunov functional exists, i.e., no energy minimization guides the dynamics. As a result, oscillatory behavior can occur, as indicated by complex eigenvalues. We will also see, that furthermore one encounters a linear complete suppression of coarsening, i.e., already the linear results can indicate that no coarsening at all may occur.

As discussed in Sec.~\ref{sec:linear} the linear behavior for weak nonvariational coupling $|\alpha|<|\rho|$ is qualitatively equal to the CH instability of the variational case [Fig.~\ref{fig:dispersion}].
The emergence of the maximum at finite $k=k_\mathrm{c}\neq 0$ in the stability border $a_+(k)$ [cf.~Eq.~\eqref{eq:kcTuring}] marks the transition from CH to a Turing instability, see Fig.~\ref{fig:lin_supp_dispersion}~(b). For $\alpha=1.4$ (red line) the linear behavior is a CH instability. Increasing the nonvariational coupling to $\alpha=1.5$ (green line) one observes a wide $k$-range where the stability border is nearly horizontal marking the transition to the Turing instability. A maximum at $k_\text{c} \neq 0$ is fully formed for $\alpha=1.6$ (blue line). 
Fig.~\ref{fig:lin_supp_dispersion}~(a) presents a corresponding dispersion relation for $a=1.44$. There, only a band of wavenumbers bound away from $k=0$ shows positive growth rates.
This linear transition can result in a suppression of coarsening. We will investigate it in Sec.~\ref{sec:nonlinear-nonvari} in its relation to the fully nonlinear dynamic behavior and the resulting steady states.

\begin{figure}
\includegraphics[width=0.45\textwidth]{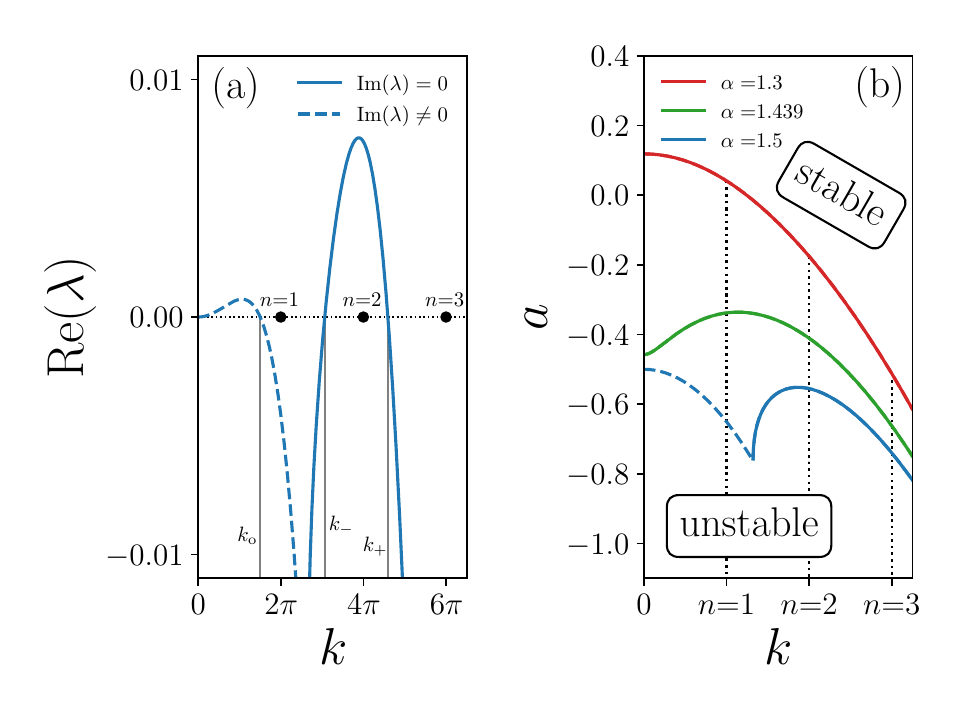}\includegraphics[width=0.45\textwidth]{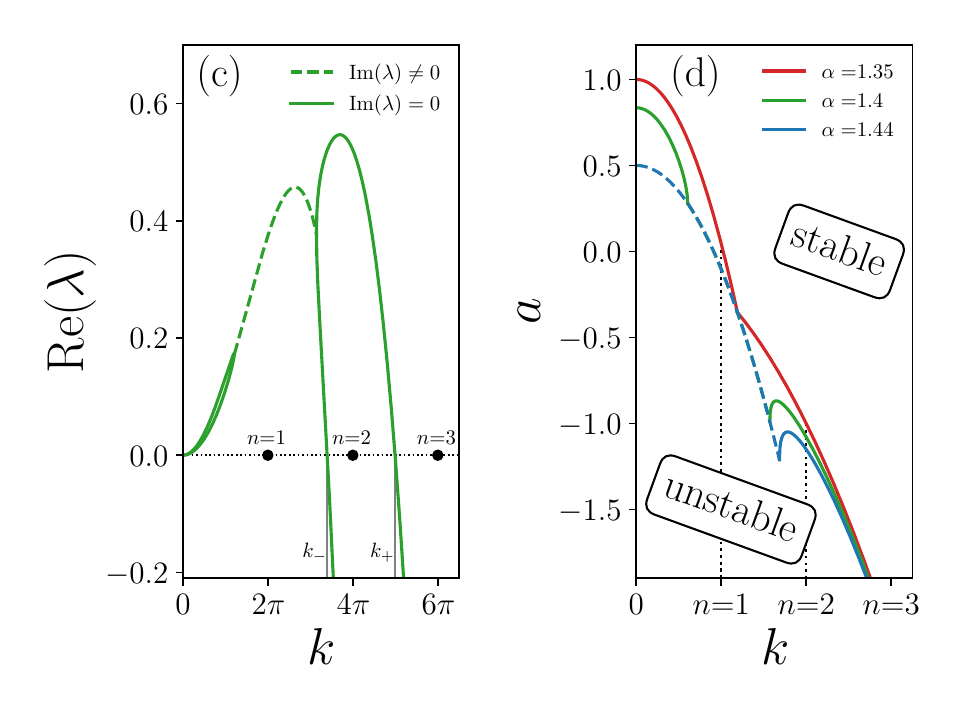}
\caption{\small \it The occurrence of oscillatory linear instability modes for two actively coupled CH equations ($\alpha\neq0$).
Panels (a) and (c) give dispersion relations [Eq.~\eqref{eq:lambda+-2}] and panels (b) and (d) the respective associated stability boundaries [Eqs.~\eqref{eq:azero},\eqref{eq:azero-os}] for two qualitatively different cases. In (a,b) complex eigenvalues occur in a band starting at $k=0$ while in (c,d) they occur in a wavenumber band away from zero [cf.~Eq.~\eqref{eq:kband_os}]. Only the real part of the eigenvalues are shown, indicating complex (real) eigenvalues by dashed (solid) lines. Panel (b) illustrates the transition from a CH instability ($\alpha=1.3$) via a Turing instability ($\alpha=1.439$) to a Hopf instability ($\alpha=1.5$). The dispersion relation in panel (a) corresponds to $\alpha=1.5$ at $a=-0.585<a^\text{T}~\text{[Eq.~\eqref{eq:a0+}]}<a^\text{H}$~[Eq.~\eqref{eq:ahopf}]. The remaining parameters for panels (a) and (b) are $\rho=1.35\,$, $a_\Delta=1\,$, $\kappa=3.82\,$, $\bar{\phi}_1=0\,$, $\bar{\phi}_2=0\,$, $\ell=8\pi$ and $Q=1$. In panel (d) a band of complex eigenvalues appears for $\alpha=\rho=1.35$ (red line) at $k\gtrapprox k_1$ and widens with increasing $\alpha$ until its left limit reaches $k=0$ at $\alpha=1.44$ (blue line). 
The dispersion relation in panel (c) corresponds to $\alpha=1.4$ at $a=-1.6<a^{\text{CH}}$~[Eq.~\eqref{eq:a_critical}]. The remaining parameters for panels (c) and (d) are $\rho=1.35\,$, $a_\Delta=-1\,$, $\kappa=3.82\,$, $\bar{\phi}_1=0\,$, $\bar{\phi}_2=0\,$, $\ell=4\pi$ and $Q=1$. 
For both dispersion relations the respective critical wavenumbers $k_\pm$ [Eq.~\eqref{eq:k_pm}] of real and $k_\text{o}$ [Eq.~\eqref{eq:k_os}] of complex roots are indicated by vertical gray lines.
The remaining symbols and lines are as in Fig.~\ref{fig:dispersion}.
}
\label{fig:cEV_dispersion} 
\end{figure}

Beside the described transition from CH to Turing instability, the nonvariational coupling can also cause oscillatory behavior if $|\alpha|>|\rho|$.
Figure \ref{fig:cEV_dispersion} shows two qualitatively different cases: Panels (a) and (b) give a dispersion relation and stability boundaries, respectively, when complex eigenvalues appear in a band starting at zero wavenumber. In particular, panel (b) shows 
how with increasing nonvariational coupling, first, a transition occurs from a CH instability ($\alpha=1.3$) as in Fig.~\ref{fig:dispersion} to a Turing instability ($\alpha=1.439$) as in Fig.~\ref{fig:lin_supp_dispersion}. Then, a further increase in $\alpha$ results in the appearance of a band of oscillatory modes at $k=0$ that extends towards larger $k$ and always represents a large-scale instability ($\alpha=1.5$). Depending on the specific value of $a$ the Hopf or the Turing instability can be dominant, i.e., have the larger maximal growth rate.
% However, these do not yet have a decisive influence on the linear stability, since the eigenvalue of the maximum of the stability limit remains real. Only with an even larger $\alpha$ this real maximum shrinks again and is suppressed by a large-scale instability with complex eigenvalues (blue line, $\alpha=1.5$).
The dispersion relation for $\alpha=1.5$ and $a=-0.585$ in Fig.~\ref{fig:cEV_dispersion}~(a) illustrates the latter case with dominant Turing instability.
%shows two maxima, one associated with the Hopf instability (dashed line) and one associated with the Turing instability (solid line).
Note that the Turing instability at intermediate $\alpha$ in Fig.~\ref{fig:cEV_dispersion}~(b) is not always part of the transition scenario from CH to Hopf instability. %The model never shows an oscillatory small-scale instability.\tobias{we have said that in the general part...delete?}

Panels (c) and (d) of Fig.~\ref{fig:cEV_dispersion} illustrate the second way how oscillatory modes can appear, namely, in a wavenumber band bound away from $k=0$. In panel (d), the red line for $\alpha=\rho=1.35$ shows the stability border when all modes are still real and the CH instability occurs. As soon as $\alpha>\rho$ , e.g., at $\alpha=1.4$ (green line), a band of oscillatory modes occurs. Since the maximum of the stability boundary remains at $k=0$ and the eigenvalues at small wavenumbers remain real, at onset (at $a\approx0.8$) one still has a CH instability. If we consider the dispersion relation in panel (c) for $\alpha=1.4$ and $a=-1.6$ (far above onset), we see that although the global maximum of the growth rate corresponds to a stationary mode, the band of oscillatory modes begins nearby and contains another (though lower) maximum. This can indicate that oscillatory behavior influences the long-time nonlinear behavior.
% Therefore, it can be expected that time-periodic behavior becomes important.\tobiasbf{Darauf beziehen wir uns nie wieder, weiß nicht wie sinnvoll der letzte Satz dann ueberhaupt ist.}
Furthermore, the band of complex eigenvalues with positive real parts causes both real eigenvalues $\lambda_\pm(k)$ to be positive at small $k$.

Further increasing $\alpha$, the band of complex eigenvalues widens. Its lower border reaches $k=0$ when $3\left(\bar{\phi}_1^2 -  \bar{\phi}_2^2\right) - a_\Delta < 2\sqrt{\Delta}$ and the CH instability becomes a Hopf instability [cf.~Eqs.~\eqref{eq:3casesA},~\eqref{eq:3cases}]. The impact of complex eigenvalues and the onset of time-periodic behavior in the fully nonlinear regime is discussed in Sec.~\ref{sec:timeperiodic}. 

%%%%%%%%%%%%%%%%%%%%%%%%%%%%%%%%%%%%%%%%%%%%%%%%%%%%%%%%%%%%%%%%%%%%%%%%%%%%%%%
\section{Variational case: Coarsening}\mylab{sec:nonlinear-vari}
%%%%%%%%%%%%%%%%%%%%%%%%%%%%%%%%%%%%%%%%%%%%%%%%%%%%%%%%%%%%%%%%%%%%%%%%%%%%%%%
%
\begin{figure}
\includegraphics[width=0.45\textwidth]{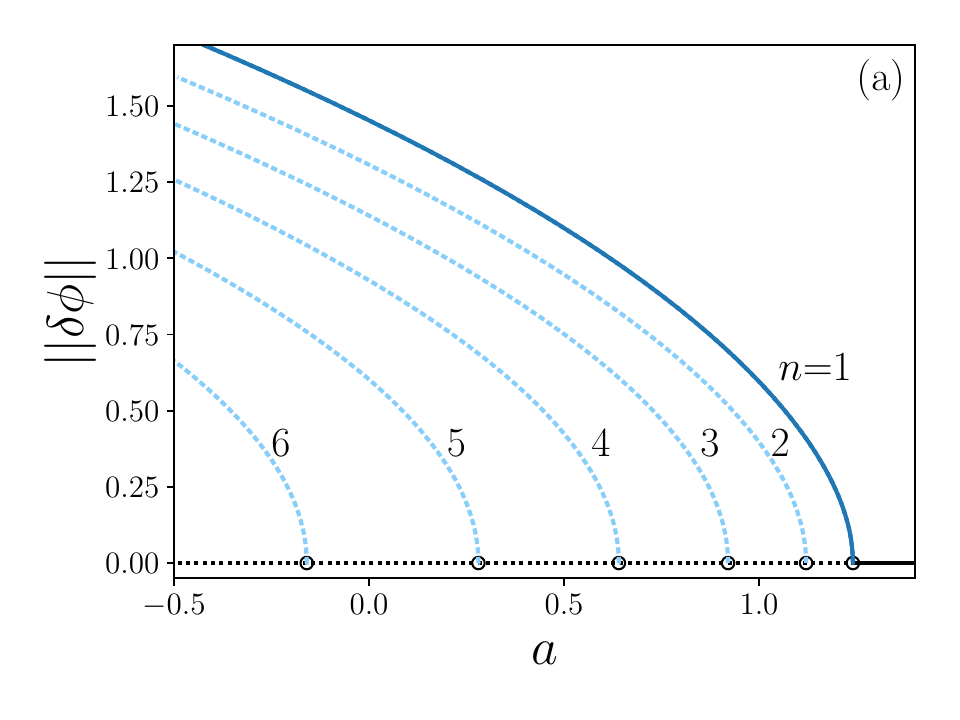}
\includegraphics[width=0.45\textwidth]{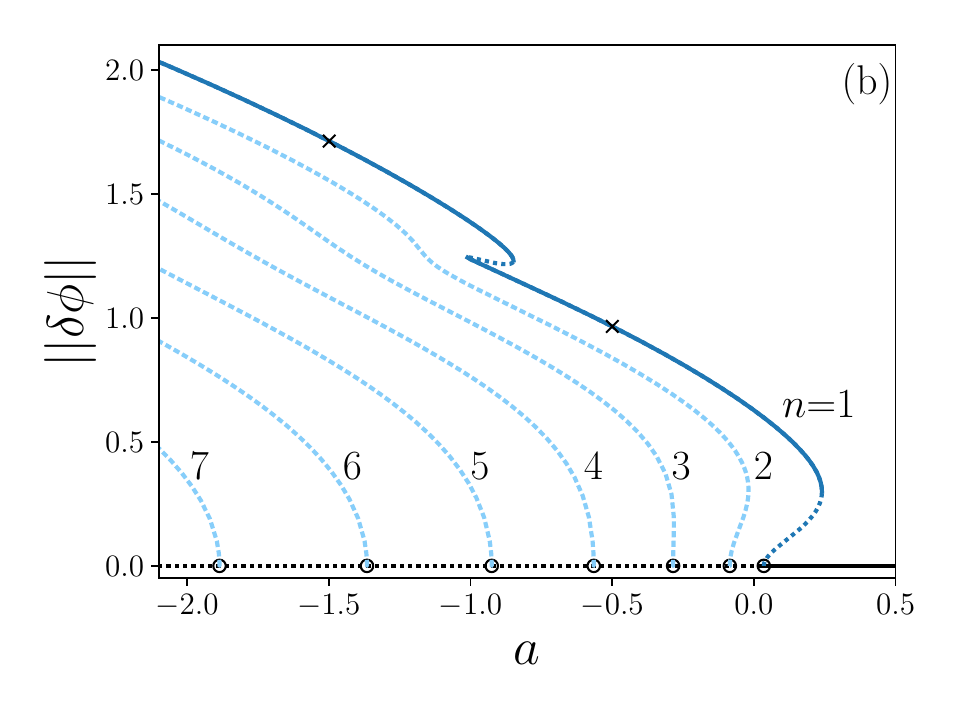}\\
\includegraphics[width=0.45\textwidth]{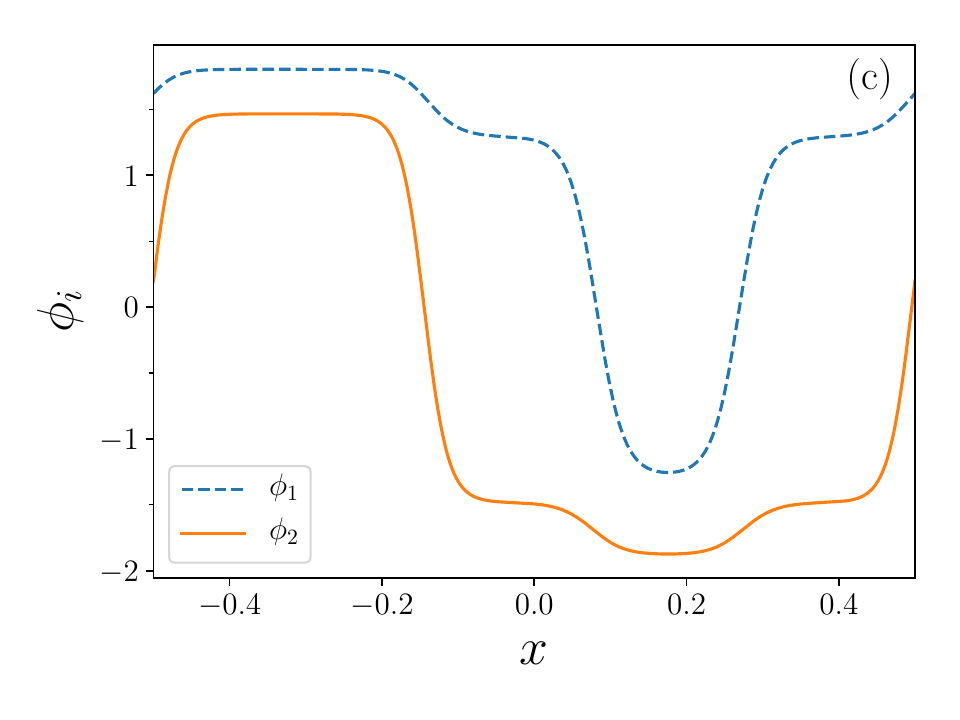}
\includegraphics[width=0.45\textwidth]{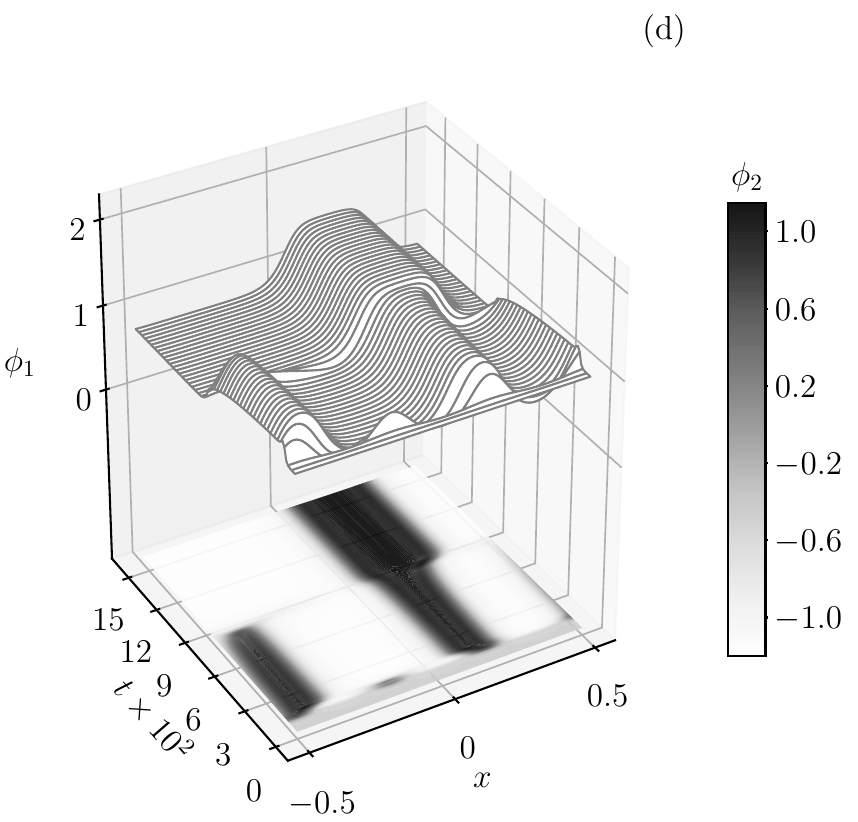}
\caption{\small \it Panels (a) and (b) show bifurcation diagrams of steady states for the ternary system in the variational case employing the effective temperature $a$ as control parameter at fixed $\ell=10\pi\,$, $\kappa=1\,$, $a_\Delta =-0.5\,$, $\rho=1$ and $Q=1$. Panel (a) represents the supercritical case for mean concentrations $\bar{\phi}_1=\bar{\phi}_2=0$, panel (b) the subcritical one for $\bar{\phi}_1=1\,$, $\bar{\phi}_2=-0.5$. Solid (dashed) lines indicate linearly stable (unstable) states. The black horizontal line represents the homogeneous state. The various blue lines represent phase-separated states with different numbers of phase-separated structures, i.e., periods. Panel (c) gives the profiles of the linearly stable state indicated by a cross symbol in (b) at $a=-1.5$. Panel (d) is a space time plot illustrating the coarsening dynamics which finally results in a fully phase-separated state indicated by a cross symbol in (b) at $a=-0.5$.}
\label{fig:GD_bif} 
\end{figure}

%\begin{figure}[h!]
%\includegraphics[width=\textwidth]{./GD-0,5}
%\includegraphics[width=\textwidth]{./GD-1,2}
%\caption{\small \it Space-time plots of selected time simulations illustrate the emergence of the $n=1$ states of the bifurcation diagram Fig.~\ref{fig:GD_bif}~(a) after some coarsening. Left and right panels show fields $\phi_1(x,t)$ and $\phi_2(x,t)$, respectively. Row~(a) shows the emergence of two-phase coexistence at $a=-0.5$ while in row (b) a three-phase state emerges at $a=-1.2$. The domain size is $L=10\pi$ and the remaining parameter are as in Fig.~\ref{fig:GD_bif}.}
%\label{fig:GD_DN} 
%\end{figure}

We begin the nonlinear study with a brief overview of typical coarsening dynamics and steady states behavior in the variational case ($\alpha=0$). 
 Along the lines of Ref.~\cite{TFEK2019njp} the behavior in finite domains presented here can be linked to the phase behavior in the thermodynamic limit, i.e.~in an infinite  domain, as considered in appendix~\ref{app:phasebehavior}. In particular, Fig.~\ref{fig:phase_diagram} presents phase diagrams in the $(\phi_2,\phi_1)$- and $(\mu_2,\mu_1)$-planes and relates them to bifurcation diagrams. Here, we focus on $a$ as main control parameter due to its importance in the nonvariational case studied below.  The bifurcation diagrams in Fig.~\ref{fig:GD_bif}~(a) and (b) show the norm [Eq.~\eqref{eq:norm}] as a function of $a$ at fixed domain size $\ell=10\pi$ and $\bar\phi_i$. 
Panel (a) considers the case $\bar\phi_1=\bar\phi_2=0$ where all primary branches emerge supercritically. It represents the simplest conceivable bifurcation behavior for the system.
 As for the one-field model \cite{TFEK2019njp}, with decreasing $a$ the uniform state becomes unstable at about $a=1.3$ where the completely phase-separated, i.e., fully coarsened, state emerges.
Decreasing $a$ further, the uniform state becomes successively unstable with respect to higher order modes and corresponding branches emerge in pitchfork bifurcations. We label the different branches by the spatial periodicity $n$ of the corresponding states. All states with $n>1$ are unstable and in time will coarsen into $n=1$ states.
Panel (b) illustrates that $\bar\phi_i\neq0$ can result in subcritical behavior: In the shown case, the $n=1$ to $n=3$ branches emerge towards larger $a$ before turning back at respective saddle-node bifurcations.
In particular, the $n=1$ branch emerges with unstable profiles (nucleation thresholds, analogue to \cite{Novi1985jsp,TNPV2002csa-pea}) and stabilizes at the saddle-node bifurcation at $a\approx0.2$. 
The resulting stable $n=1$ states first show a coexistence of two phases related to the binodals discussed in appendix \ref{app:phasebehavior}. The time simulation at $a=-0.5$ in panel~(d) illustrates how such a phase-separated state is reached dynamically when starting with a homogeneous state with added white noise of small amplitude $5 \cdot 10^{-3}$. %If not stated otherwise this is done in all simulations throughout the paper.
The two-phase $n=1$ state develops after coarsening via volume modes from an $n=3$ state. Comparing with the phase diagram in Fig.~\ref{fig:phase_diagram}~(c) the fully phase-separated state is identified as a coexistence of high-$\phi_1$, high-$\phi_2$ phase~I and the high-$\phi_1$, low-$\phi_2$ phase~IV.

Further following the $n=1$ branch with decreasing $a$, it eventually undergoes another pair of saddle-node bifurcations, thereby passing through a short sub-branch of unstable states ($-1.1<a<-0.6$) before stabilizing again. The corresponding hysteresis loop is related to the nucleation of a third phase (here, phase~III: low-$\phi_1$, low-$\phi_2$) that emerges at the center of the phase~IV plateau. The remaining part of the $n=1$ branch shows well-developed three-phase coexistence of phases~I, III and IV, illustrated by the solution profiles in Fig.~\ref{fig:GD_bif}~(c) for $a=-1.5$. Here, after coarsening an $n=1$ three-phase state emerges as the system is in the parameter region corresponding to the triple point [Fig.~\ref{fig:phase_diagram}~(c)]. We note that the plateau concentrations are already relatively close to the corresponding values in the thermodynamic limit. An increase in domain size will result in full convergence. The unstable state existing in the hysteresis range corresponds to a threshold state that has to be overcome to switch between the two- and three-phase coexistence.
The existence of such a hysteresis loop depends on the other parameters, e.g., decreasing the domain size it eventually vanishes in a hysteresis bifurcation.
\section{Nonvariational case: Suppression of coarsening}\mylab{sec:nonlinear-nonvari}
%%%%%%%%%%%%%%%%%%%%%%%%%%%%%%%%%%%%%%%%%%%%%%%%%%%%%%%%%%%%%%%%%%%%%%%%%%%%%%%
%
Next, we increase the strength of nonvariational coupling $\alpha$ from zero and investigate how breaking the gradient dynamics structure changes the coarsening behavior.
For clarity, we first define the three different types of suppression of coarsening that are discussed in this Section:
\begin{itemize}
\item \textbf{Linear complete suppression of coarsening}: Already the linear stability analysis (Section~\ref{sec:linear}) indicates that stable patterns of finite wavelengths emerge and no coarsening takes place. 
This occurs if only certain modes with $n>1$ are present in the band of unstable wavenumbers of a Turing instability. 
%This occurs if only certain modes with $n>1$ are present in the band of unstable wavenumbers of a Turing instability. Furthermore, the fully phase-separated state is linearly unstable \tobias{near its birth, i.e., when the $n=1$ mode enters the unstable band}, resulting in ``reverse coarsening''. %also called ``mesa splitting''.
\item \textbf{Nonlinear complete suppression of coarsening}: The linearly fastest growing mode is of finite wavelength and dynamically develops into a stable pattern of the same spatial periodicity without any coarsening ($n\neq1$). This occurs in regions of multistability where several steady states are linearly stable including the fully phase-separated ($n=1$) state. It can be observed for both Turing and CH instabilities. Ultimately, the behavior can only be predicted in a fully nonlinear analysis.
%\tobias{Delete:As with the first mechanism,} the linearly fastest growing mode is of finite wavelength and dynamically develops into a stable pattern of the same spatial periodicity without any coarsening. However, the linear instability of the homogeneous state is either small-scale with the $n=1$ mode within the unstable wavenumber band or large-scale \tobias{which means that in both cases the $n=1$ mode has entered the unstable band before the now fast growing mode.} Ultimately, the behavior can only be predicted in a fully nonlinear analysis.
\item \textbf{Nonlinear partial suppression of coarsening}: Here, the linearly fastest growing mode develops but is unstable with resepct to (w.r.t.) coarsening. However, coarsening is arrested before reaching the $n=1$ state. The conditions of multistability and stable $n=1$ state are as in the second case. 
Again, the behavior can only be predicted in a fully nonlinear analysis.
%The linear behavior is as in the second case, but the linearly fastest growing mode undergoes some coarsening steps, then coarsening is arrested before reaching the fully phase-separated state. \tobias{Primary?} and secondary bifurcations have to be known to predict this case.
\end{itemize}
These three types stand for three different mechanisms that can transform perpetual coarsening into pattern formation.

Our analyses reveal that most qualitative changes as compared to the variational case occur for a nonvariational coupling stronger than the variational one. Therefore, we now focus on $|\alpha|>|\rho|$, i.e., $\Delta>0$. In appendix~\ref{sec:app:small} we consider the instructive limiting case without variational coupling, i.e., $\rho=0$ and $\alpha\neq0$.

The linear analysis in Section \ref{sec:act} shows that the nonvariational coupling in a two-field CH model can induce a Turing instability that does neither occur for variational CH models nor for the studied nonvariational one-field CH models. Next, we employ time simulations and a bifurcation analysis to explore the resulting consequences for the fully nonlinear regime. As super- and subcritical behaviors at primary bifurcations differ in their influence on the coarsening behavior we consider these cases separately in Sections~\ref{sec:super} and \ref{sec:subcrit}, respectively. In passing, we show that subcritical primary bifurcations may occur even without quadratic nonlinearity, i.e., here at mean concentrations $\bar{\phi}_1=\bar{\phi}_2=0$. Such behavior is unknown for the classical one-field CH equation. Surprisingly, we find that subcritical behavior acts as stepping stone to time-periodic behavior discussed below in Section \ref{sec:OnsetLargeOscill}.
%Note that in the nonvariational case we use smaller domain sizes than in the previous Section which reduces the number of branches. Hence, the bifurcation diagrams become clearer and it is easier to recognize the qualitative change of behavior. Nevertheless, we expect that the following results also hold for larger systems.

%%%%%%%%%%%%%%%%%%%%%%%%%%%%%%%%%%%%%%%%%%%%%%%%%%%%%%%%%%%%%%%%%%%%%%%%%%%%%%%
\subsection{The supercritical case} \label{sec:super}
%%%%%%%%%%%%%%%%%%%%%%%%%%%%%%%%%%%%%%%%%%%%%%%%%%%%%%%%%%%%%%%%%%%%%%%%%%%%%%%

\begin{figure}
\includegraphics[width=0.45\textwidth]{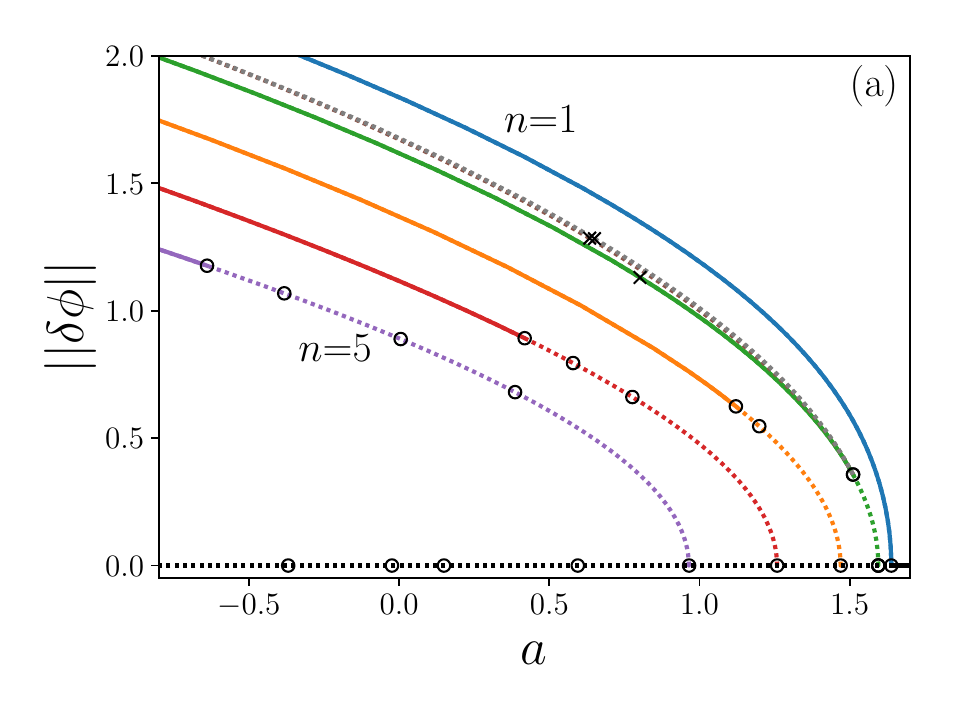}
\includegraphics[width=0.45\textwidth]{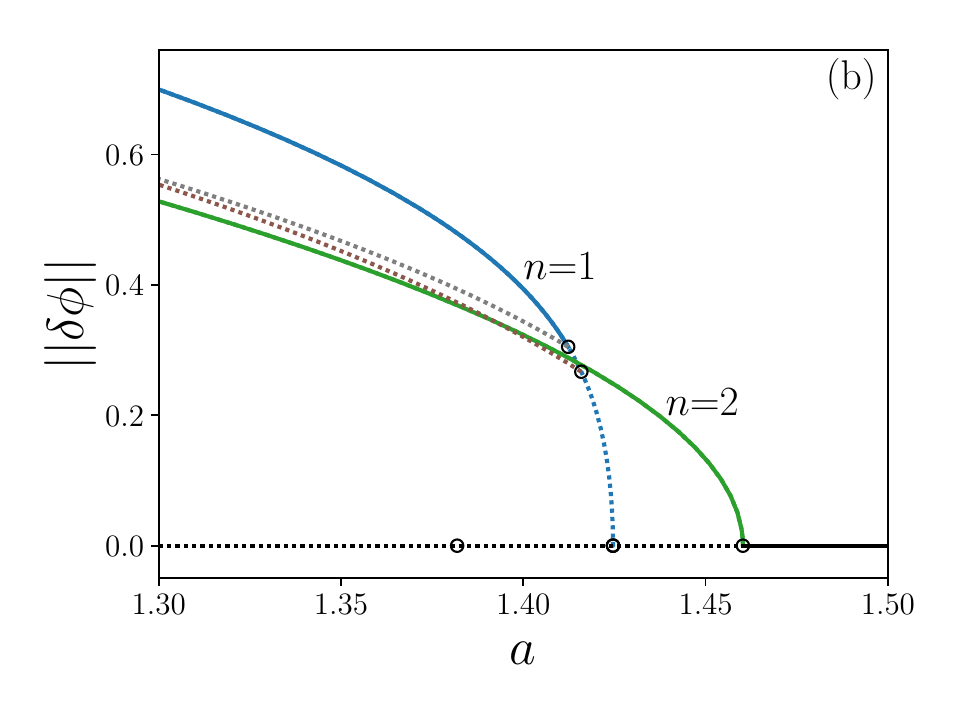}~\\
\includegraphics[width=0.3\textwidth]{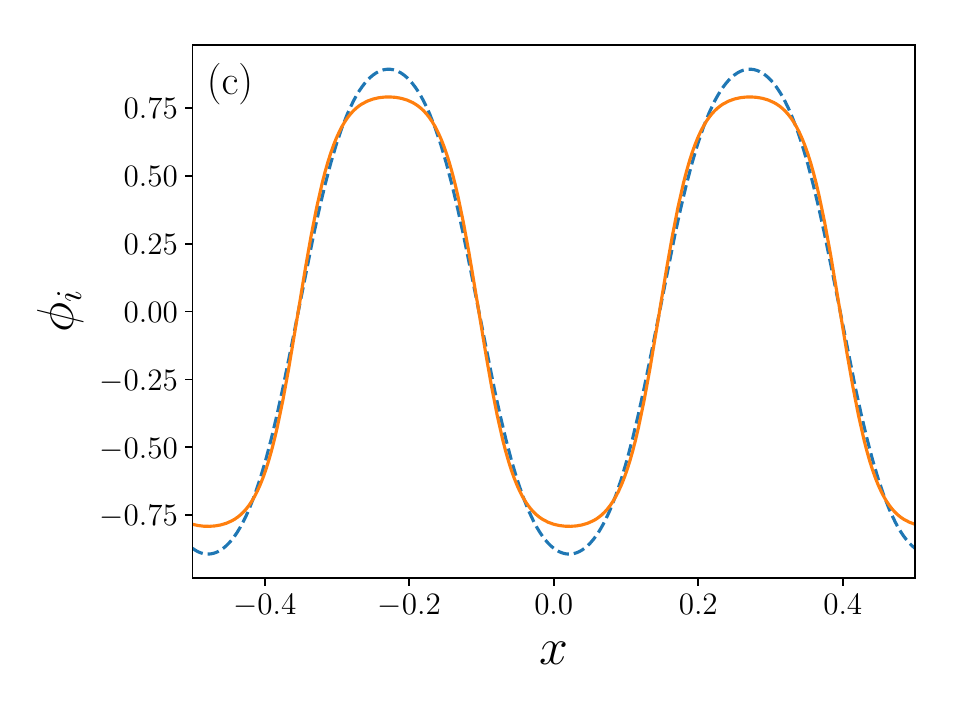}
 \includegraphics[width=0.3\textwidth]{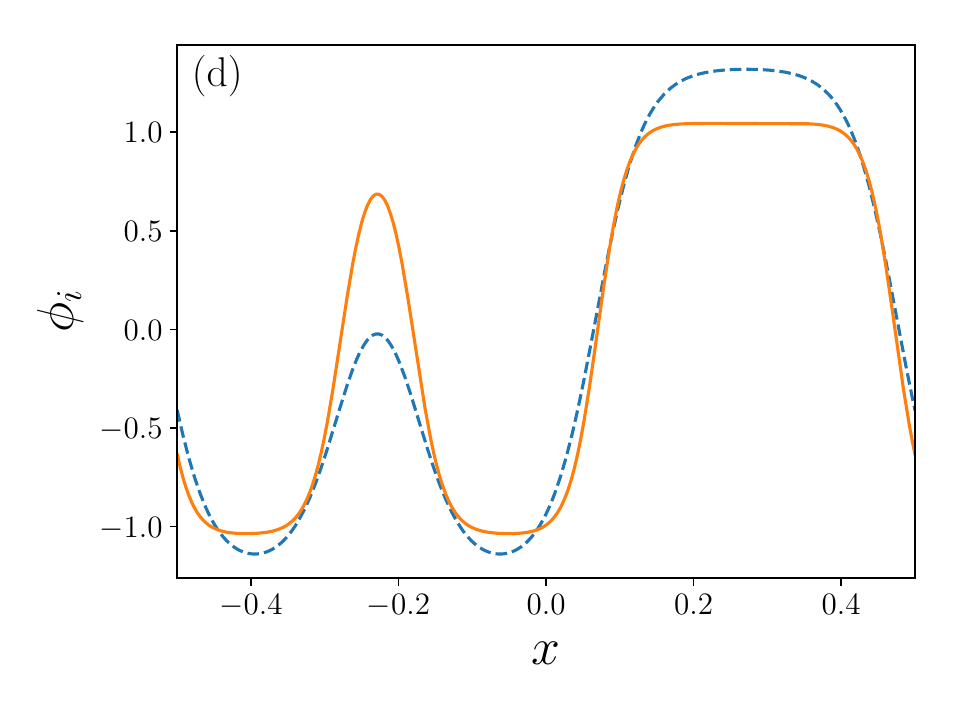} 
 \includegraphics[width=0.3\textwidth]{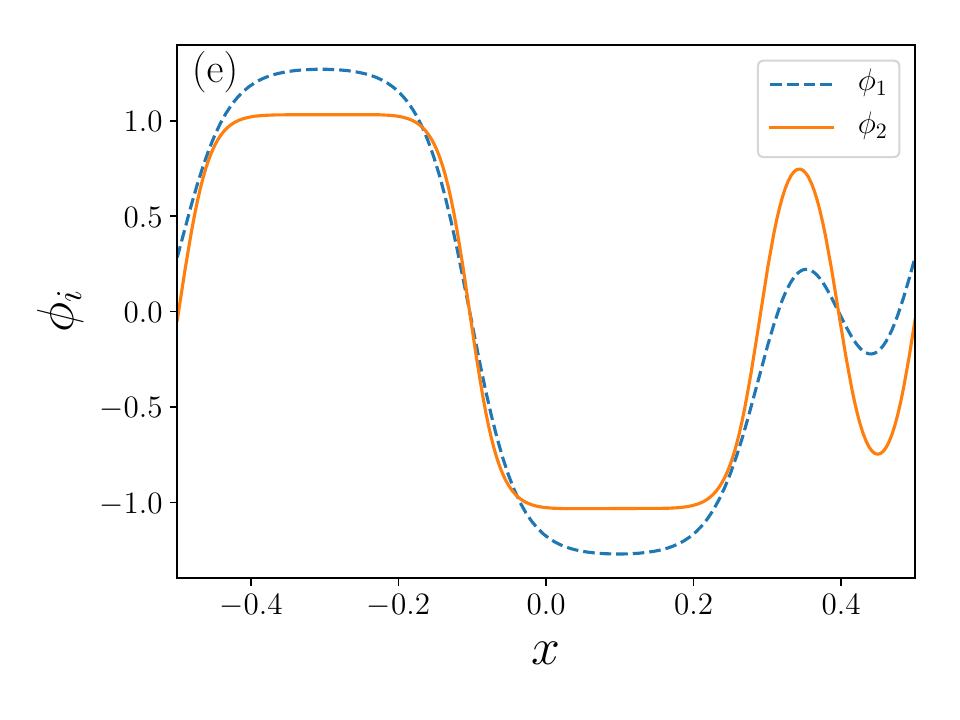}
\caption{\small \it Bifurcation diagrams related to the suppression of coarsening for the nonvariationally coupled CH model [Eq.~\eqref{eq:nondim_bG_final} with $\alpha\neq0$]. Shown is the norm as a function of the parameter $a$ in the supercritical case. The nonvariational coupling strengths (a) $\alpha=1.5$ and (b) $\alpha=1.6$ are larger than the variational one $\rho=1.35$.
In (b) only the fully phase-separated state ($n=1$, blue line) and the two-period state ($n=2$, green line) are shown while in (a) states up to $n=5$ are included. Selected side branches are also given. Circle symbols mark primary and secondary pitchfork bifurcations. The remaining line styles and parameters are as in Fig.~\ref{fig:lin_supp_dispersion}. The lower panels show profiles of (c) stable and (d,e) unstable steady states at loci marked by crosses in panel (a). }
\label{fig:linsupp_bif_super} 
\end{figure}
It is instructive to first consider the bifurcation behavior of steady states in the purely CH and Turing cases in Figs.~\ref {fig:dispersion} and \ref{fig:lin_supp_dispersion}, respectively. In both cases, we use $\alpha>0$ and consider parameter values where both eigenvalues are still real. 

The resulting bifurcation behavior for two values of $\alpha$ is presented in Fig.~\ref{fig:linsupp_bif_super} again using $a$ as control parameter. Fig.~\ref{fig:linsupp_bif_super}~(a) with $\alpha=1.5$ corresponds to parameters corresponding to the green line in Fig.~\ref{fig:lin_supp_dispersion}~(b), and Fig.~\ref{fig:linsupp_bif_super}~(b) with $\alpha=1.6$ belongs to the dispersion curve in Fig.~\ref{fig:lin_supp_dispersion}~(a) and the blue line in Fig.~\ref{fig:lin_supp_dispersion}~(b). Branches emerging at primary bifurcations from the uniform state are named by the periodicity $n$ of the corresponding decomposition pattern as before.

As expected based on the linear result, when decreasing $a$ in Fig.~\ref{fig:linsupp_bif_super}(a) the $n=1$ state bifurcates first, corresponding to a CH instability. The bifurcation is a supercritical pitchfork as all other considered primary bifurcations. In consequence, the shown $n=2$ (green line) to $n=5$ (purple line) states inherit two to eight unstable eigenvalues from the uniform state since the eigenvalues of the uniform state are all double (note that there is no translation mode as the uniform state itself is translational invariant). Then when an inhomogeneous state emerges a double eigenvalue crosses zero and the emerging branch acquires a zero eigenvalue (due to translation symmetry) beside the inherited negative (supercritical) or positive (subcritical). In contrast to the variational case, where no secondary bifurcations exist and all $n>1$ states are always unstable, here, they eventually stabilize at secondary pitchfork bifurcations. In the weakly nonvariational case which we define as $|\alpha|<|\rho|$ we do observe secondary bifurcations (not shown). They always occur in pairs of one destabilizing and one stabilizing bifurcation related to higher order modes and do not result in the appearance of further stable states as observed for $|\alpha|>|\rho|$.

In particular, the $n=2$ state [cf.~Fig.~\ref{fig:linsupp_bif_super}~(c)] stabilizes through a degenerate pitchfork bifurcation where two real eigenvalues cross zero and two distinct subcritical branches (brown and gray lines) simultaneously emerge towards smaller values of $a$. Note that on the scale of Fig.~\ref{fig:linsupp_bif_super}~(a) the two curves can not be distinguished by eye. Also, each branch corresponds to four states related by symmetry (see below).  Example profiles on the two secondary branches are given in Fig.~\ref{fig:linsupp_bif_super}~(d) and~(e), respectively. Both states break the discrete translational symmetry of the primary $n=2$ branch, i.e., they correspond to a spatial period doubling.
%The symmetry group of this degenerate pitchfork bifurcation is Z$_2 \times $Z$_2$ what we want to examine briefly.
The bifurcation structure can be understood considering reflection symmetries: States on the $n=2$ primary branch have two independent reflection symmetries, one with respect to their minima and another one with respect to their maxima.
  % (for periodic boundary conditions only pairs of two reflection symmetry axes can occur).
 For nonzero mean concentrations, two distinct pitchfork bifurcations correspond to the respective breaking of these symmetries (not shown). In Fig.~\ref{fig:linsupp_bif_super}~(a), $\bar\phi_1=\bar\phi_2=0$ ensures inversion symmetry and the two reflection symmetries can be identified via an inversion. That is, they are always broken together in a degenerate pitchfork (also termed Z$_2 \times $Z$_2$ bifurcation \cite{Hoyle2006}) with normal form
\begin{align*}
\dot{x}_1= \mu x_1 - b_1 x_1^3 - b_2 x_2^2 x_1~\\
\dot{x}_2 = \mu x_2 -b_1 x_2^3 - b_2 x_1^2 x_2\,.
\end{align*}
Here $x_1$ and $x_2$ refer to the two modes of symmetry breaking, e.g.~$x_1$ [$x_2$] breaks the reflection symmetry w.r.t.~the minima [maxima]. The primary $n=2$ branch (green line) is represented by $(x_1,x_2)=(0,0)$, see example profile in Fig.~\ref{fig:linsupp_bif_super}~(c). Then, there are two pairs of branches which keep either the reflection symmetry w.r.t.\ the minima or w.r.t.\ to the maxima with representations $(0,\pm\sqrt{\frac{\mu}{b_1}})$ and $(\pm\sqrt{\frac{\mu}{b_1}},0)$. One of these pairs corresponds to the profile in Fig.~\ref{fig:linsupp_bif_super}~(d) and the other one to its inversion. In Fig.~\ref{fig:linsupp_bif_super}~(a) these states correspond to the brown line. Furthermore, there are four branches which break both reflection symmetries, however keep full inversion symmetry, i.e., $(x,\phi_i) \to (-x,-\phi_i)$, see example profile in Fig.~\ref{fig:linsupp_bif_super}~(e). Their representations are $(\pm\sqrt{\frac{\mu}{b_1+b_2}},\pm\sqrt{\frac{\mu}{b_1+b_2}})$ and correspond to the gray line in panel~(a). In total there are eight simultaneously emerging secondary branches, i.e., each of the two distinct secondary branches in Fig.~\ref{fig:linsupp_bif_super}~(a) is four-fold and can be ``unfolded'' choosing adequate parameters and model amendments.

We note that a consequence of the degenerate pitchfork bifurcation is the simultaneous stabilization of both coarsening modes (volume transfer and translation). Similar stabilizations are observed for the branches of larger $n$ where, however, a sequence of several bifurcations is needed. Namely, two, three and four degenerate pitchfork bifurcations on the $n=3$, $4$ and $5$ branch, respectively, ensure that for $a\lesssim -0.6$ all $n\leq 5$ branches are linearly stable. It is intriguing that the simultaneous stabilization of translation and volume coarsening modes is generic in a wide range of parameters. Again this is a consequence of the choice $\bar\phi_1=\bar\phi_2=0$. Note that none of the studied emerging secondary branches reconnects to the primary branch.

Comparing Figs.~\ref{fig:linsupp_bif_super}~(a) at $\alpha=1.5$ and (b) at $\alpha=1.6$ we see that with the increase of $\alpha$ the first two primary bifurcations have swapped position reflecting the transition from CH to Turing instability [cf.~Section~\ref{sec:linear}]. In consequence, at the first primary bifurcation the (now linearly stable) $n=2$ state emerges supercritically. Thus, in accordance with the linear result only the patterned $n=2$ state exists. This corresponds to the linear complete suppression of coarsening. The fully phase-separated ($n=1$) state only emerges at the second primary bifurcation, supercritical but twice unstable. There, coarsening is still suppressed, expanding the concept of linear complete suppression to the case where the fully phase-separated ($n=1$) state exists but is linearly unstable. This enables ``reverse coarsening'' of the $n=1$ state into the $n=2$ state. This extended $a$-range of linear complete suppression ends where the $n=1$ state gains stability at secondary bifurcations.

It is noteworthy that when the primary bifurcations switch places, the above discussed secondary bifurcations move from the $n=2$ branch onto the $n=1$ branch [Fig.~\ref{fig:linsupp_bif_super}~(b)]. In  consequence, the two primary and two secondary bifurcations all coincide at the crossover. Four parameters, $\alpha$, $a$ and both mean concentrations $\bar\phi_1$, $\bar\phi_2$ need to be adjusted to pinpoint the corresponding codimension-4 bifurcation point. However, when the two secondary pitchfork bifurcations have switched onto the $n=1$ branch, they do not coincide anymore. The reason is that one can not anymore independently break the reflection symmetries with respect to the minimum and the maximum. As a result, the breakings of the reflection and the full inversion symmetry occur independently.
%As the $n=1$ branch  has fewer symmetries than the $n=2$ branch, the symmetry relations between primary and emerging secondary branches have changed after the switchover. The $n=1$ primary branch is reflection symmetric w.r.t.\ its single maximum and minimum, and also has the inversion symmetry. In contrast to the former case where both symmetries of the $n=2$  primary branch were related by inversion, here they are unrelated. \ttuwe{I do not see a difference to the previous case concerning the relation of reflection and inversion.}
%
Hence, the degeneration of the secondary bifurcations is lifted. The first [second] pitchfork bifurcation breaks reflection [full inversion] symmetry and pairs of branches with solutions similar to Fig.~\ref{fig:linsupp_bif_super}~(e) [Fig.~\ref{fig:linsupp_bif_super}~(d)] emerge. For any nonzero mean concentration the inversion symmetry is broken for all patterned states, and the second pitchfork bifurcation unfolds into a saddle-node bifurcation and a continuous branch (not shown).

\begin{figure}
\begin{minipage}[c]{0.3\textwidth}
% {\small (a) Linear complete suppression}\\
  \includegraphics[width= \hsize]{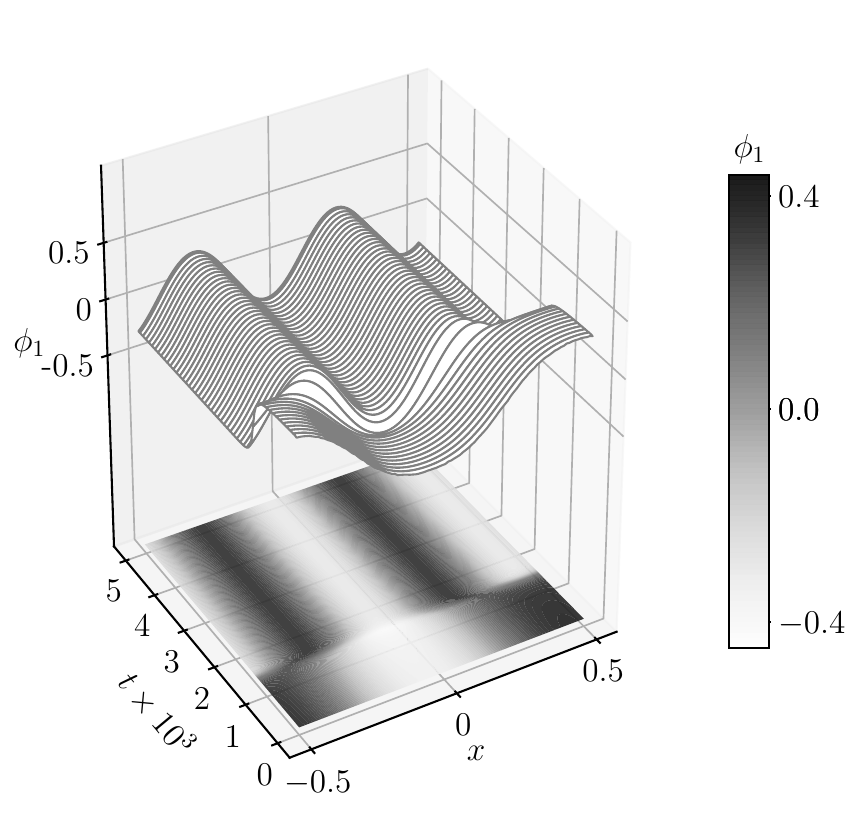}
\end{minipage}
  % \label{fig:DN_1_subfig1}
\begin{minipage}[c]{0.3\textwidth}
% {\small  (b) Nonlinear partial suppression}\\
  \includegraphics[width= \hsize]{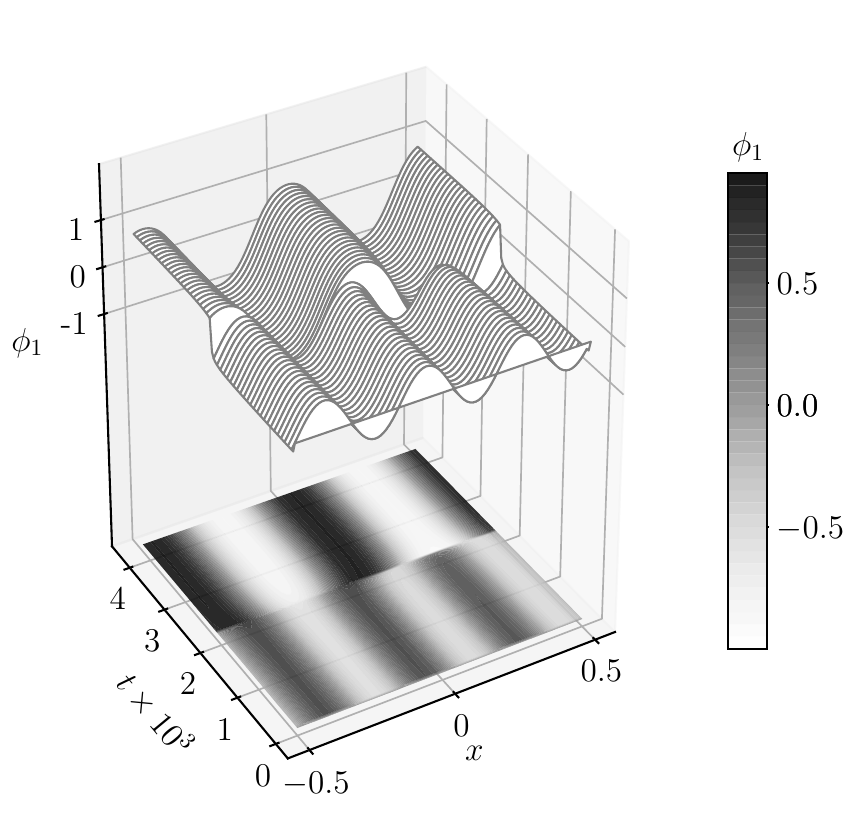}
\end{minipage}
 %   \label{fig:DN_1_subfig2}
\begin{minipage}[c]{0.3\textwidth}
 %{\small  (c) Nonlinear complete suppression}\\
  \includegraphics[width=\hsize]{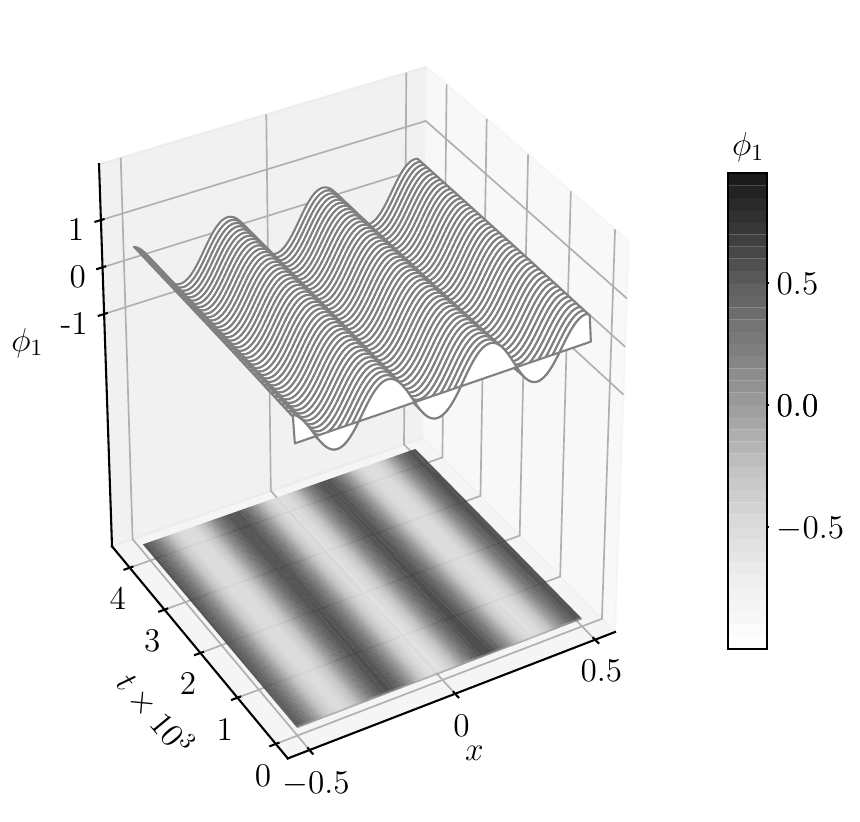}
\end{minipage}
%    \label{fig:DN_1_subfig3}
%    \includegraphics[width= 0.49 \textwidth]{./Diagramm_whitenoise_stop_coarsening_a1,15_v1.pdf}
%    \includegraphics[width=0.49 \textwidth]{./Diagramm_whitenoise_stop_coarsening_a1,1_v1.pdf}\\
%    \includegraphics[width=0.49 \textwidth]{./Diagramm_whitenoice_onestable_state.pdf}
%    \includegraphics[width=0.49 \textwidth]{./Diagramm_epsilon16_epsilon15.pdf}
\caption{\small \it Space-time plots obtained by direct numerical simulation of structuring processes in the nonvariational case. They illustrate three qualitatively different behaviors that replace the classical coarsening of the variational case: (a) splitting of the fully phase-separated state due to linear complete suppression of coarsening at $\alpha=1.6$ and $a=1.413$ [cf.~Fig.~\ref{fig:linsupp_bif_super}~(b)], (b) nonlinear partial suppression of coarsening at $\alpha=1.5$ and $a=1.12$ [cf.~Fig.~\ref{fig:linsupp_bif_super}~(a)], and (c) nonlinear complete suppression of coarsening at $\alpha=1.5$ and $a=1.11$ [cf.~Fig.~\ref{fig:linsupp_bif_super}~(a)]. For details see main text.
  % Each panel shows staggered profiles $\phi_1(x,t)$ over a gray-scale contour plot of $\phi_2(x,t)$.
}\label{fig:DN_1}
\end{figure}

As explained above, the linear suppression of coarsening in Fig.~\ref{fig:linsupp_bif_super}~(b) is only valid until the $n=1$ branch stabilizes via the two secondary pitchfork bifurcations at $a\approx 1.416$ and $a\approx 1.412$. Before this occurs, the $n=1$ branch is unstable to the $n=2$ mode resulting in splitting of the fully phase-separated state (see Fig.~\ref{fig:DN_1}~(a) as explained below). One may call the dynamical process ``reverse coarsening'' in analogy to the ``reverse Ostwald ripening'' in Ref.~\cite{TjNC2018prx}. A similar process is called ``mesa splitting'' in Ref.~\cite{BWHY2021prl}.
At lower $a\lesssim 1.412$, multistability with higher-$n$ states arises as before resulting in nonlinear partial or complete suppression.

Fig.~\ref{fig:DN_1} uses space-time plots to illustrate the discussed consequences of multistability for the coarsening dynamics. Panel~(a) focuses on a region in Fig.~\ref{fig:linsupp_bif_super}~(b) where $n=1$ and $n=2$ state both exist, but only the patterned $n=2$ state is stable. The chosen $a=1.413$ lies between the two secondary bifurcations, i.e., the $n=1$ state has one unstable eigenvalue. Starting with the $n=1$ state with added noise, we observe reverse coarsening via the mass transfer mode converging to the patterned $n=2$ state. This clearly illustrates that the nonvariational coupling can reverse the original coarsening process of a phase separating system. It is a direct result of the linear complete suppression of coarsening discussed above, because the stability of the relevant branches is a direct consequence of the linear stability of the uniform state.

Next, we consider a time evolution in the multistable $a$-range of Fig.~\ref{fig:linsupp_bif_super}~(a). % In the present nonvariational case there exists no global selection criterion like the minimization of an free energy that allows one to predict which of the linearly stable states is selected. Therefore, we resort to direct numerical simulations and give in
Figs.~\ref{fig:DN_1}~(b) and~(c) present results for $a=1.12$ and $a=1.11$, respectively. In both cases, first an $n=3$ state develops corresponding to the fastest growing linear mode. As at $a=1.12$ the $n=3$ state is still unstable, a single coarsening step occurs in Fig.~\ref{fig:DN_1}~(b). It results in the linearly stable $n=2$ state where coarsening is arrested. This corresponds to the nonlinear partial suppression of coarsening.
%\footnote{%
%Note that at the parameters of Fig.~\ref{fig:DN_1}~(b) one may also start with a large-amplitude $n=1$ mode. Then the system evolves into the linearly stable $n=1$ state (not shown) as expected in a multistable region.}
In contrast, at the slightly smaller $a=1.11$ [Fig.~\ref{fig:DN_1}~(c)] the now linearly stable $n=3$ state forms and no coarsening occurs. This corresponds to nonlinear complete suppression of coarsening as it depends on the sequence of secondary bifurcations. In contrast to the linear suppression it can not be deduced from a linear analysis of the homogeneous state and does not cause reverse coarsening.

To summarize, the bifurcation diagram and simulation results show that partial or complete suppression of coarsening can occur even if the dispersion relation for the uniform state indicates a CH instability and one would naturally predict coarsening. The underlying mechanism is nonlinear and can be characterized as follows. In the common coarsening process in the variational system, clusters of the same phase merge over time minimizing the overall interface energy. Their number successively decreases until the fully phase-separated state is reached. 
%The coarsening process slows down with increasing cluster size that follows a power law - the Lifshitz-Slyozow-Wagner law $t^{\nu}$ with $\nu=1/3$ for two- and three-dimensional systems. For one-dimensional systems the growth is logarithmic \ttuwe{need specific citations}. 
%
This implies that the eigenvalues of all coarsening modes become very small for states with a small number of clusters, but they always remain positive. Here, the nonvariational coupling disrupts the coarsening before the $n=1$ state is reached because all relevant eigenvalues have become negative. Thus, the onset of multistability marks the partial or complete suppression of coarsening depending on the fastest growing linear mode. 

The same mechanism also acts for large-$n$ states. In Fig.~\ref{fig:linsupp_bif_super}~(a) we observe it up to the $n=5$ branch where the fourth degenerate secondary pitchfork bifurcation marks the arrest of coarsening at the state with five peaks. Note that below in Section~\ref{sec:Hopf_fully_nonlinear} we discuss more intricate, time-periodic behavior. Then our simple explanation how coarsening is suppressed is not valid anymore. 
However, next we focus on another interesting system property related to the behavior of the primary bifurcations that forms a stepping stone to time-periodic behavior. The considerations in Section~\ref{sec:super} have focused on situations where all primary bifurcations are supercritical. For phase separation phenomena this is often not the case. Therefore, we next consider how the suppression of coarsening is amended if primary bifurcations are subcritical. 

%%%%%%%%%%%%%%%%%%%%%%%%%%%%%%%%%%%%%%%%%%%%%%%%%%%%%%%%%%%%%%%%%%%%%%%%%%%%%%%
\subsection{The subcritical case} \label{sec:subcrit}
%%%%%%%%%%%%%%%%%%%%%%%%%%%%%%%%%%%%%%%%%%%%%%%%%%%%%%%%%%%%%%%%%%%%%%%%%%%%%%%

\begin{figure}[tbh]
\includegraphics[width=0.45\textwidth]{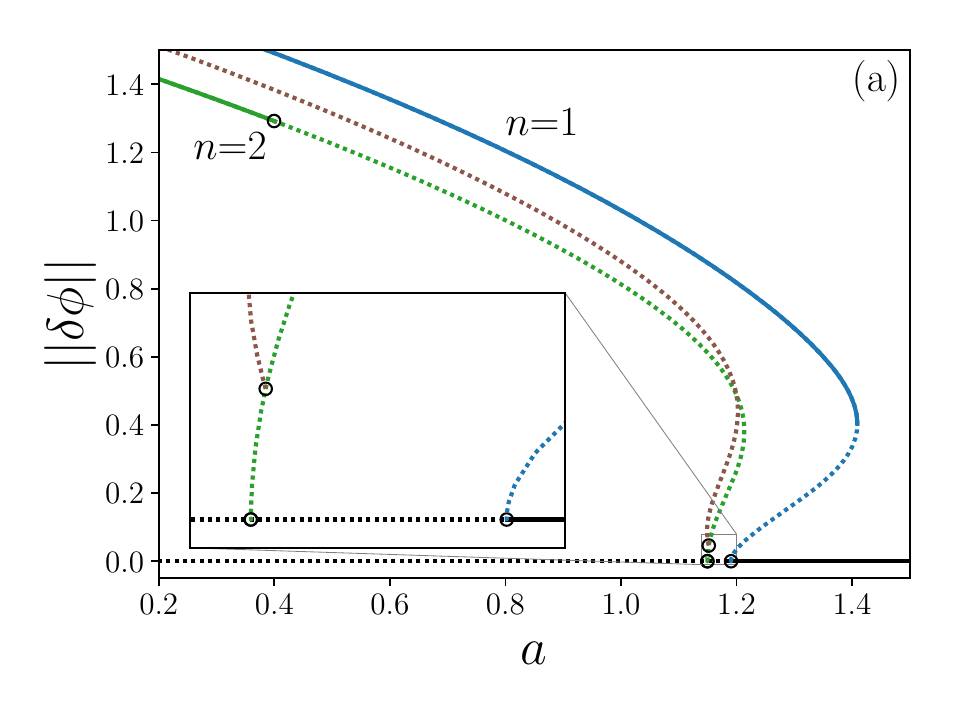}
\includegraphics[width=0.45\textwidth]{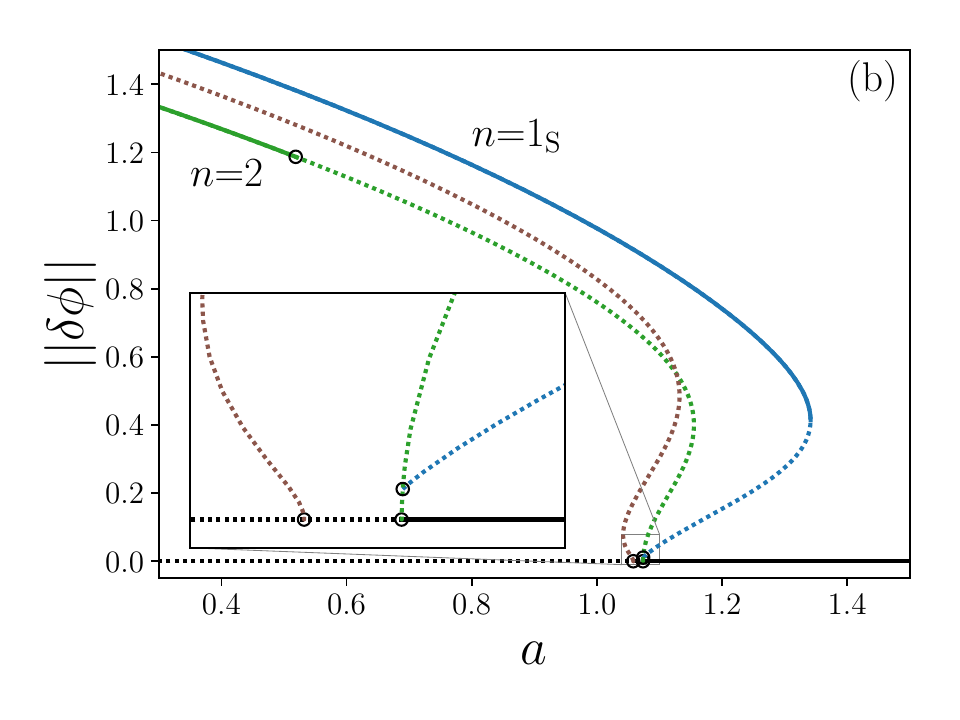}
\includegraphics[width=0.45\textwidth]{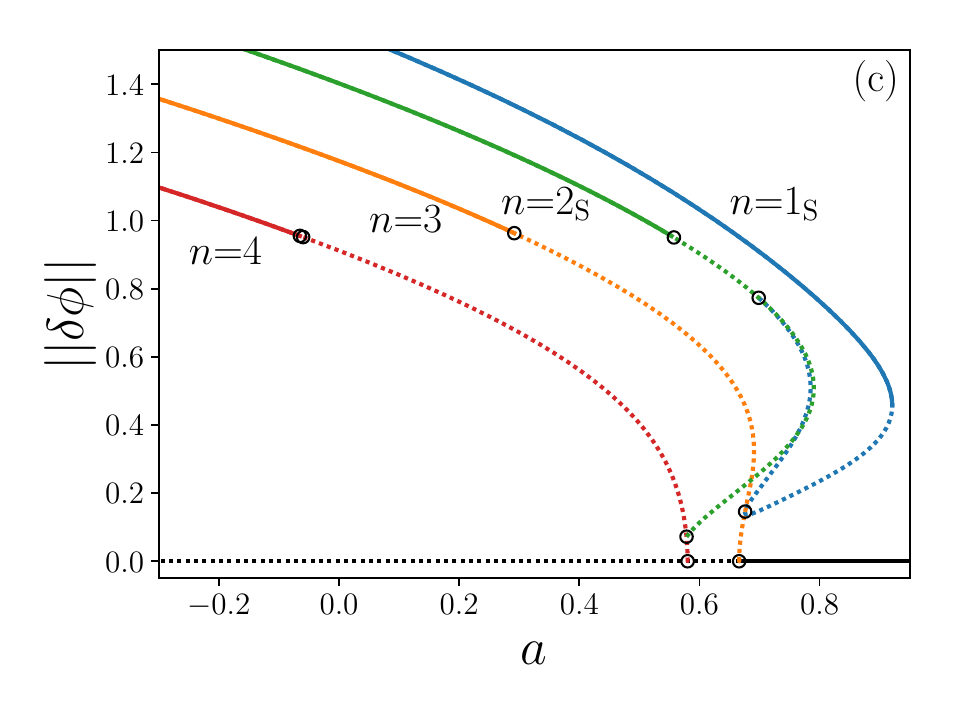}
\includegraphics[width=0.45\textwidth]{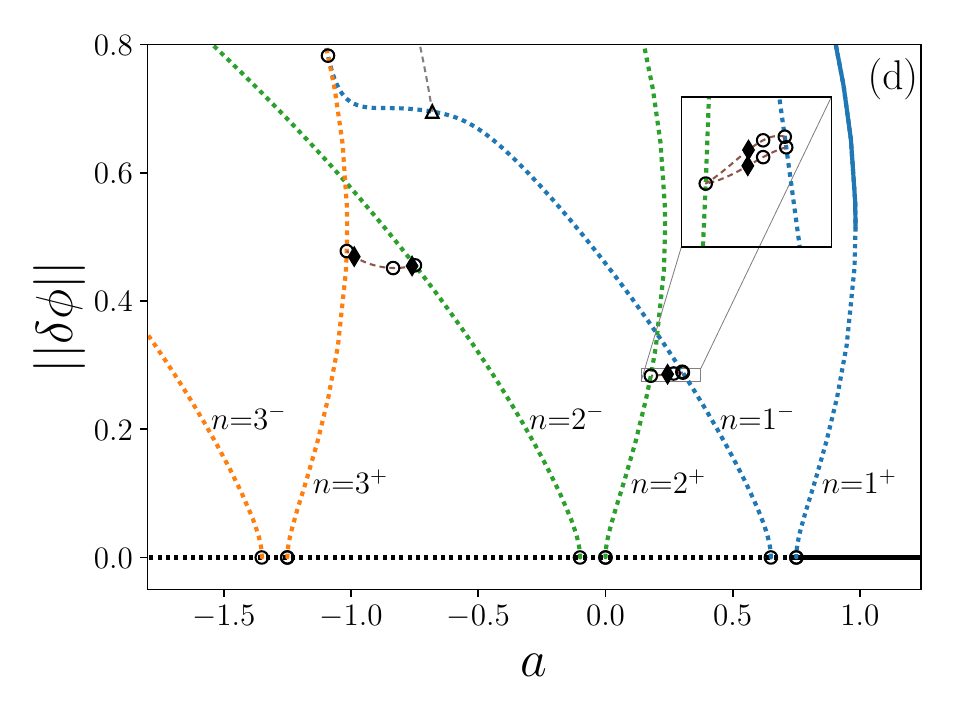}

\caption{\small \it Subcritical bifurcation behavior of steady states for the nonvariationally coupled CH model. Panels (a) and (b) show for $\rho=1.35$ cases of CH and Turing instability at $\alpha=1.45$ and $\alpha=1.5$, respectively. Mean concentrations are $\bar{\phi}_1 = 0$ and $\bar{\phi}_2 = 0.4 \neq 0$, with remaining parameters as in Fig.~\ref{fig:linsupp_bif_super}. Panel (c) gives more intricate behavior at $\rho=1.4$ and $\alpha=1.8$, parameters otherwise as (a,b). Relevant branches are labeled by their periodicity $n$ and a subscript ``S'' if they emerge in a secondary pitchfork bifurcation.
Panel (d) shows that subcritical behavior may for $\kappa=1$ and $\alpha=1.65$ even arise at $\bar{\phi}_1 = \bar{\phi}_2 = 0$; other parameters are as in (a,b). Circles, triangles and diamonds indicate pitchfork, drift pitchfork and Hopf bifurcations, respectively.}
\label{fig:linsupp_bif_sub} 
\end{figure}

In the passive one-field CH equation, subcritical primary bifurcations at $a=-1$ occur for mean concentrations $|\phi|>1/\sqrt{5}$ \cite{Novi1985jsp} (for details, use $D=0$ in the derivation in the appendix of \cite{TALT2020n} or consider Appendix D). In general, it is known that quadratic nonlinearities (in general, nonlinearities of even power) break the field inversion symmetry and lead to subcritical behavior \cite{CrHo1993rmp}\footnote{Depending on their sign, nonlinearities of odd power can also act destabilizing and lead to subcritical behavior while keeping field inversion symmetry.}. In the CH case, moving at least one mean concentration away from zero indeed breaks the field inversion symmetry and facilitates the occurrence of subcritical bifurcations. This can be clearly seen when transforming Eqs.~\eqref{eq:nondim_bG_final} using shifted concentration fields such that the new homogeneous state is always at zero. The original mean concentrations then appear as parameters and the original purely cubic nonlinearities unfold into a cubic polynomial containing quadratic and linear terms.

If the quadratic term passes a certain threshold, e.g., for a range of nonzero $\bar{\phi}_2$, primary bifurcations can become subcritical. Here, we choose $\bar{\phi}_2=0.4$ and accordingly adapt $\alpha$ to investigate the transition from CH to Turing instability. The linear behavior is similar to the case discussed at Fig.~\ref{fig:lin_supp_dispersion} in Section~\ref{sec:act}.
Bifurcation diagrams characterizing the nonlinear behavior near the transition are shown in Fig.~\ref{fig:linsupp_bif_sub}. Panel~(a) and (b) give results for $\alpha=1.45$ and $\alpha=1.5$, respectively, showing all branches that eventually connect to the homogeneous state at the first or second primary bifurcation. Between the two panels a transition occurs analog to the one between Figs.~\ref{fig:linsupp_bif_super}~(a) and (b) for the supercritical case. 

In Fig.~\ref{fig:linsupp_bif_sub}~(a) the $n=1$ branch (blue line) bifurcates first and coarsening can proceed unhindered as all other states are unstable in a large $a$-range (CH instability). The branch bifurcates subcritically and gains stability at a saddle-node bifurcation at about $a\sim 1.41$.
At the second primary instability the $n=2$ state (green line) emerges subcritically with three unstable eigenvalues. They are stabilized through two secondary pitchfork bifurcations and a saddle-node bifurcation, finally resulting in linear stability for $a \lesssim 0.4$. At the two well-separated secondary bifurcations the $n=2$ state is stabilized with respect to the two coarsening modes. 
The secondary branch which emerges in Fig.~\ref{fig:linsupp_bif_sub}~(a) at the first secondary bifurcation very close to the second primary bifurcation [see inset] emerges due to the stabilization of the volume mode of the primary branch.

In contrast, Fig.~\ref{fig:linsupp_bif_sub}~(b) at $\alpha=1.5$ illustrates a case beyond the transition where the linear analysis of the uniform state indicates the occurrence of a Turing instability. Although, overall the appearance and stability are rather similar to Fig.~\ref{fig:linsupp_bif_sub}~(a), inspection of the inset shows that the local bifurcation behavior has strongly changed: At the first primary bifurcation, now the $n=2$ state subcritically emerges carrying one unstable eigenvalue. Shortly after, a secondary supercritical pitchfork bifurcation occurs, where the blue $n=1_\text{S}$ branch supercritically emerges inheriting the one unstable eigenvalue.
Comparing to panel (a), we still consider it as the fully phase-separated $n=1$ state but indicate by the subscript ``S'' the qualitative different emergence in a secondary instead of a primary bifurcation. Nevertheless, as before, the $n=1_\text{S}$ branch fully stabilizes at the saddle-node bifurcation and in a wide $a$-range it is the only stable state. In the second primary bifurcation, the $n=1$ branch (brown line) emerges supercritically carrying two and, after a nearby saddle-node bifurcation, three unstable eigenvalues, i.e.,~it has similar properties as in Fig.~\ref{fig:linsupp_bif_sub}~(a) where it emerges at the first secondary bifurcation of the $n=2$ state. One may say that the primary $n=1$ bifurcation and the first secondary bifurcation on the $n=2$ branch exchange their roles at the transition from CH to Turing instability. 
Only two parameters ($\alpha$ and $a$) are adjusted to reach the transition point that displays properties of a higher codimension point as two primary and one secondary bifurcations coincide. However, since the latter breaks the reflection symmetry of the states, an additional restriction is provided by the reflection symmetry of the model. Here, the particular choice $\bar\phi_1=0$ does not qualitatively influence the described transition. It does not make the case nongeneric since $\bar \phi_2 \neq 0$. However, adding a symmetry-breaking term to the model lifts the degeneracy and decreases the codimension (not shown).

%Only two parameters ($\alpha$ and $a$) are adjusted to reach the transition point that again displays properties of a higher codimension point as three bifurcations coincide, namely two primary ones and a secondary one. This is due to the additional special condition provided by $\bar\phi_1=0$.

% However, we do not call the transition point a bifurcation of higher codimension, since we do not need to adjust any further parameter for this seemingly special bifurcation behavior. 
% Rather, it is generic in a way that the nonlinear and in this case stable areas of the branches force this local bifurcation behavior. \tobias{agree?}
% Indeed, the apparently simpler case, that the primary bifurcations exchange positions and nothing else happens, is not compatible with the independence of the fully nonlinear behavior from the local stabilities.

We see that the merely local changes at the transition are largely overshadowed by mainly undisturbed global behavior related to the subcriticality. Hence, due to the branches which emerge from secondary bifurcations no linear complete suppression of coarsening occurs. Only for supercritical primary bifurcations, a switch from CH to Turing instability directly results in the linear complete suppression of coarsening. In contrast, the nonlinear effects of partial and complete suppression of coarsening are unaffected by the subcriticality as they depend on secondary bifurcations. For instance, the final secondary bifurcation of the $n=2$ branch (where it becomes linearly stable) still marks the onset of the nonlinear partial or complete suppression of coarsening as in the supercritical case.
% \tobias{see Fig.~\ref{fig:symmetry_sub} to consider the instability modes at the bifurcations, should we include it into the paper?}
% \begin{figure}
% \includegraphics[width=\textwidth]{./symmetry_subcritical}
% \caption{Panels show the steady state at the bifucation (black) and the instability modes (red).
% (a) and (b): States at first (second) pitchfork bifurcation on second branch in Fig.\ref{fig:linsupp_bif_sub}~(a))
% (c) and (d): First and second pitchfork bifurcation on second branch in Fig.\ref{fig:linsupp_bif_super}~(b).
% Instability modes in (a) and (c) are volume modes. The ones in (b) and (d) are translation modes. Following our argument of locality modes in (b) and (d) are not influenced by the transition from large- to small-scale instability
% }\label{fig:symmetry_sub}
% \end{figure}

Fig.~\ref{fig:linsupp_bif_sub}~(c) illustrates more extensive reordering of the primary bifurcations. Increasing $\rho$ and $\alpha$ as compared to Figs.~\ref{fig:linsupp_bif_sub}~(a) and (b), now at the first primary bifurcation the $n=3$ branch emerges subcritically. It carries a secondary degenerate bifurcation where the $n=1_\text{S}$ branch emerges as well as another branch that connects to the first secondary branch of the second primary branch. The latter is actually the $n=4$ branch from which the $n=2_\text{S}$ branch emerges. However, when crossing the stable parts at large norm the branches are still well ordered: from right to left $n=1_\text{S},2_\text{S},3,4,\dots$.

\begin{figure}[tbh]
\includegraphics[width=0.7\textwidth]{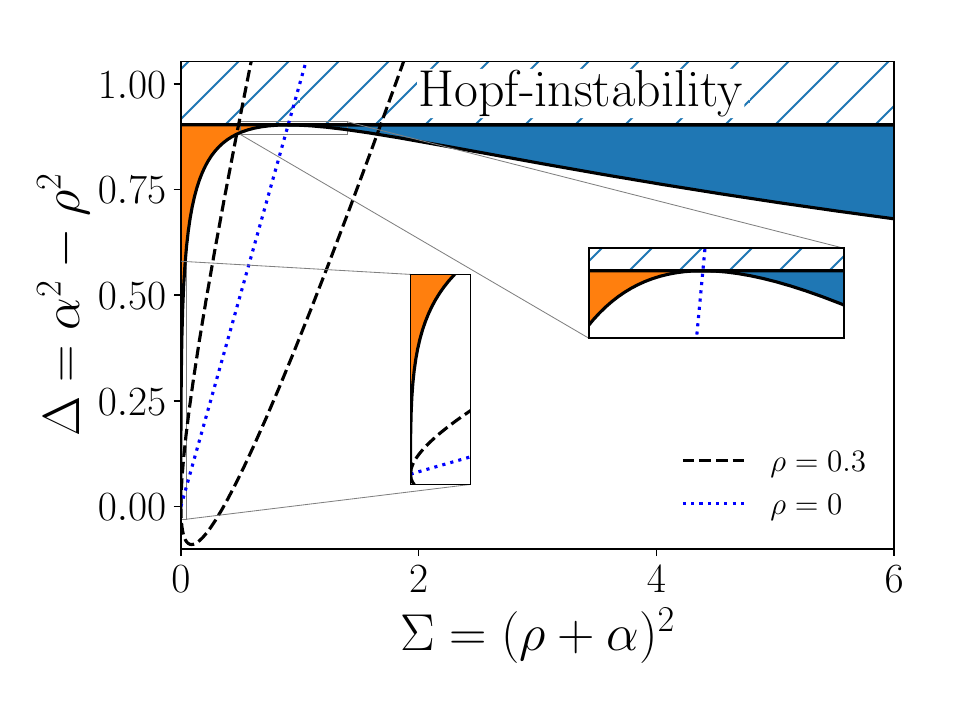}
\caption{\small \it Illustration of criteria for subcriticality [cf.~Eq.~\eqref{eq:sub_criterion}] and Hopf instability [cf.~Eq.~\eqref{eq:hopf_onset2}] in the plane spanned by $\Delta$ and $\Sigma$ at $\bar\phi_1=\bar\phi_2=0$. Shown is the special case of equal mobilities and rigidities, $Q=\kappa=1$, and $M=-a_\Delta = 1.9$. Blue [orange] regions imply subcriticality of $n^+$[$n^-$] branches. The hatched region indicates the occurrence of a Hopf instability, i.e.~where $\Delta> a_\Delta^2/4$ [cf.~Eq.~\eqref{eq:hopf_onset2}]. The dashed and dotted lines indicate how the parameter plane is crossed when changing $\alpha$ at fixed $\rho$ (as given in the legend).
  The left inset suggests that subcriticality can be observed even in the immediate vicinity of $\alpha\approx-\rho$. The right inset shows that the colored regions do not overlap and only without variational coupling, i.e.~for $\rho=0$ (blue dotted line), subcritical behavior does not occur.}
\label{fig:sub_ranges}
\end{figure}

It is remarkable, that in the present nonvariationally coupled system subcritical behavior can even occur at zero mean concentrations, i.e., where the above argument regarding the quadratic nonlinearity does not hold. This is shown in Fig.~\ref{fig:linsupp_bif_sub}~(d) and mathematically illuminated by weakly nonlinear analysis in appendix~\ref{sec:app-wna}. The derived amplitude equations [see Eq.~\eqref{eq:AE_zero}] define the parameter ranges illustrated in Fig.~\ref{fig:sub_ranges} where this unexpected behavior occurs. At the core of the argument is a projection that is performed when applying the Fredholm alternative. If the necessary criterion $\Delta>0$ is fulfilled this projection can produce nonlinearities that act destabilizing to leading order and result in subcritical behavior (even if the nonlinearities in the original equations appear stabilizing). Such projections can only occur if the model couples at least two fields.

% There we see that the origin of the subcritical behavior is the projection of the residual of leading order onto the kernel of the adjoint linear operator, following the Fredholm alternative.
% If the necessary criterion $|\alpha|>|\rho|$ is fulfilled, this projection can cause unstable nonlinearities at leading order, i.e. subcritical behavior. This means that even if the nonlinearities in the original equations appear stabilizing $\sim \partial_{xx} \phi_i^3 $, they can have a destabilizing effect if projected onto a subspace with a fixed amplitude ratio between the fields.
% In one-component fields, this projection does not occur, so the character of the nonlinearities can be easily read from the sign and the exponent of each nonlinearity.

The bifurcation diagram in Fig.~\ref{fig:linsupp_bif_sub}~(d) for $\kappa=1$ and $\bar\phi_1=\bar\phi_2=0$ shows six primary bifurcations, three being subcritical. The various branches are marked by their periodicity $n$ and a superscript ``$+$'' or ``$-$'' that indicates which eigenvalue [$\lambda_+$ or $\lambda_-$ in Eq.~\eqref{eq:lambda+-}] crosses zero at the corresponding primary bifurcation. In previous diagrams the distinction was not needed since all $n^-$ branches emerged far away from the instability onset and were not further considered.
Here, however, for each $n$ a supercritical $n^-$ and a subcritical $n^+$ branch emerge close to each other. Since $\lambda_+>\lambda_-$ for all real eigenvalues, the first bifurcation of each pair is always the $n^+$ state.

Since the necessary conditions for subcritical behavior and for primary Hopf bifurcations are identical, $\Delta>0$ (see Sec.~\ref{sec:linear}), it is not surprising that pairs of structured states emerge close together. To create a primary Hopf bifurcation two pitchfork bifurcations belonging to the same $n$ (i.e., $n^+$ and $n^-$) have to collide. For all $\rho \neq 0$ one of these two branches displays subcritical behavior right before collision. Note that Fig.~\ref{fig:linsupp_bif_sub}~(d) shows the particular case $\kappa=1$. Then the onset of subcritical behavior as well as the creation of the primary Hopf bifurcations is independent of $n$ [cf.~discussion in appendix \ref{sec:app-wna} and Eq.~\eqref{eq:hopf_onset2}]. Although all primary bifurcations are still stationary we already observe time-periodic behavior at secondary and tertiary bifurcations. The inset shows two secondary bifurcations on the $n=1^-$ branch which are connected to one degenerated pitchfork bifurcation on the $n=2^+$ branch. Again the degeneracy is caused by additional symmetries resulting from zero mean concentrations [cf.~discussion of Fig.~\ref{fig:linsupp_bif_super}~(a)]. On both connecting branches (brown dashed lines) Hopf bifurcations marked by filled diamonds occur. Similar bifurcation structures are found on all branches which connect an $n^-$ branch with an $(n+1)^+$ branch (see e.g.~connecting branches between $n=2^-$ and $n=3^+$ branch). Furthermore, on the $n=1^-$ branch a drift pitchfork bifurcation marked by a triangle occurs. The emerging branch (gray dashed line) represents stationary drifting states. All of these time-dependent states are unstable, at least in the vicinity of their emergence. Summarized, Fig.~\ref{fig:linsupp_bif_sub}~(d) implies that time-periodic behavior can arise in various ways when the nonvariational coupling strength is increased.
Interestingly, the subcritical behavior observed at stationary bifurcations provides us with two different scenarios for the emergence of time-dependent states close to the Hopf instability. This is further investigated in Section~\ref{sec:OnsetLargeOscill}.

First, we return to the subcritical primary branches for zero mean concentrations and consider in Fig.~\ref{fig:sub_ranges} the parameter plane spanned by $\Delta=\alpha^2 -\rho^2$ and $\Sigma=(\rho+\alpha)^2$. The orange [blue] shaded region indicates where the $n^-$ [$n^+$] branch shows subcritical behavior. At large $\Delta$ both regions are limited by the horizontal Hopf threshold [Eq.~\eqref{eq:hopf_onset2}], otherwise their shape only depends on the composed parameter $M_n=\frac{k^2_n}{\ell^2}\left(1-\kappa\right)- a_\Delta$.\footnote{For general functions $f_1(\phi_1)$ and $f_2(\phi_2)$, the relevant parameter is $M_n= \frac{k^2_n}{\ell^2}\left(1-\kappa\right) + f_1'' - f_2''$. It is similarly valid for the present coupled CH equations, as for coupled Swift-Hohenberg and coupled conserved Swift-Hohenberg equations.} For the special case $\kappa=1$ presented in Fig.~\ref{fig:sub_ranges} simply $M_n=M=-a_\Delta$, i.e., it is independent of the periodicity of the linear mode. Then, for $a_\Delta<0$ all $n^+$ branches in Fig.~\ref{fig:linsupp_bif_sub}~(d) emerge subcritically. 

The analysis in appendix~\ref{sec:app-wna} reveals two further remarkable features: First, at $M=0$, i.e.,~for identical subsystems, no subcritical regions exist. Second, for purely nonvariational coupling (i.e.~$\rho=0$) no subcritical behavior precedes the appearance of primary Hopf bifurcations. This is indicated by the dotted blue line in Fig.~\ref{fig:sub_ranges} which passes the Hopf threshold without crossing the shaded regions. 
This emphasizes the nongeneric character of systems with purely nonvariational coupling.
For any fixed $\rho$ and increasing $|\alpha|$ the system follows curves in Fig.~\ref{fig:sub_ranges} given by 
\begin{align}
% \Delta = \rho^2 - \alpha^2~ \quad \Sigma = (\rho+ \alpha)^2
% ~\\
% \Rightarrow \alpha= \pm \sqrt{\Sigma} - \rho
% ~\\
% \Rightarrow 
\Delta(\Sigma) = \mp 2 \rho \sqrt{\Sigma} + \Sigma \quad \text{for}\, \alpha \gtrless - \rho\,.
\end{align}
For instance, the dashed black lines at fixed $\rho=0.3$ demonstrate that either the orange (for $\alpha<-\rho$) or the blue (for $\alpha>\rho$) shaded region is crossed before passing the Hopf threshold. These pathways represent two different scenarios. We call them \textbf{``SubMinus''} and \textbf{``SubPlus''} as passing the orange and blue region indicates subcritical $n^-$ and $n^+$ branches, respectively. The scenarios occur if:
\begin{alignat}{1}
\begin{aligned}
\text{\textbf{SubMinus}:}\, \, M_n \gtrless 0 \, \, \text{and} \, \, \alpha \,\mathrm{passes}\,\mp\rho
\\
\text{\textbf{SubPlus}:}\, \, M_n \gtrless 0 \, \, \text{and} \, \, \alpha \,\mathrm{passes}\,\pm\rho
\end{aligned}
\label{eq:scenarios}
\end{alignat} 
These scenarios are of great importance for the time-dependent behavior of the fully phase-separated (i.e.~$n=1$) states focused on in Section \ref{sec:OnsetLargeOscill}.
%\textbf{SubMinus} and \textbf{SubPlus} 

%%%%%%%%%%%%%%%%%%%%%%%%%%%%%%%%%%%%%%%%%%%%%%%%%%%%%%%%%%%%%%%%%%%%%%%%%%%%%%%
\section{Nonvariational case: Emergence of time-periodic states}\label{sec:timeperiodic}
%%%%%%%%%%%%%%%%%%%%%%%%%%%%%%%%%%%%%%%%%%%%%%%%%%%%%%%%%%%%%%%%%%%%%%%%%%%%%%%

\subsection{Hopf bifurcations in the strongly nonlinear regime}\label{sec:Hopf_fully_nonlinear}
\begin{figure} 
\includegraphics[width= 0.5 \textwidth]{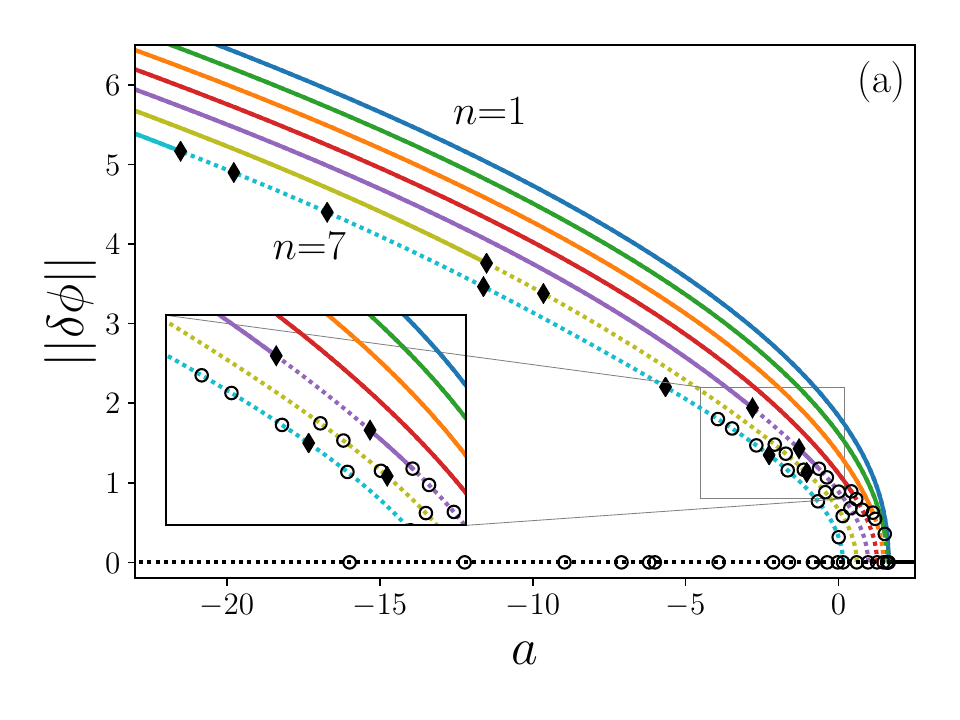}
\includegraphics[width= 0.35 \textwidth]{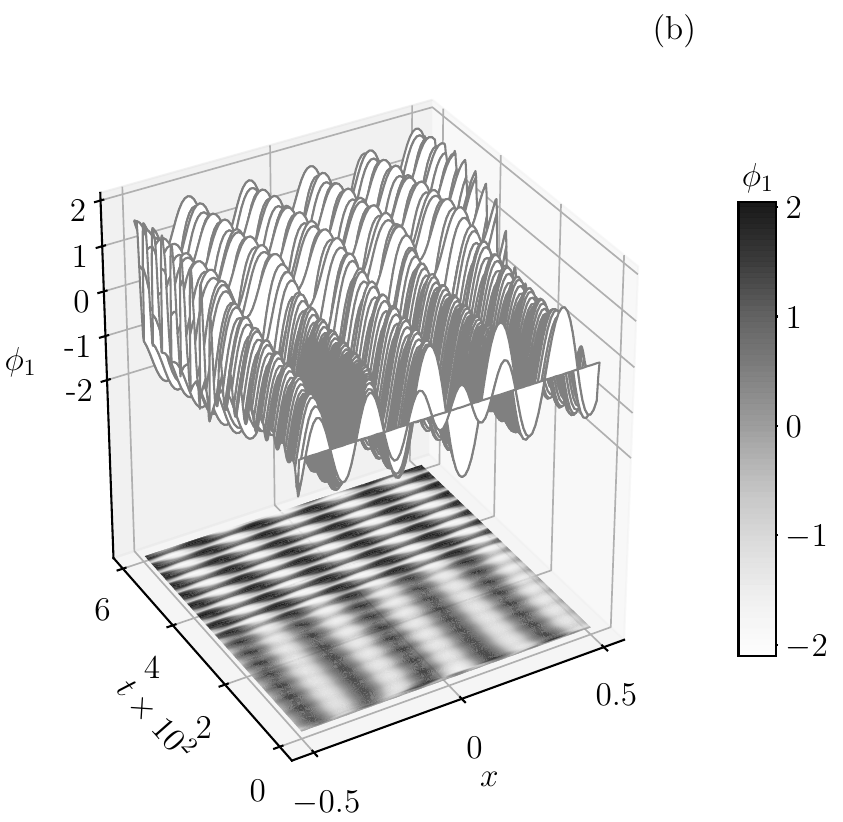}
\caption{\small \it (a) Bifurcation diagram showing the trivial state (black horizontal line) and seven primary branches ($n=1\dots7$) for a large $a$-range at parameters identical to Fig.~\ref{fig:linsupp_bif_super}~(a). Pitchfork and Hopf bifurcations are marked by circles and filled diamonds, respectively. Solid [dashed] lines indicate stable [unstable] states. The inset magnifies the region where the first four Hopf bifurcations occur. Panel (b) presents a space-time plot of a simulation at $a=-2.3$, i.e. between the two stable regions of the $n=5$ state. It is initialized with white noise and after a transient steady state converges to a drifting oscillatory state.
  %\ttuwe{I would like to see more detailed s-t plots for the oscillating phase and the drifting phase.}\tobias{what do you mean? Extra plots?}\ttuwe{not plots in the paper: I just wanted to see zooms into smaller $t$-windows.}
  \label{fig:TPB_nonlinear}}
\end{figure}
%\newpage

After having discussed suppression of coarsening due to nonvariational coupling, we next analyze under which conditions such coupling causes time-periodic behavior like traveling and standing waves. Again, we normally consider situations with variational and nonvariational coupling both present.
As before, we employ numerical path continuation and direct time simulation to characterize the fully nonlinear behavior. In addition to path continuation for steady states employed before, here, we also track time-periodic states. For a description of these techniques see Ref.~\cite{EGUW2019springer}.

The linear considerations in Section~\ref{sec:linear} and appendix \ref{app:linear} have shown that $\Delta>0$ is a necessary condition for a Hopf instability of the uniform state and that oscillatory modes may occur in a wavenumber band $[k^o_-,k^o_+]$ with $k^o_-$ either zero or nonzero. However, the Hopf instability is always large-scale and occurs at ${f''_1}^H = - Q f''_2$ given that $\Delta>Q {f''_2}^2$ [Eqs.~\eqref{eq:a11ho2} and \eqref{eq:a11hofreq}]. In general, we find that also in the nonlinear regime time-periodic behavior only occurs for $\Delta>0$. However, nonlinearly it can emerge at lower activity than in the linear regime.

First, we revise the case in Fig.~\ref{fig:linsupp_bif_super}~(a) where we have found nonlinear suppression of coarsening. We explained that all steady $n>1$ states are stabilized by $n-1$ secondary degenerate pitchfork bifurcations. This is the complete picture for $a>-0.6$, the range presented in Fig.~\ref{fig:linsupp_bif_super}~(a). In contrast, Fig.~\ref{fig:TPB_nonlinear}~(a) presents a much larger $a$-range down to $a\approx -23$. Shown are the branches of homogeneous states and of structured states with $n=1$ to $n=7$. We note that a number of Hopf bifurcations (marked by filled diamonds) exist on the $n=5, 6$ and $7$ branches. This implies that the simplified picture of successively extended multistability and related nonlinear partial or complete suppression of coarsening has to be amended as time-periodic behavior occurs for structured states of larger $n$.
%
%Further, there are some additional pitchfork bifurcations, e.g., on the $n=3$ branch (at $a\approx -16$ and $a\approx -18$), that \bfuwe{result in windows of destabilization after all coarsening modes had been stabilized as described in Section~\ref{sec:super}.} This means in the strongly nonlinear region the suppression of coarsening is amended by other phenomena.

The inset of Fig.~\ref{fig:TPB_nonlinear}~(a) magnifies the $a$-range where the first four Hopf bifurcations occur. We focus on the $n=5$ branch (purple line). Starting at the primary bifurcation where it emerges, subsequently four stabilizing degenerate pitchfork bifurcations occur that eventually stabilize the branch in full accordance with Section~\ref{sec:super}. Then, after a small range of stability a window of oscillatory instability occurs framed by two Hopf bifurcation. Beyond the stabilizing Hopf bifurcation the branch remains stable.
 The time-periodic behavior found in the unstable window is illustrated in Fig.~\ref{fig:TPB_nonlinear}~(b). Initialized at $a=-2.3$ with white noise of small amplitude, first, the fastest linear mode grows and the steady $n=6$ state develops [barely visible in the Fig.~\ref{fig:TPB_nonlinear}~(b)]. Being unstable, it remains a transient and coarsens into the steady $n=5$ state ($t\approx 0.1\times10^2$). There, the spatial coarsening is arrested. However, as also the steady $n=5$ state is linearly unstable, temporal oscillations in the form of a standing wave develop ($t=2\times10^2$). Then, even the standing wave turns out to be only a transient and at $t\approx 3\times10^2$ an additional slow drift develops. Finally, a drifting oscillating $n=5$ state develops, that represents a modulated wave. This shows that even for parameters where the linear analysis of the uniform state only shows a CH instability [cf.~green line in Fig.~\ref{fig:lin_supp_dispersion}~(b) and remember that changing the value of $a$ can not render the eigenvalues complex], oscillatory instabilities of nonlinear states may occur that result in stable time-dependent patterned states.

\subsection{Onset of a large-scale time-periodic behavior}\label{sec:OnsetLargeOscill}
Next we scrutinize the onset of such time-dependent behavior focusing on the fully phase-separated ($n=1$) state. In particular, we consider the case of zero mean concentrations at parameter values where the instability of the uniform state changes from stationary (CH) to oscillatory (Hopf) [cf.~Fig.~\ref{fig:cEV_dispersion}]. The two scenarios described next also occur in the general case of nonzero mean concentrations (not shown).
%More intricate behavior may be expected in a more general case, however, this shall not concern us here.

\begin{figure}[tbh]
\includegraphics[width=0.45\textwidth]{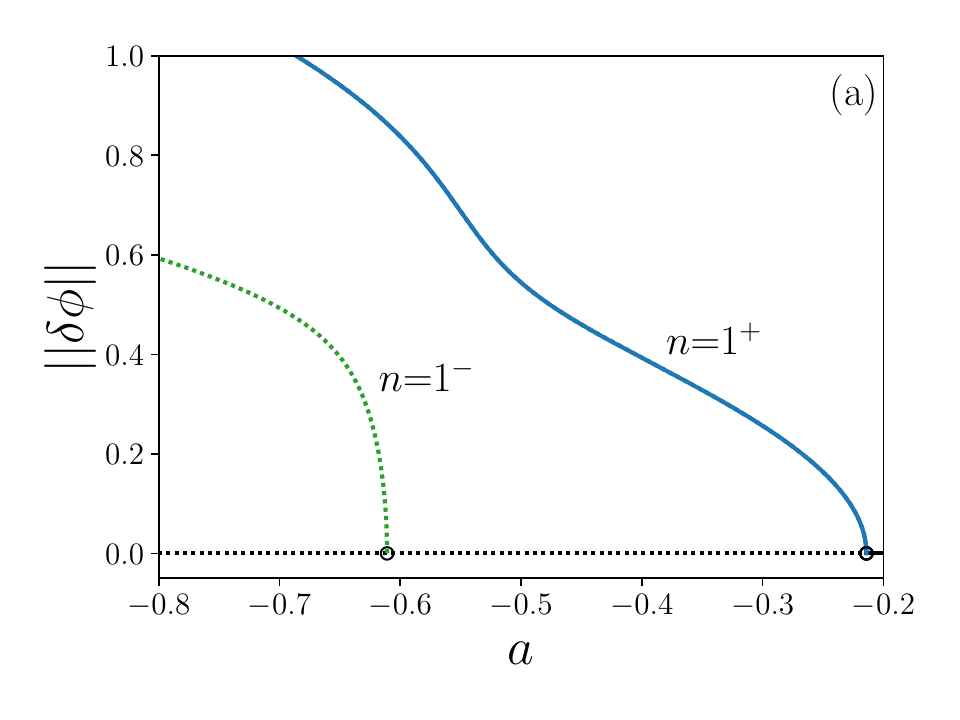}
\includegraphics[width=0.45\textwidth]{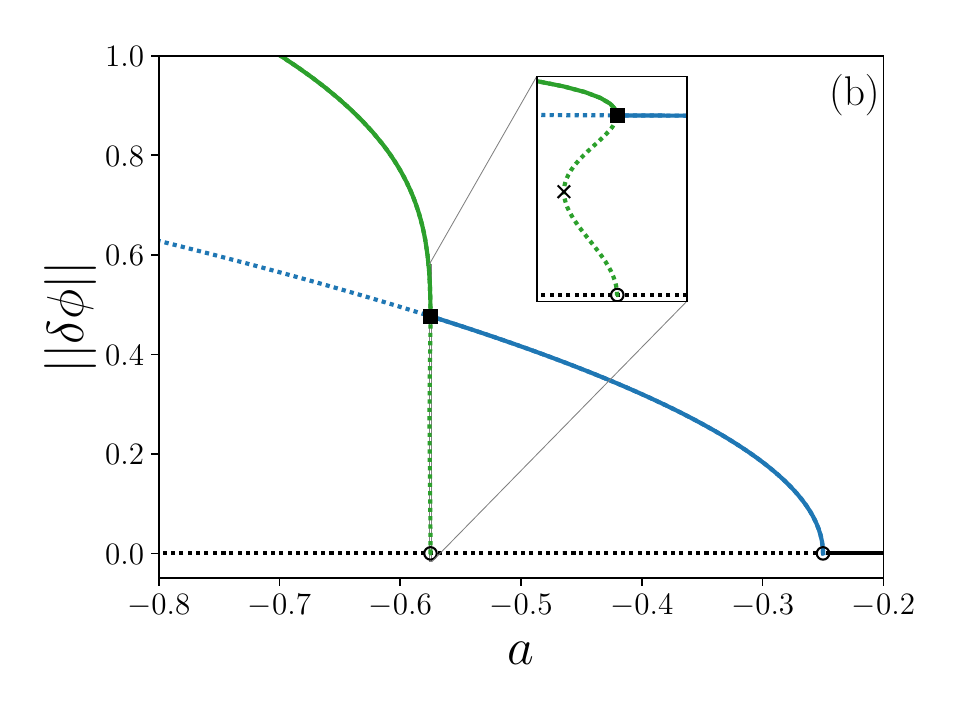}\\[-2ex]
\includegraphics[width=0.45\textwidth]{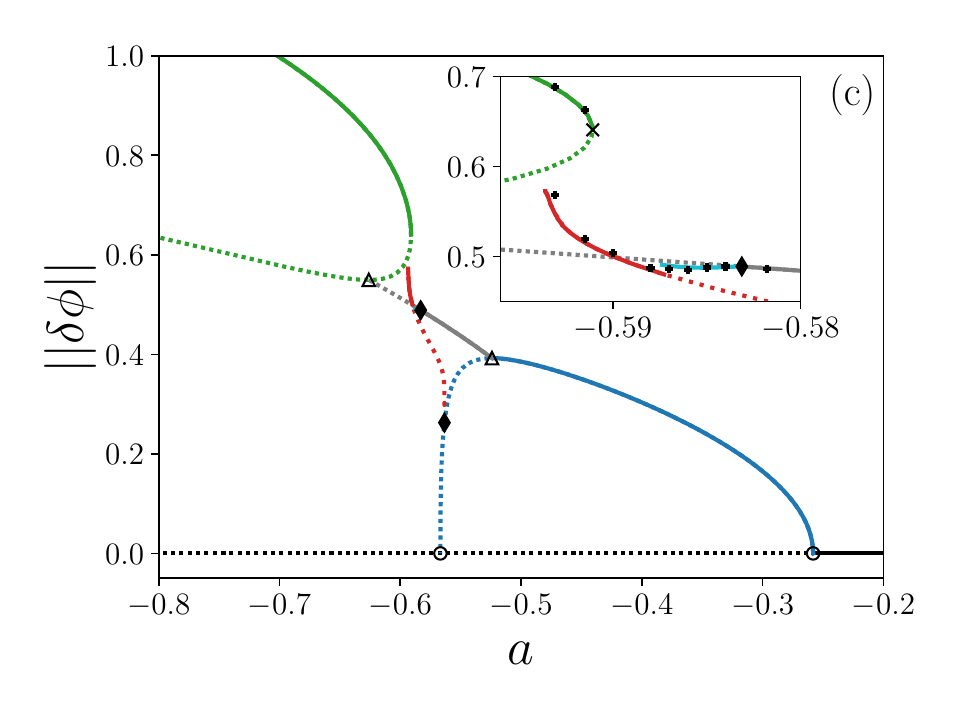}
\includegraphics[width=0.45\textwidth]{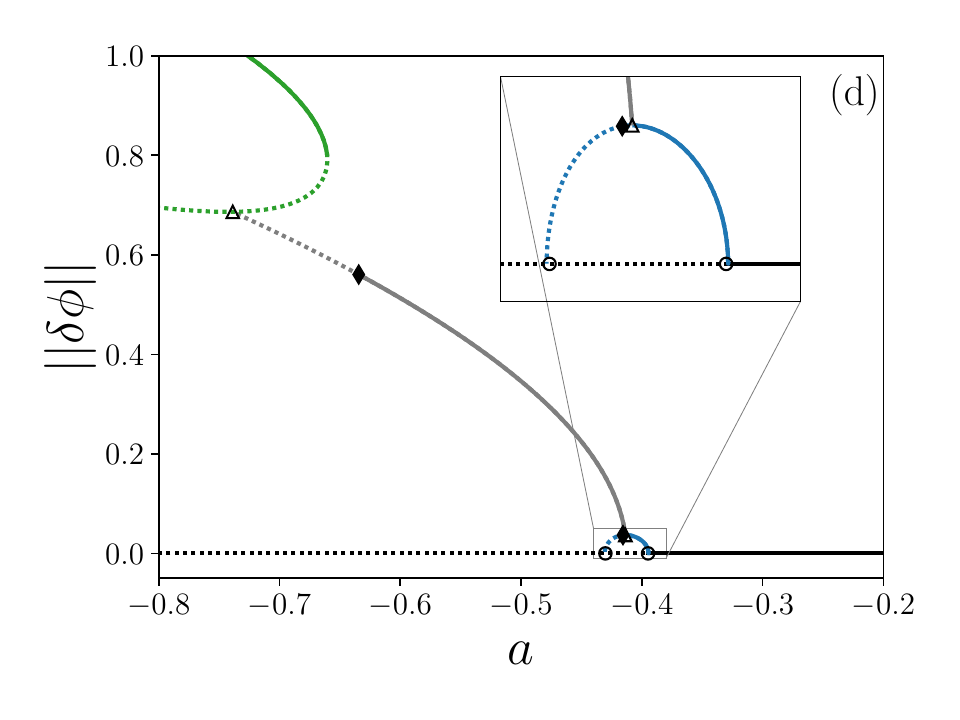}\\[-2ex]
\includegraphics[width=0.45\textwidth]{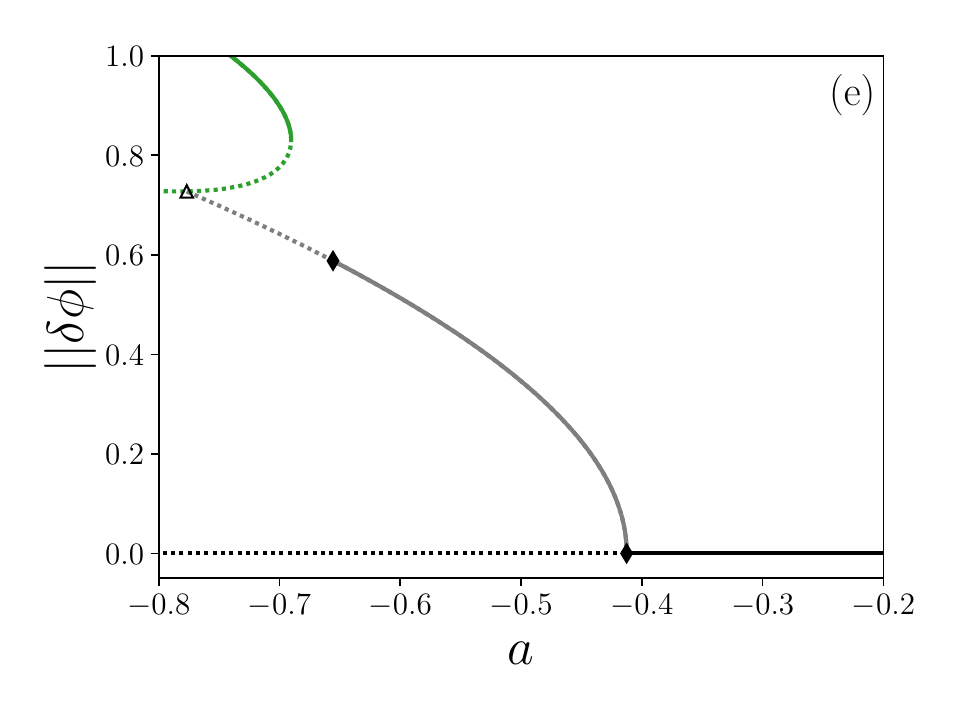}
\includegraphics[width=0.45\textwidth]{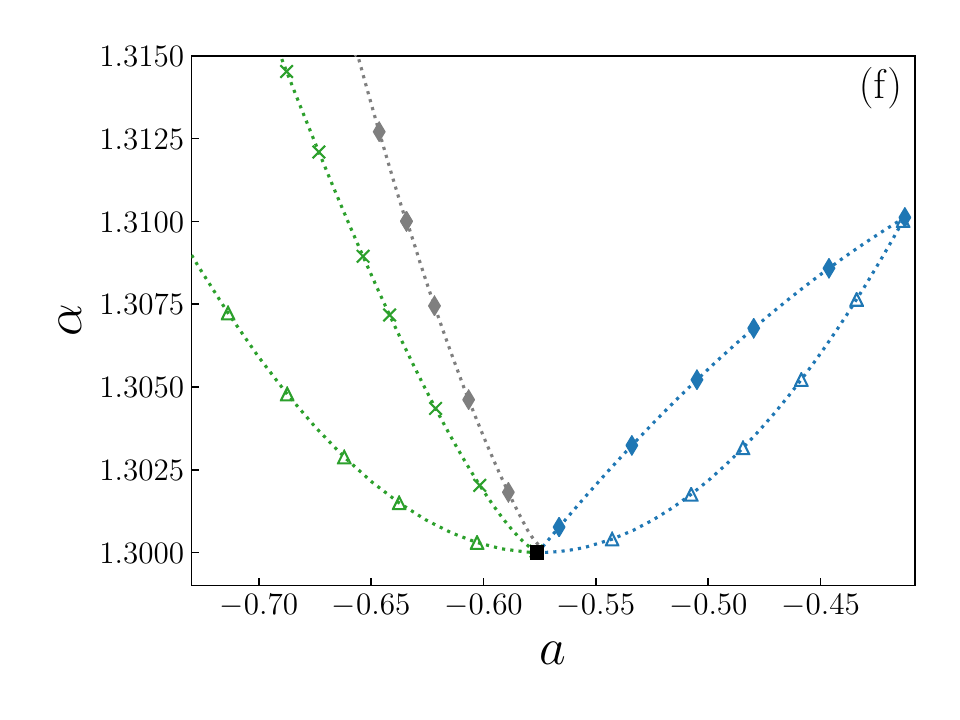}
\caption{\small \it Panels (a)-(e) show a sequence of bifurcation diagrams for scenario \textbf{SubMinus} of the emergence of subcriticality and time-dependent behavior for increasing nonvariational coupling $\alpha=1.295\,, \,\,1.3\,,\,\, 1.301\,,\,\, 1.31$ and $1.315$ at $\rho=1.3$ and $M_n<0$. Solid [dotted] lines represent linearly stable [unstable] states. Pitchfork, drift pitchfork and Hopf bifurcations are marked by circle, triangle and filled diamond symbols, respectively. The saddle node bifurcations referred to in the main text are indicated in the insets in (b) and (c) by cross symbols. The inset in (c) further marks by plus symbols states emerging in time simulations, see e.g.\ Fig.~\ref{fig:DN_2}. The remaining parameters are $\ell=4\pi\,$, $a_\Delta = -0.38\,$, $\bar{\phi}_1=\bar{\phi}_2=0\,$, $\kappa=3.82$ and $Q=1$. Panel~(f) displays in the ($\alpha,a$)-plane the loci of all five local secondary bifurcations that emerge from the high codimension bifurcation marked by the square symbol in~(b). The used colors correspond to the ones in (c).}
\label{fig:bGD_left_bif} 
\end{figure}

\begin{figure}[tbh]
\includegraphics[width=0.35\textwidth]{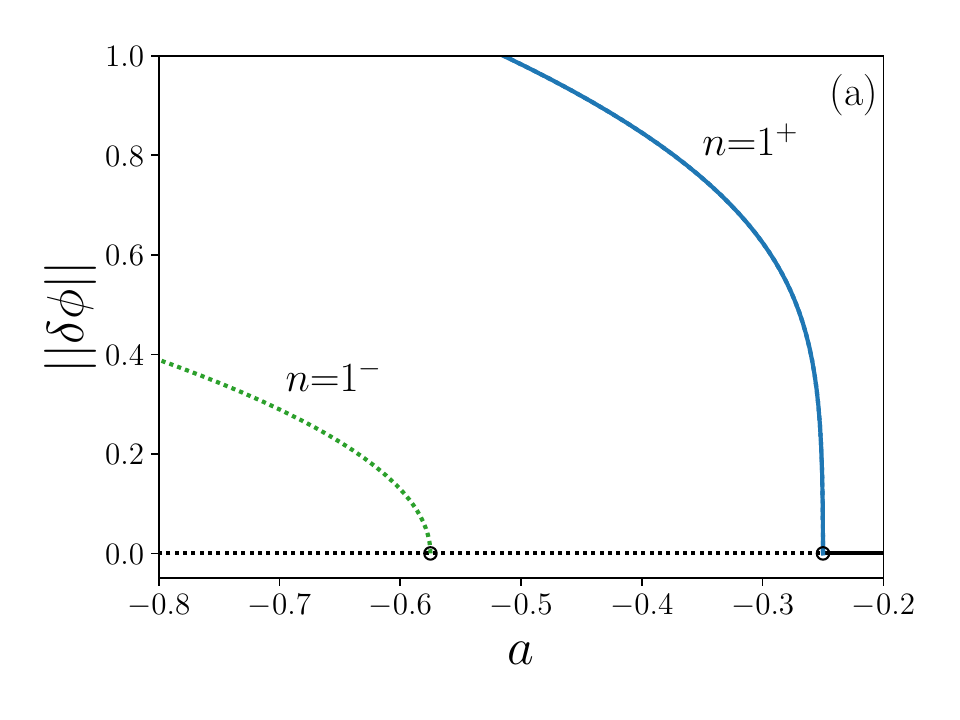}
\includegraphics[width=0.35\textwidth]{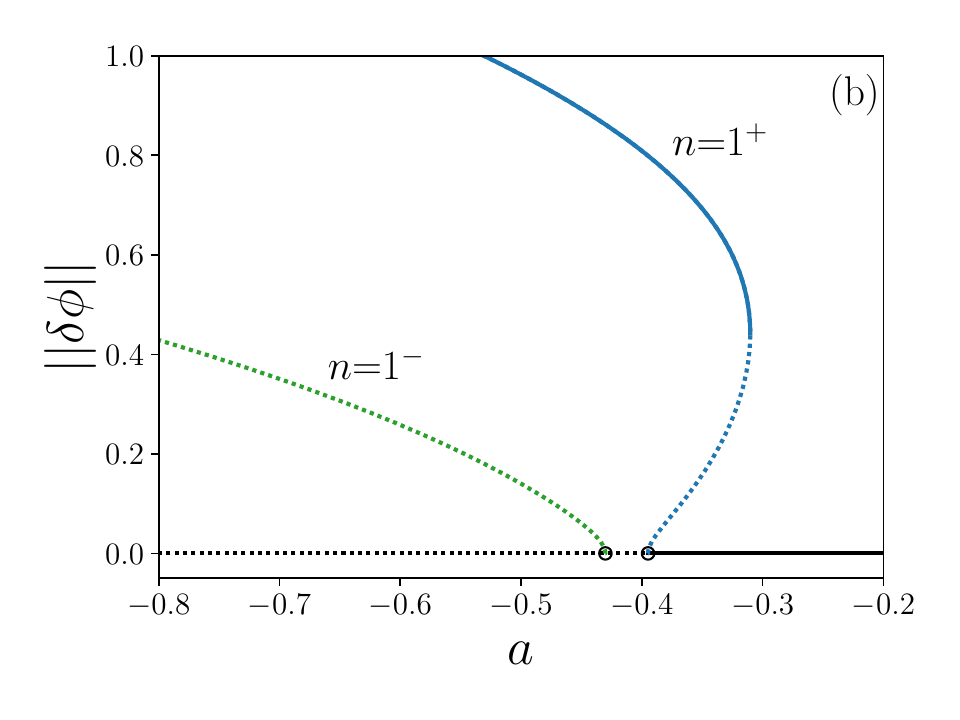}~\\
\includegraphics[width=0.5\textwidth]{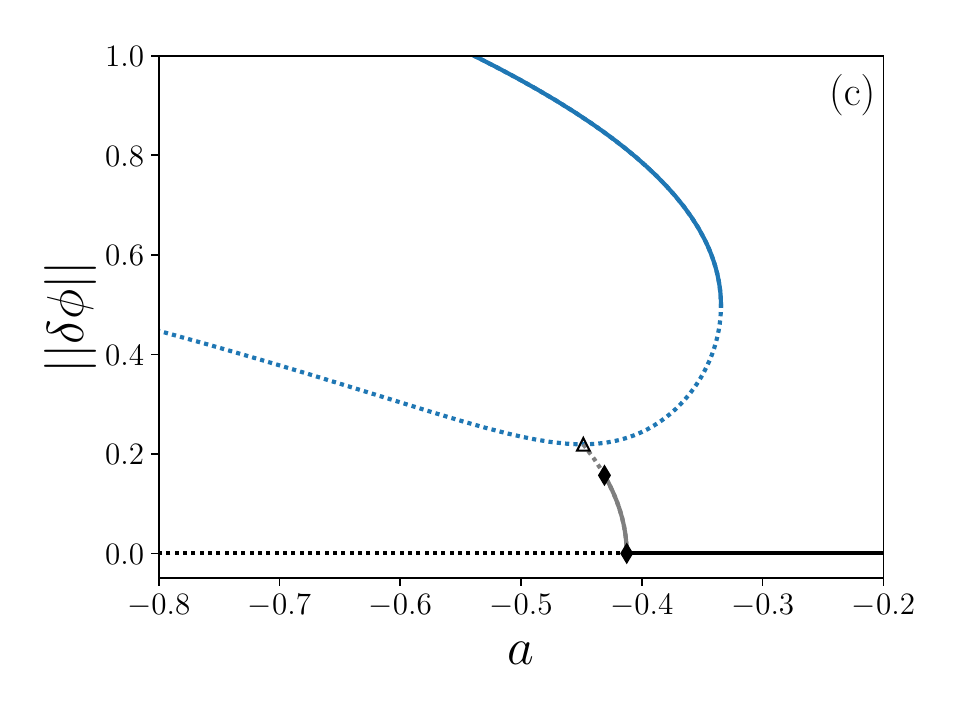}
\caption{\small \it Panels (a)-(c) show  a sequence of bifurcation diagrams that illustrates scenario \textbf{SubPlus} of the emergence of subcriticality and subsequent time-periodic behavior in the fully nonlinear regime for decreasing nonvariational coupling $\alpha= -1.3\,,\,\, -1.31$ and $-1.315$ at $\rho=1.3$ and $M_n<0$. Linestyles, symbols and remaining parameters are as in Fig.~\ref{fig:bGD_left_bif}.}
\label{fig:bGD_right_bif} 
\end{figure}

As explained in Section~\ref{sec:subcrit}, the onset of primary time-periodic behavior is for all $\rho\neq0$ preceded by the occurrence of subcritical primary bifurcations. Namely, before two primary pitchfork bifurcations can collide to form a Hopf bifurcation one of them has to become subcritical.
Sequences of bifurcation diagrams for increasing nonvariational coupling (passing $\Delta=0$) that detail the intricacies of this transition are shown for the $n=1$ branch in Figs.~\ref{fig:bGD_left_bif} and~\ref{fig:bGD_right_bif} in the two qualitatively different cases of scenario~\textbf{SubMinus} (subcriticality of the $n^-$ branch)
  and scenario~\textbf{SubPlus} (subcriticality of the $n^+$ branch), respectively [cf.~Fig.~\ref{fig:sub_ranges} and Eqs.~\eqref{eq:scenarios}].
  
\begin{figure}[h]
    \includegraphics[width= 0.3 \textwidth]{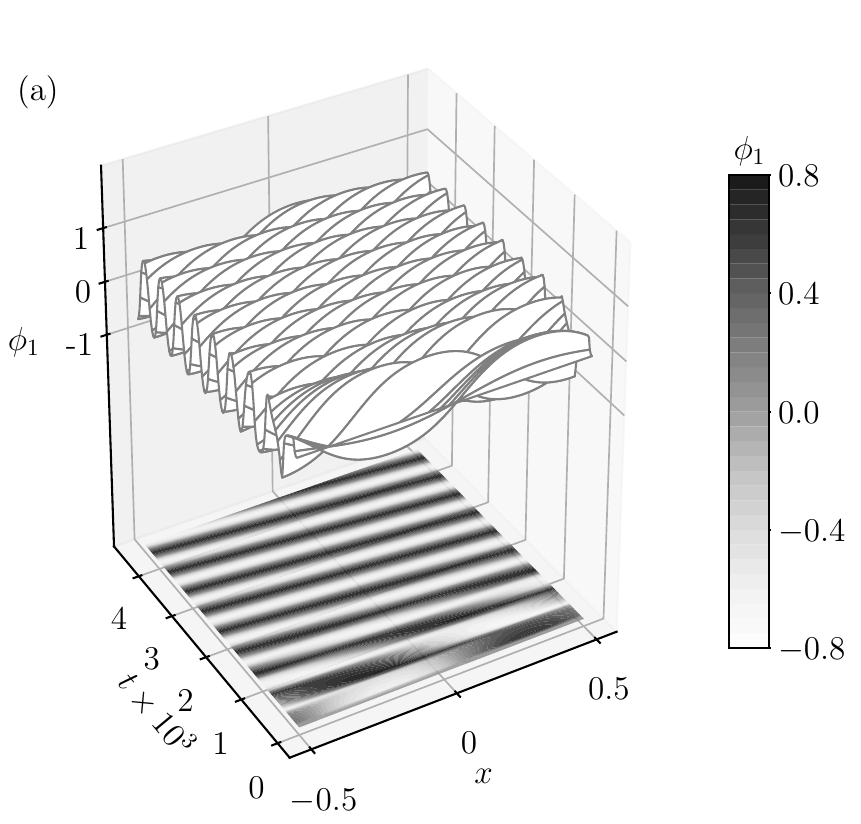}
    \includegraphics[width=0.3 \textwidth]{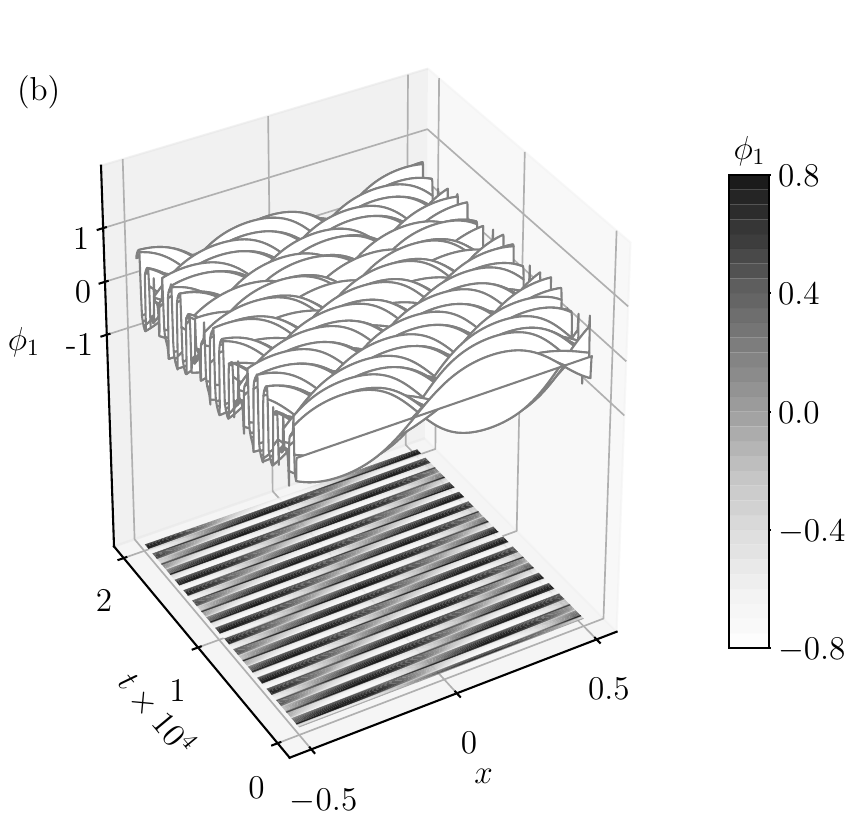}
    \includegraphics[width=0.3 \textwidth]{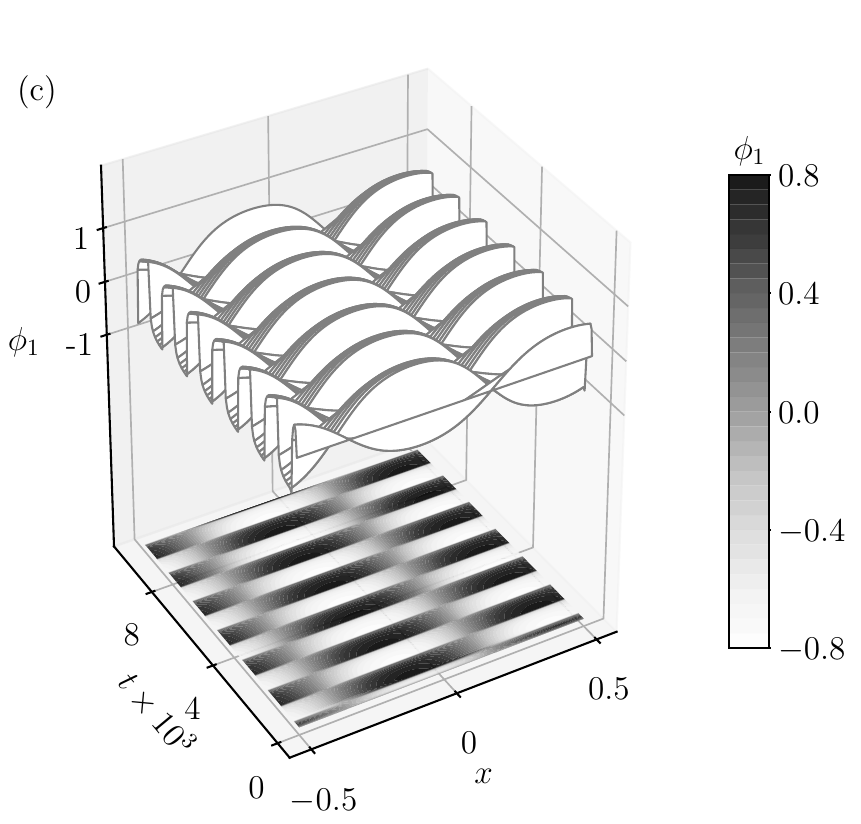}
    \caption{\small \it Space-time plots illustrating selected time-periodic behavior emerging in certain $a$-ranges for the bifurcation diagram of Fig.~\ref{fig:bGD_left_bif}~(c). We find (a) a stationary traveling wave at $a=-0.57$ [i.e., on solid part of gray branch in Fig.~\ref{fig:bGD_left_bif}~(c)]; (b) a modulated traveling wave at $a=-0.585$ [i.e., on light blue branch in Fig.~\ref{fig:bGD_left_bif}~(c)]; and (c) a standing wave at $a=-0.59$ [i.e., on solid part of red branch in Fig.~\ref{fig:bGD_left_bif}~(c)].
    %(d) at $a=-59305$\tobias{left from the fold} a standing waves of larger period; and (e) at $a=-0.6$ a steady phase-separated state [i.e., on upper part of green solid line in Fig.~\ref{fig:bGD_left_bif}~(c)].
    } \label{fig:DN_2}
\end{figure}

%The linear stability of time-periodic states has not been calculated, so we plot the corresponding Hopf branch dashed-dotted.

We begin with scenario \textbf{SubMinus} shown in Fig.~\ref{fig:bGD_left_bif} where $M_n<0$ for all $n \geq  1$, and $\alpha$ is increased from panel (a) ($\alpha<\rho$) to (e) ($\alpha>\rho$). To better understand the bifurcation behavior we first develop an argument from the linear analysis at approximately equal coupling strengths: For $\alpha \approx \rho$ the coupling term in the $\phi_2$-equation approaches zero, i.e., $\phi_2$ decouples from $\phi_1$ (but not $\phi_1$ from $\phi_2$). %and $\phi_1$ even at static $\phi_2$\tobiasbf{den letzten Teil versteh ich gar nicht. Was meinen wir damit?}.
Hence, one eigenfunction of the uniform state has amplitudes $(1,0)$ and the linear regime within this subspace is equivalent to the one for a one-field CH equation for $\phi_1$ with eigenvalue $\lambda_1$. The other is $\lambda_2$, the eigenvalue of the decoupled CH equation for $\phi_2$, with eigenvector $(-2 k^2 \rho/(\ell^2 \lambda_1),1)$. That is, both eigenvalues corresponds to decoupled CH equations, but one of the eigenvectors is not decoupled if $\rho\neq 0$. For the present $M_n<0$, then $\lambda_+=\lambda_1$ and $\lambda_-=\lambda_2$.

In Fig.~\ref{fig:bGD_left_bif}~(a) for $\alpha=1.295<\rho=1.3$ %corresponds to a completely real dispersion relation (not shown).
the steady $n=1^+$ and $n=1^-$ branches both emerge at supercritical pitchfork bifurcations at $a$-values where the real $\lambda_+$ and $\lambda_-$ cross zero at $k=k_{n=1}=2\pi$ [cf.~Eq.~\eqref{eq:lambda+-2}], respectively. The stable $n=1^+$ branch features fully phase-separated states dominated by field $\phi_1$, while the unstable $n=1^-$ branch consists of states where both fields have similar amplitudes. At first sight, the behavior is qualitatively similar to phase separation in the purely variational case although $\alpha$ is already quite large. Note, however, that at the chosen concentration values, the passive system would separate into phases~I and~III [not shown, cf.~Fig.~\ref{fig:phase_diagram}~(c)]. Here, this is not the case as the nonvariational coupling effectively decouples $\phi_2$ from $\phi_1$ as discussed above. However, no oscillatory states appear.

Increasing $\alpha$, the two primary bifurcations slowly move towards each other, while the $n=1^+$ branch develops a bulge that extends towards the $n=1^-$ branch, that itself increases the curvature of its leftward bend. Eventually, at $\alpha=\rho$ the $n=1^+$ bulge touches the $n=1^-$ bend and a bifurcation of higher codimension forms at the point of contact, see Fig.~\ref{fig:bGD_left_bif}~(b). It is noteworthy that the second primary bifurcation occurs at exactly the same value of $a$ as the high-codimension point. Caused by the complete decoupling at $\Delta=0$, $\phi_2$ is exactly zero on the complete $n=1^+$ branch. Furthermore the eigenvalue $\lambda_-=\lambda_2$ does not depend on the $\phi_1$-component of the corresponding steady state, i.e., it does not make any difference whether $\phi_1$ is uniformly zero (black branch) or structured (blue branch). In consequence, the second primary bifurcation and the first secondary bifurcation occur at identical $a$. %since both related eigenvalues only depend on the $\phi_2$ component of the steady state which is zero in both cases.
This implies that at smaller $a$ there will exist many further pairs and even groups of simultaneous bifurcations.
%We also expect them to be similarly connected via nearly vertical lines as we observe it in panel (b). 
The inset of Fig.~\ref{fig:bGD_left_bif}~(b) shows that the dotted line connecting the two bifurcations is not exact vertical. Instead it bifurcates supercritically (i.e., to the
left), folds to the right in a saddle-node bifurcation before becoming vertical again at the crossing point that coincides with its second saddle-node bifurcation. 

Slightly increasing $\alpha$ further, we obtain the diagram in Fig.~\ref{fig:bGD_left_bif}~(c) that shows very rich behavior in the region of the crossing point in Fig.~\ref{fig:bGD_left_bif}~(b). Now, the two primary bifurcations are directly linked by an $n=1$ branch of steady states. The $n=1^+$ part emerges supercritically and becomes unstable via a secondary drift pitchfork bifurcation exactly at the apex of the branch. Following Ref.~\cite{OpGT2018pre} one can derive a condition, $0=\int \left( \phi_1^2 + \frac{\rho + \alpha}{\rho- \alpha}\phi_2^2 \right) {\rm d} x$, for drift pitchfork bifurcations to occur.
%The $n=1^-$ part emerges subcritically since the chosen parameters correspond to a locus inside the orange shaded region of Fig.~\ref{fig:sub_ranges}\tobiasbf{maybe confusing da $M$ hier ein unterschiedliches Vorzeichen im Vergleich zur Fig. 9 hat, sodass blaue und orange Gebiete Farben tauschen, vielleicht besser: 
The $n=1^-$ part emerges subcritically since the chosen parameters qualitatively correspond to a locus inside the orange shaded region of Fig.~\ref{fig:sub_ranges}. When comparing, note that different parameters were used in Fig.~\ref{fig:sub_ranges}, in particular, there $M_n>0$ unlike Fig.~\ref{fig:bGD_left_bif}. In addition, the branch features a secondary Hopf bifurcation. 

The branch of traveling $n=1$ states that emerges at the drift pitchfork bifurcation is first linearly stable [cf.~Fig.~\ref{fig:DN_2}~(a)], then destabilizes in a Hopf bifurcation before finally ending in another drift pitchfork bifurcation on the unstable part of the ``upper left part'' of the steady $n=1$ branch (green dotted line). The latter then stabilizes in a saddle-node bifurcation at $a \approx -0.591$ (green solid line).

Tracking the branches of time-periodic states emerging at the Hopf bifurcations until their termination is numerically rather challenging. Therefore we accompany the continuation results with results of direct time simulations [marked by bold ``+''-symbols in the inset of Fig.~\ref{fig:bGD_left_bif}~(c)]. Fig.~\ref{fig:DN_2} shows a selection of space-time plots which illustrate the various qualitatively different behaviors at different values of $a$. All time evolutions are initialized with a noisy homogeneous state. Drawing on both sets of results proposes the following bifurcation behavior: At the Hopf bifurcation on the stationary $n=1$ branch (blue line) a branch of standing waves (red dotted line) emerges supercritically, i.e., towards smaller $a$, and carries one unstable eigenvalue. A branch of modulated waves (light blue line) emerges supercritically at the Hopf bifurcation of the traveling wave state (gray line) and is at first stable. An example of such a state is given in Fig.~\ref{fig:DN_2}~(b). The magnification in Fig.~\ref{fig:bGD_left_bif}~(c) focuses on the region where both branches of time-periodic states approach each other. 
Taking results from continuation and time simulations into account one can discern that the branch of modulated waves terminate on the branch of standing waves at $a \approx -0.588$. At the corresponding drift bifurcation, the standing waves gain stability [transition from dotted to solid line, cf.~Fig.~\ref{fig:DN_2}~(c)]. The corresponding branch continues toward a global homoclinic bifurcation on the unstable part of the $n=1$ branch of steady states (green dotted line). In particular, we find a narrow window of multistability of standing waves and steady states.
 
A further increase of $\alpha$, gives Fig.~\ref{fig:bGD_left_bif}~(d), where the half-loop of $n=1$ states connected to the primary bifurcations has shrunk. Note that we do not include the time-periodic states but only the drifting ones (gray line).
%\ttuwe{are both Hopf branches still there? Can you get them? Also for (e)?}\tobias{to be done}
%
With further increasing $\alpha$ the two primary pitchfork bifurcations move closer together and eventually fuse into a primary Hopf bifurcation when the two eigenvalues form a complex conjugate pair. The result is a bifurcation diagram as in Fig.~\ref{fig:bGD_left_bif}~(e), where the branch of traveling states directly emerges in a primary Hopf bifurcation. Note that at the transition between the structure of primary bifurcations in Figs.~\ref{fig:bGD_left_bif}~(d) and (e) more than two bifurcations fuse to become the Hopf bifurcation.
% higher codimension?
% \tobias{ Or can a Hopf Bif. occur on the tivial branch in a different way as presented here? } % GOOD QUESTION, I DO NOT HAVE THIS SCENARIO IN MY LIST OF STANDARD SCENARIOS.

Finally, we briefly discuss the high codimension point in Fig.~\ref{fig:bGD_left_bif}~(b): If all the structure described for Fig.~\ref{fig:bGD_left_bif}~(c) emerges at the high codimension point of Fig.~\ref{fig:bGD_left_bif}~(b), only considering secondary bifurcations this point ``contains'' one Bogdanov-Takens bifurcations, a double drift pitchfork bifurcation and an inverse necking bifurcation, i.e., three standard codimension-2 bifurcations. To test this, we present in Fig.~\ref{fig:bGD_left_bif}~(f) the loci of all local secondary bifurcations visible in Fig.~\ref{fig:bGD_left_bif}~(c) in the ($\alpha$, $a$)-plane. They are obtained by two-parameter continuations. Indeed, the picture indicates that all five tracked bifurcations emerge from the single point of high codimension marked by the square symbol in Fig.~\ref{fig:bGD_left_bif}~(b). Note that this seemingly strongly nongeneric behavior is also observed for nonzero mean concentrations (not shown) and is a consequence of the decoupling at $\Delta=0$. Replacing the linear coupling by a nonlinear one incorporates further parameters decreases the nongenericity. 
% we expect the high codimension point to only arise for zero mean concentrations. 

The second scenario for the onset of primary time-periodic behavior is \textbf{SubPlus}: In agreement with conditions~\eqref{eq:scenarios}, we need to change the sign of $M_n$ or of $\alpha$: In Fig.~\ref{fig:bGD_right_bif} we use $M_n<0$ for all $n\geq 1$ and decrease the nonvariational coupling in two steps from $\alpha > -\rho$ to $\alpha < -\rho$ while keeping the remaining parameters as in Fig.~\ref{fig:bGD_left_bif}. We find, that in contrast to the rich transition behavior in scenario \textbf{SubMinus}, scenario \textbf{SubPlus} is less intricate.

In  Fig.~\ref{fig:bGD_right_bif}~(a) for $\alpha=-\rho=-1.3$ the systems shows a CH instability and both, $n=1^+$ and $n=1^-$, branches emerge supercritically. As $\Delta=0$, again one field is decoupled, here it is $\phi_1$ (due to the switched sign of $\alpha$). In contrast to scenario \textbf{SubMinus}, where the $n^+$ branch is characterized by $\phi_2=0$, here the $n^-$ branch features a zero $\phi_1$-field. Thus, the argument for the simultaneous occurrence of a pair of bifurcations on the trivial branch and the $n=1^+$ branch does not apply. Instead the $n=1^+$ branch stays stable and no point of higher codimension appears.
Decreasing $\alpha$, the primary bifurcations approach each other, see Fig.~\ref{fig:bGD_right_bif}~(b). 
Furthermore, the $n=1^+$ branch becomes subcritical, i.e., a parameter range is reached that is similar to the blue shaded region in Fig.~\ref{fig:sub_ranges}.

Finally, the two primary bifurcations collide at the Hopf threshold [see Eq.~\eqref{eq:hopf_onset2}] 
%the nonlinear parts of the two steady $n=1$ branches connect in a saddle-node bifurcation 
and with further decreasing $\alpha$ a branch of time-dependent states emerges not unlike a zipper. It connects the primary Hopf bifurcation via a branch of stationary traveling states with a drift pitchfork bifurcation on the unstable part of the branch of steady $n=1$ states. There exists a further Hopf bifurcation where a branch of modulated traveling states emerges (not shown). 

In summary, in scenario \textbf{SubPlus} stable time-periodic behavior only arises when the primary pitchfork bifurcations collide at the onset of a Hopf instability and the appearance of the related Hopf bifurcation. In contrast, in the earlier considered scenario \textbf{SubMinus}, stable time-periodic states directly emerge nonlinearly when the nonvariational coupling dominates the variational one ($|\alpha|>|\rho|$). Then they determine the behavior for a wide parameter range even before a Hopf instability occurs.

It is remarkable that in the case of a purely nonvariational coupling time-periodic behavior may occur at arbitrarily small nonvariational coupling. This is analyzed in appendix~\ref{sec:app:small}.

%%%%%%%%%%%%%%%%%%%%%%%%%%%%%%%%%%%%%%%%%%%%%%%%%%%%%%%%%%%%%%%%%%%%%%%%%%%%%%%
\section{Conclusion}\label{sec:conc}
%%%%%%%%%%%%%%%%%%%%%%%%%%%%%%%%%%%%%%%%%%%%%%%%%%%%%%%%%%%%%%%%%%%%%%%%%%%%%%%
%
We have systematically analyzed the influence of nonvariational (or active, or nonreciprocal) coupling in a generic two-field Cahn-Hilliard (CH) model describing, e.g., phase separation in a ternary mixture by the coupled evolution of two concentration fields. This has shed light on activity-induced transitions from large-scale phase decomposition (mediated by coarsening) to the formation of steady patterns with finite typical length scales on the one hand and to time-periodic and  drifting behavior on the other hand. We particularly emphasize that the chosen coupling does not affect the conservation properties, i.e. both fields stay conserved in the passive and the active case. This qualitatively differs from Ref.~\cite{SATB2014c} where the coupling breaks both conservation laws.

The employed linear coupling between the two species corresponds to cross-diffusion and has a symmetric (variational) and an antisymmetric (nonvariational) contribution. The variational part corresponds to simple thermodynamic Fickian cross-diffusion. The antisymmetric part represents the active element of the model as it breaks the gradient dynamics structure of the passive case. The corresponding inter-species interaction is nonreciprocal as it breaks the third law of Newtonian mechanics \cite{IBHD2015prx}.

On the one hand, we have studied the active two-field CH model as generic model for the influence of activity on structure formation when the full conservation properties are kept. This contrasts most other active models in the literature. For instance, reaction-diffusion (RD) models do normally not feature conserved quantities \cite{Turi1952ptrslsbs,CrHo1993rmp,Liehr2013}. When recently the role of conservation laws in such systems attracted increasing attention \cite{JoBa2005pb,PeZi2006pre,YEMC2015sr,HaFr2018np,BrHF2018arxiv} - normally, one conservation law was considered in a multi-species model.
On the other hand, the model shall allow one to discuss the behavior of particular active systems where all relevant species are conserved and (molecular) interactions may result in phase decomposition. For instance, the model is well suited to describe the dynamics of different chemical or biological species that show nonreciprocal interactions but do not transform into each other or otherwise change their number on the considered time scales. This includes catalytic species whose chemical interaction is mediated via other species not explicitly described by the model, or bacteria/cell populations with a predator-prey type attraction-repulsion pattern, e.g., mediated via chemicals. At sufficiently large densities attractive and repulsive interactions may occur, e.g., as in active emulsions \cite{WZJL2019rpp}.

In the case of catalytic reactions, it is found that enzyme clustering can increase the efficiency of a two-step RD system where two enzymes are involved and one of them processes a substance into an unstable intermediate and the other one transforms the intermediate into a product \cite{CWXJ2014nb}. Interestingly, optimal separation distance and size of clusters where both enzymes are co-located are predicted which are similarly found in nature \cite{AKSB2008science}. Such patterns of species location optimize the so-called proximity channeling (two catalysts positioned sufficiently close to each other) \cite{CWXJ2014nb}. In this context our results suggest that the co-location in large coarsening clusters could be obtained via the (thermodynamic) reciprocal coupling. However, an effective nonreciprocal couplings will then favor patterns of clusters preventing coarsening toward the fully phase-separated state. In the case of populations of bacteria or cells, see Ref.~\cite{AgGo2019prl} for a discussion of phase separation in a microscopic model of two species of interacting particles (of conserved numbers) where nonreciprocity is established via nonequilibrium chemical interactions. Mean-field models such as the one studied here may then be derived by coarse-graining, as done, for example, for mixtures of active and passive Brownian particles (see SI Section VI.B in \cite{YoBM2020pnas}) or in mixtures of colloids with competing repulsive and attractive interactions (see SI Section VI.A in \cite{YoBM2020pnas}). An alternative  model describes self-propulsion of active fluids modeled by two conserved scalar fields which represent the interior/exterior of the droplet and the amount of active material, respectively \cite{SiTC2020prr}. Their dynamics is determined by CH type free energy functionals, (passive) linear coupling and advection. In contrast to the present case, there activity enters via an active stress in the Stokes description of the hydrodynamic flow velocity.

Already the linear stability analysis of the uniform state in Sec.~\ref{sec:linear} has uncovered a surprisingly rich behavior. In particular, we have shown that Turing and Hopf instabilities may occur if the nonvariational coupling dominates the variational one ($|\alpha|>|\rho|$). Notably, we could establish a direct relation to the linear stability analysis of the classical Turing system of two coupled RD equations \cite{Turi1952ptrslsbs}. In consequence, parallels are drawn between the respective parameters. Most importantly, the product of the ratio of rigidities $\kappa$ and the ratio of mobilities $Q$ in our case takes the important role of the ratio of diffusion constants in the RD system. This directly implies that much of the more intricate behavior is only found if $Q \kappa\neq1$ (and finite). This is important as for simplicity sometimes such ratios are set to one or zero \cite{SaAG2020prx,YoBM2020pnas}.

In particular, $Q \kappa\neq1$ is a necessary condition for a Turing instability. It can not be found otherwise. Here, one focus has been on the transition from CH to Turing instability and the resulting changes in coarsening behavior. We have provided an analysis based on the bifurcation behavior for relatively small systems. For larger systems an occurring Turing instability implies that further features can be expected like the existence of localized states with their slanted snakes-and-ladders bifurcation structure \cite{TARG2013pre,Knob2016jam}. This aspect is pursued in Ref.~\cite{FrTh2020arxiv}.

%After the linear analysis, we have briefly treated the passive limit of the model where well-studied phase separation and coarsening dynamics characterizes the system that ultimately approaches an equilibrium state. We have related common phase diagrams to corresponding bifurcation diagrams which have then served as the main tool to discuss the behavior in the nonvariational case.

After a brief discussion of coarsening dynamics and the corresponding bifurcation structure in the variational case, Sec.~\ref{sec:nonlinear-nonvari} has presented our first main results, namely, the dramatic effects of nonvariational coupling on the coarsening dynamics. We have uncovered three different mechanisms of suppression of coarsening, which we termed (i) linear complete suppression, (ii) nonlinear complete suppression and (iii) nonlinear partial suppression of coarsening. We have shown how the linear complete suppression relates to the Turing instability and have explained why it only occurs for supercritical primary bifurcations. In consequence, it may result in reverse coarsening dynamics. Further, we have related the nonlinear complete and partial suppression to secondary bifurcations where particular patterns stabilize. This may occur in the case of supercritical as well as subcritical primary bifurcations.

Suppression of coarsening is also found in two-field CH type models where the coupling breaks the conservation property \cite{PoTo2015jsm,SATB2014c}. Similarly, coarsening is suppressed in a general two-component RD system with one conservation law when the conservation property is relaxed \cite{BWHY2021prl}. Non-massconserving bacterial proliferation terms have a similar effect in a one-field CH type model \cite{CMPT2010pnasusa}, a behavior also found for a standard CH model with a linear nonconserved term \cite{PoTo2015jsm}. A similar transition is described by a thin-film model of a Rayleigh-Taylor unstable heated evaporating liquid film \cite{BeMe2006prl} - another model of CH type. There, one may call the term that drives the transition ``nonvariational evaporation''. Several studies consider active phase separation employing models for active media that obey a conservation law where the species show coarsening dynamics \cite{WTSA2014nc,CaTa2015arcmp,SBML2014prl,BeRZ2018pre,BeZi2019po}. We emphasize that in contrast to all these cases, we have shown that activity can suppress coarsening without relaxation of mass conservation.

Our analysis has further highlighted that subtle simplifications as employed in Refs.~\cite{SaAG2020prx,YoBM2020pnas} can already have dramatic consequences. Ref.~\cite{SaAG2020prx} considers the case of equal rigidities ($\kappa=1$) and mobilities ($Q=1$). This has turned out to be a nongeneric case where the Turing instability and all stabilizing stationary secondary pitchfork bifurcations are absent. Ref.~\cite{YoBM2020pnas} mostly investigates a limiting case where one rigidity is set to zero ($\kappa=0$) and also includes a few results for $\kappa=Q=1$. In consequence, no suppression of coarsening is observed. %It also seems that the amendment does not need to be nonvariational, the CH model with a non-massconserving term in Ref.~\cite{PoTo2015jsm} can actually be written as a conserved gradient dynamics for a nonlocal energy functional.
  Future work should investigate if there are qualitative differences between the various cases of coarsening suppression in dependence of variational and conservation properties of the models. The differences could concern the prevalence of linear vs.\ nonlinear mechanisms of the suppression of coarsening or systematic changes to underlying scaling laws \cite{KoOt2002cmp,ACRT2005pre}.
An analysis of the case of subcritical primary bifurcations has revealed another intriguing feature of the model. The standard scenario for phase separation modeled by the CH equation is the occurrence of subcriticality only beyond a certain nonzero mean concentration \cite{Novi1985jsp}. However, here we have shown based on a weakly nonlinear analysis that, surprisingly, in a two-field model a nonvariational coupling can cause subcriticality even at zero mean concentrations. This indicates that common symmetry arguments indicating supercritical behavior have to be amended.

Our second main result concerns the emergence of time-periodic and drifting states. For simplicity, in this point we have focused on the case of zero mean concentrations, however, the described behavior also occurs for nonzero mean concentrations. First, we have found that in the strongly nonlinear regime where multistability of many different steady patterns occurs, one may also find stable time-periodic states. This is not indicated by the linear analysis of uniform states as in the corresponding parameter range it only shows a stationary instability. The time-periodic behavior emerges via the appearance of windows of oscillatory patterns on the primary branches via secondary Hopf bifurcations.
Second, we have focused on the parameter range where the Hopf instability of the uniform state is approached. There, two distinguished scenarios for the emergence of time-dependent fully phase-separated states occur. % that we have named \textbf{SubPlus} and \textbf{SubMinus}.
%  Both scenarios appear to be rather generic for nonreciprocally coupled fields since in both the emergence of time-periodic states is preceded by subcritical behavior. The latter we could clearly relate to the nonreciprocal coupling.
  In one scenario (\textbf{SubPlus}), branches of traveling and standing waves directly emerge from the uniform state when its instability changes from stationary (CH) to oscillatory (Hopf). In the other scenario (\textbf{SubMinus}), time-periodic states already occur when the uniform state is still unstable with respect to a stationary mode related to the emergence of an inhomogeneous steady state. This state then starts to move triggered by a secondary drift-pitchfork bifurcation. Similar behavior is also found for other active models, e.g., an active phase-field-crystal model \cite{OpGT2018pre}. At the drift-pitchfork bifurcation the parity symmetry of the steady state is broken and the newly emerging state drifts with a velocity that shows the typical square-root behavior. This agrees with the result of a one-mode approximation done for the simplified coupled CH model in Ref.~\cite{YoBM2020pnas}. Further secondary Hopf and drift-pitchfork bifurcations create a rich variety of time-dependent states. %Note that multistable regions exist where steady and time-dependent states are both linearly stable.
Time-dependent states are reported in the related models studied in Refs.~\cite{SATB2014c,SaAG2020prx,YoBM2020pnas} based on linear analyses and time simulations. However, to compare the emergence of time-periodic behavior to our cases, one first needs to supplement the literature studies by investigations of the underlying bifurcation structure.

Our model features a simple linear coupling of the two fields. It is chosen as a simple and transparent option natural for weakly coupled systems. However, this surely corresponds to a limitation that should be lifted when considering strong (variational and/or nonvariational) coupling. In the future, it might then be interesting to study how the interplay of nonlinear variational and nonvariational couplings alters the observed behavior when for instance general cubic polynomials in both fields are used. First results for a nonlinear variational coupling are given in the linear analysis in Ref.~\cite{SaAG2020prx}. We conclude from their results that nonlinear coupling may suppress the Hopf instability. However, local nonlinear coupling terms alone cannot cause instabilities qualitatively different from the case of linear coupling. %To reach a full understanding of the nonlinear impact a bifurcation analysis in the generic case of different rigidities and mobilities should be performed.

  Note finally, that our results are relevant for a wider class of CH type systems. This includes thin-film models (long-wave hydrodynamics) \cite{OrDB1997rmp} for two-layer films of nonvolatile immiscible liquids on heated or cooled substrates \cite{PBMT2005jcp,NeSi2007pf}. There, the heating takes the role of the nonvariational coupling and the passive CH type dynamics results from an interplay of interfacial tensions and wettability. Notably, in this case variational and nonvariational coupling are both highly nonlinear. There are as well matrices of nonlinear mobilities replacing the present constant diagonal matrix. Our results suggest that this system will also exhibits a short-scale stationary instability not yet reported in the literature.
    % Another approach where amplitude equations near the Hopf instability are derived from the full hydrodynamic equations of the two liquid layer system yields nonvariationally coupled CH and diffusion equations \cite{KNSZ1998pre}. A similiar analysis near the codimension-2 point where both layers become unstable might lead to two coupled CH equations.
The investigation of other nonvariational contributions as known from one-field models is also an option, see e.g.\ the classification in the introduction of Ref.~\cite{EGUW2019springer}. Further, the bifurcation structure of the presented generic model should also be investigated for two-dimensional systems.

The data that support the findings of this study and selected computer codes are openly available \cite{FrWT2021zenodo}.

\newpage
%%%%%%%%%%%%%%%%%%%%%%%%%%%%%%%%%%%%%%%%%%%%%%%%%%%%%%%%%%%%%%%%%%%%%%%%%%%%%%%
\appendix
\section{Nondimensionalization}
\label{sec:app-nondim}
%%%%%%%%%%%%%%%%%%%%%%%%%%%%%%%%%%%%%%%%%%%%%%%%%%%%%%%%%%%%%%%%%%%%%%%%%%%%%%%
%
This appendix discusses our nondimensionalization of the coupled CH model and thereby elucidates the physical meaning of the various nondimensional parameters of the model~\eqref{eq:nondim_bG_final}. We start with the dimensional coupled system
\begin{alignat}{2}
\begin{aligned}
\qquad \frac{\partial \phi_1}{\partial t}  &= Q_1 \frac{\partial^2}{\partial x^2} \left(-\kappa_1 \frac{\partial^2 \phi_1}{\partial x^2}  + \zeta_1 f'_1(\phi_1) - \gamma_1 \phi_2 \right)\\
\qquad  \frac{\partial \phi_2 }{\partial t}&= Q_2 \frac{\partial^2}{\partial x^2} \left(-\kappa_2\frac{\partial^2 \phi_2}{\partial x^2}  + \zeta_2 f'_2(\phi_2)  - \gamma_2 \phi_1 \right) \ ,
\end{aligned}
\label{eq:gekoppelteCHNondim0}
\end{alignat}
and express the dimensional fields as $\phi_1 = \hat{\phi}_1 \tilde \phi_1$ and $\phi_2 = \hat{\phi}_2 \tilde{\phi}_2$, where a hat indicates a fixed scale (to be determined) and a tilde the nondimensional quantity. Furthermore, we introduce energy scales via $f_i=\hat f_i \tilde f_i$, and time and length scales $\tau$ via $t = \tau\tilde{t}$ and $L$ via $x = L \tilde{x}$, respectively. Here $L$ is the dimensional physical domain size, i.e., the scaled domain size that corresponds to our computational domain size is fixed to one. However, as we want to discuss heterogeneous states of different mode numbers $n$, we write $L = \ell L_0$ where $L_0$ is a typical length scale, e.g., given by the critical wavelength for a one-field CH equation at a reference temperature and $\ell$ remains as a nondimensional parameter named ``nondimensional domain size''. Then, choosing it as $\ell=2\pi n_\mathrm{max}$ allows one to consider the interaction of different numbers of modes as at the reference state only modes with $n=1\dots n_\mathrm{max}$ occur.
% For practical reasons we introduce the (nondimensional) ratio $L=\frac{\hat{L}}{\ell}$ with $\ell$ being the computational domain size \tobias{on which we apply the continuation method}.
%In this way we can fix the computational domain size $\ell=1$ and use the nondimensional ratio $L$ to control the physical domain size. \tobias{In the main text we call $L$ the nondimensional domain size which can be used to consider continuation in domain sizes.}
After introducing the scales and re-grouping parameters, the system of nondimensional equations is 
% \begin{alignat}{2}
% \begin{aligned}
%  \frac{\partial  \tilde \phi_1}{\partial \tilde{t}}\, &= 
% \frac{Q_1 \tau}{L^2} \frac{\partial^2}{\partial \tilde{x}^2} \left(-\frac{\kappa_1}{L^2} \frac{\partial^2 \tilde \phi_1}{\partial \tilde{x}^2} + \frac{\zeta_1}{\hat{\phi}_1} \, f'( \hat{\phi}_1 \tilde{\phi}_1 )  - \frac{\hat{\phi}_2}{\hat{\phi}_1}\gamma_1 \tilde \phi_2 \right)\\
%  \frac{\partial \tilde \phi_2}{\partial \tilde{t}}\,  &= 
% \frac{Q_2 \tau}{L^2} \frac{\partial^2}{\partial  \tilde{x}^2} \left(-\frac{\kappa_2 }{L^2}  \frac{\partial^2 \tilde \phi_2}{\partial \tilde{x}^2} + \frac{\zeta_2}{\hat{\phi}_2} \,g'( \hat{\phi}_2 \tilde{\phi}_2  )  - \frac{\hat{\phi}_1}{\hat{\phi}_2}\gamma_2 \tilde\phi_1 \right). 
% \end{aligned}
% \label{eq:gekoppelteCHNondim1}
% \end{alignat}
%
\begin{eqnarray}
\qquad \frac{\partial}{\partial \tilde{t}}\, \tilde \phi_1 &=& 
\frac{Q_1 \tau \kappa_1}{L_0^4} \frac{1}{\ell^2} \frac{\partial^2}{\partial \tilde{x}^2} \left( -\frac{1}{\ell^2} \frac{\partial^2 \tilde \phi_1}{\partial \tilde{x}^2} +   \frac{\zeta_1L_0^2}{\hat{\phi}^2_1\kappa_1} \, \hat f_1 \tilde f'_1(\tilde{\phi}_1)  - \frac{\hat{\phi}_2 \gamma_1 L_0^2}{\hat{\phi}_1 \kappa_1} \tilde \phi_2 \right)\nonumber\\
\qquad \frac{\partial}{\partial \tilde{t}}\, \tilde \phi_2 &=&
\frac{Q_2}{Q_1} \frac{Q_1 \tau \kappa_1}{L_0^4}\frac{1}{\ell^2} \frac{\partial^2}{\partial  \tilde{x}^2} \left(-\frac{\kappa_2 }{\kappa_1 \ell^2} \frac{\partial^2 \tilde \phi_2}{\partial \tilde{x}^2} + \frac{\zeta_2L_0^2}{\hat{\phi}^2_2\kappa_1} \,\hat f_2 \tilde f'_2(\tilde{\phi}_2)  - \frac{\hat{\phi}_1 \gamma_2 L_0^2}{\hat{\phi}_2 \kappa_1} \tilde\phi_1 \right) \, .
\label{eq:gekoppelteCHNondim2}
\end{eqnarray}
It contains nondimensional combinations of physical parameters and of the scales $\tau, \hat{\phi}_1$, and $\hat{\phi}_2$ that still need to be chosen. We define
\begin{align}
Q \equiv \frac{Q_2}{Q_1} \,\text{,} \quad
\tilde \gamma_1 \equiv \frac{\hat{\phi}_2 \gamma_1 L_0^2}{\hat{\phi}_1 \kappa_1} \,\text{,} \quad
  \tilde \gamma_2 \equiv \frac{\hat{\phi}_1 \gamma_2 L_0^2}{\hat{\phi}_2 \kappa_1} \,\text{,} \quad
  \kappa \equiv \frac{\kappa_2 }{\kappa_1 } \,\text{,} \quad
  \tilde b_1 \equiv \frac{\zeta_1 \hat f_1 L_0^2}{ \kappa_1 \hat{\phi}^2_1} 
  \,\text{,} \quad \tilde b_2 \equiv \frac{\zeta_2 \hat f_2 L_0^2}{\kappa_1 \hat{\phi}^2_2}\, ,
 \end{align}
set $\tau =L_0^4/(Q_1 \kappa_1)$,  and obtain
  \begin{alignat}{2}
\begin{aligned}
  \qquad \frac{\partial}{\partial \tilde{t}}\, \tilde \phi_1 &= 
\frac{1}{\ell^2} \frac{\partial^2}{\partial \tilde{x}^2} \left(-\frac{1}{\ell^2}\frac{\partial^2 \tilde \phi_1}{\partial \tilde{x}^2} +  \tilde b_1\, \tilde f_1'(\tilde{\phi}_1)  - \tilde \gamma_1 \tilde \phi_2 \right)\\
\qquad \frac{\partial}{\partial \tilde{t}}\, \tilde \phi_2 &= \frac{Q}{\ell^2} \frac{\partial^2}{\partial  \tilde{x}^2} \left(-\frac{\kappa }{\ell^2}  \frac{\partial^2 \tilde \phi_2}{\partial \tilde{x}^2} + \tilde b_2 \,\tilde f_2'(\tilde{\phi}_2)  - \tilde \gamma_2 \tilde\phi_1 \right) \, .
\end{aligned}
\label{eq:gekoppelteCHNondim3}
\end{alignat}
We assume the bulk energies to be double-well potentials, i.e. $f_i(\phi_i)= \frac{a_i}{2}\phi_i^2 + \frac{b_i}{4}\phi_i^4$.
We use $\hat f_i = b_i \hat \phi_i^4$ and obtain 
\begin{equation}
\tilde f’_i = \tilde a_i \tilde\phi_i + \tilde\phi_i^3\, \quad\text{with} \quad \tilde a_i =\frac{a_i}{b_i \hat \phi_i^2}\,.
\end{equation}
Furthermore we demand $\tilde b_i =\frac{\zeta_i \hat f_i L_0^2}{ \kappa_1 \hat{\phi}^2_i} = \frac{\zeta_i b_i \hat \phi_i^2 L_0^2}{ \kappa_1}= 1$, i.e. we set
\begin{align}
\hat{\phi}_1= \sqrt{\frac{\kappa_1}{ \zeta_1 b_1 L_0^2}} \,\text{,} \qquad \hat{\phi}_2=\sqrt{\frac{\kappa_1}{ \zeta_2 b_2 L_0^2}}\, ,
\end{align}
and obtain
\begin{alignat}{2}
\begin{aligned}
\qquad \frac{\partial}{\partial \tilde{t}}\, \tilde \phi_1 &= 
\frac{1}{\ell^2} \frac{\partial^2}{\partial \tilde{x}^2} \left(-\frac{1}{\ell^2}\frac{\partial^2 \tilde \phi_1}{\partial \tilde{x}^2} +  \tilde a_1 \tilde{\phi}_1 + \tilde{\phi}^3_1  - \tilde \gamma_1 \tilde \phi_2 \right)\\
\qquad \frac{\partial}{\partial \tilde{t}}\, \tilde \phi_2 &= \frac{Q}{\ell^2} \frac{\partial^2}{\partial  \tilde{x}^2} \left(-\frac{ \kappa }{\ell^2}  \frac{\partial^2 \tilde \phi_2}{\partial \tilde{x}^2} + \tilde a_2 \tilde{\phi}_2 + \tilde{\phi}^3_2    - \tilde \gamma_2 \tilde\phi_1 \right) \, .
\end{aligned}
\label{eq:gekoppelteCHNondim4}
\end{alignat}
The parameter $a$ in the linear term of $f'$ for the one-field CH equation is often referred to as effective temperature. Here, we define $\tilde a_2 = \tilde a_1 + a_\Delta \equiv a + a_\Delta$.  The effective temperature $a$ is used as a main control parameter and $a_\Delta$ represents the shift in critical temperature between the two decoupled CH equations. Furthermore, the two coupling parameters $\gamma_i$ are split into symmetric ($\rho$) and antisymmetric ($\alpha$) contributions
$$
\rho \equiv \frac{\tilde{\gamma}_1 + \tilde{\gamma}_2}{2}\,,  \quad 
\alpha \equiv \frac{\tilde{\gamma}_1 - \tilde{\gamma}_2}{2}\, .
$$
They represent variational (reciprocal) and nonvariational (nonreciprocal) coupling between the two fields, respectively.
Dropping the tildes, one obtains the nondimensional system
\begin{alignat}{2}
\begin{aligned}
\qquad \frac{\partial}{\partial t} \phi_1 &= \frac{1}{\ell^2} \frac{\partial^2}{\partial x^2} \left(- \frac{1}{\ell^2} \frac{\partial^2 \phi_1}{\partial x^2} + f'_1(\phi_1) - \left( \rho+ \alpha\right) \phi_2 \right)\\
\qquad \frac{\partial}{\partial t} \phi_2 &= \frac{Q}{\ell^2} \frac{\partial^2}{\partial x^2} \left(- \frac{\kappa}{\ell^2} \frac{\partial^2 \phi_2}{\partial x^2} + f'_2(\phi_2)- \left( \rho - \alpha\right) \phi_1 \right) \, ,
\end{aligned}
\label{eq:nondim_final}
\end{alignat}
with $f'_1(\phi_1) = a \phi_1  + \phi_1^3$ and $ f'_2(\phi_2) = (a+a_\Delta) \phi_2  + \phi_2^3$. It corresponds to Eqs.~\eqref{eq:nondim_bG_final} of the main text.

% Furthermore, the conservations of the mean concentrations are also expressed in nondimensionalised variables:
% \begin{align}
% \int \textrm{d}x \phi_1 = \bar{\phi}_{1}
%  ~\\
%   \int \textrm{d}x \phi_2 = \bar{\phi}_{2}.
% \end{align}

\section{Linear stability results for specific $f_i$}\label{app:linear}
In Section~\ref{sec:linear} we have analyzed the linear stability of homogeneous states for the model~\eqref{eq:nondim_bG_final}. The stability diagrams in Fig.~\ref{fig:stab-diagram} summarize the linear results for general local energies $f_i$. 
Here, we specify the discussion for our case where $f''_1=a + 3 \bar{\phi}_1^2\, , \, \, f''_2=a+a_\Delta + 3 \bar{\phi}_2^2 \, , \, \,Q=1$ and use the effective temperature $a$ as main control parameter. The dispersion relations \eqref{eq:lambda+-} become
\begin{align}\label{eq:lambda+-2}
 \lambda_\pm(q)=&\frac{1}{2}q^2  \Bigg(-\Big[q^2\left(1+\kappa\right)+2a+a_\Delta+3 \left(\bar \phi_1^2+\bar \phi_2^2\right)\Big]
 \nonumber~\\
 & \pm\sqrt{\Big[q^2\left(1-\kappa\right)+3\left(\bar \phi_1^2- \bar \phi_2^2\right) - a_\Delta\Big]^2 - 4 \Delta}\Bigg)\,.
\end{align}
Note that we use the abbreviation $q=k/\ell$ throughout the appendix.
Since the discriminant is independent of $a$, the occurrence of complex eigenvalues does not depend on $a$.
In contrast, the coupling strengths $\rho$ and $\alpha$ only appear in the combination $\Delta= \alpha^2 - \rho^2$ and only enter the discriminant. Complex eigenvalues occur if
\begin{equation}\label{eq:hopf_onset2}
 \Delta > \frac14 \Big[q^2\left(1-\kappa\right)+3\left(\bar \phi_1^2- \bar \phi_2^2\right) - a_\Delta\Big]^2\,.
\end{equation}
Then, Hopf bifurcations of modes with wavenumber $q$ occur at [cf.~\eqref{eq:hopfonset}]
\begin{equation}
a_\mathrm{o}(q) =  -\frac{1}{2}\Big[q^2\left(1+\kappa\right)+a_\Delta+3 \left(\bar{\phi}_1^2+\bar{\phi}_2^2\right)\Big]
\label{eq:azero-os}
\end{equation}
independently of both coupling strengths. In consequence, the onset of the Hopf instability occurs at
\begin{equation}
a^\text{H} = a_\mathrm{o}(0)=  -\frac{1}{2}\Big[a_\Delta+3 \left(\bar{\phi}_1^2+\bar{\phi}_2^2\right)\Big]\,.
\label{eq:ahopf}
\end{equation}
For stationary instabilities we use Eq.~\eqref{eq:alltu} to obtain the critical values
\begin{align}
a_\pm(q)= & \frac{1}{2}\Bigg(-\Big[q^2\left(1+\kappa\right)+a_\Delta+3 \left(\bar{\phi}_1^2+\bar{\phi}_2^2\right)\Big]~\nonumber ~\\
 & \pm\sqrt{\Big[q^2\left(1-\kappa\right)+3\left(\bar{\phi}_1^2- \bar{\phi}_2^2\right) - a_\Delta\Big]^2 - 4 \Delta }\Bigg)\,.\label{eq:azero}
\end{align}
Since the dispersion relation has two branches, $\lambda_\pm(q)$, we obtain two critical $a$. The stability border in the $(q,a)$-plane is then represented by the $a_+(q)$ curve where $\lambda_+(q)$ changes sign. In particular, the onset of a CH instability is at
\begin{align}\label{eq:a_critical}
  a^{\text{CH}} = a_+(q_\mathrm{c}=0) = \frac{1}{2}\bigg(-\Big[a_\Delta+3 \left(\bar{\phi}_1^2+\bar{\phi}_2^2\right)\Big] +\sqrt{\Big[3\left(\bar{\phi}_1^2- \bar{\phi}_2^2\right) - a_\Delta\Big]^2 - 4 \Delta }\bigg)\,.  
\end{align}
For a Turing instability the onset is at nonzero $q_\text{c}$ [cf.~Eq.~\eqref{eq:kcTuring}] at
\begin{equation}
a^\text{T}\equiv a_+(q_\textrm{c})=\frac{1}{\kappa-1}\left[ a_\Delta + 3 \left(\bar \phi_2^2 - \kappa \bar \phi_1^2\right) \mp 2 \sqrt{\kappa \Delta} \right]
\label{eq:a0+}
\end{equation} 
[cf.~Eq.~\eqref{eq:a11tu2}]. 
Stability borders $a_\mathrm{o}(q)$ [Eq.~\eqref{eq:azero-os}] for oscillatory and $a_+(q)$ [Eq.~\eqref{eq:azero}] for stationary instabilities are plotted for different cases in Figs.~\ref{fig:dispersion},~\ref{fig:lin_supp_dispersion} and~\ref{fig:cEV_dispersion}. 

An alternative approach is to consider the shape of the dispersion relations [Eq.~\eqref{eq:lambda+-2}] at fixed parameters.
For $q=0$ both eigenvalues $\lambda_\pm$ are always zero as expected for two conservation laws.
Setting $\lambda_\pm=0$, we can determine other real roots. Due to parity, the dispersion relation crosses zero at 0, 1 or 2 nonzero and positive wavenumbers 
\begin{equation}
q_\pm=\sqrt{\frac{-f_2'' - \kappa f_1'' \pm \sqrt{(f_2'' - \kappa f_1'')^2 - 4\kappa \Delta}}{2 \kappa}}\,,
\label{eq:k_pm}
\end{equation}
with $f_2''$ and $f_1''$ as given above. If $q_+$ becomes real it corresponds to a nontrivial root of $\lambda_+$ (note that $\lambda_+> \lambda_-$).
If both, $q_+$ and $q_-$ are real, two nontrivial roots exist. Again $q_+$ corresponds to a root of $\lambda_+$.
The second root $q_-$ can correspond to a root of $\lambda_-$ or to a second root of $\lambda_+$. The latter case corresponds to the appearance of a Turing instability.
An oscillatory instability has its threshold when the real part of $\lambda_\pm(q)$ is zero. This gives 0 or 1 nonzero and positive wavenumber values
\begin{equation}
q_\text{o}=\sqrt{-\frac{2a + a_\Delta + 3\left( \bar\phi_1^2 + \bar\phi_2^2 \right)}{1+ \kappa}}\,,
\label{eq:k_os}
\end{equation}
implying that it is always a large-scale oscillatory (Hopf) instability.

Independently of the onset of instabilities we can determine the band of wavenumbers $\left[q^o_-,q^o_+\right]$ %\tobias{We should use different symbols, maybe $q_\pm^\text{imag}$?}
where complex eigenvalues occur by setting the discriminant in Eq.~\eqref{eq:lambda+-2} to zero. 
This yields
\begin{equation}\label{eq:kband_os}
 q^o_\pm =  \sqrt{\frac{ \pm 2\sqrt{\Delta} - 3 \left(\bar{\phi}_1^2 -  \bar{\phi}_2^2\right) + a_\Delta}{1-\kappa}}\, .
 \end{equation}
Requesting that the limiting values $q^o_\pm$ have to be real implies that for
\begin{align}\label{eq:3casesA} 
\left( \begin{array}{lcl}
 3\left(\bar{\phi}_1^2 -  \bar{\phi}_2^2\right) - a_\Delta & >  2\sqrt{\Delta}~\\
  | 3\left(\bar{\phi}_1^2 -  \bar{\phi}_2^2\right) - a_\Delta| & < 2\sqrt{\Delta} \\
 3\left(\bar{\phi}_1^2 -  \bar{\phi}_2^2\right) - a_\Delta & < -2\sqrt{\Delta}
\end{array}
\right)
\end{align}
complex eigenvalues occur in the bands
 \begin{align}        \label{eq:3cases}                                      
&  \left( \begin{array}{lcl}
        \left[q^o_-,q^o_+\right] \text{if }\kappa>1 ~\\
        \left[0,q^o_\mp\right]\text{if }\kappa \gtrless 1 ~\\
        \left[q^o_-,q^o_+\right] \text{if }\kappa<1
       \end{array}\right)\, .
\end{align}
In the special case of identical interface rigidities $\kappa_1=\kappa_2$ (i.e., $\kappa=1$), the occurrence of complex eigenvalues does not depend on  wavenumber as $q^o_\pm\rightarrow \infty$ for $\kappa \rightarrow 1$. Then for
\begin{align}
%\left[3(\bar{\phi}_1^2 - \bar{\phi}_2^2) - a_\Delta\right]^2  < 4 \Delta \, .~\\
|3\left(\bar{\phi}_1^2 -  \bar{\phi}_2^2\right) - a_\Delta| < 2\sqrt{\Delta}
\end{align}
eigenvalues are complex at any $q$.

\section{Phase behavior in variational case}
\mylab{app:phasebehavior}
Here, we briefly review the phase behavior of the coupled CH model in the variational case ($\alpha=0$). 
It can be applied to describe passive phase separation of a ternary mixture, e.g., in a liquid-liquid-gas system \cite{WiSN1998pre}.
In particular, we discuss the phase diagram in the thermodynamic limit, i.e., for an infinite domain where interfaces between phases can be neglected. The phase behavior is then related to bifurcation diagrams determined for finite systems.

\begin{figure}
 \includegraphics[width=0.8\textwidth]{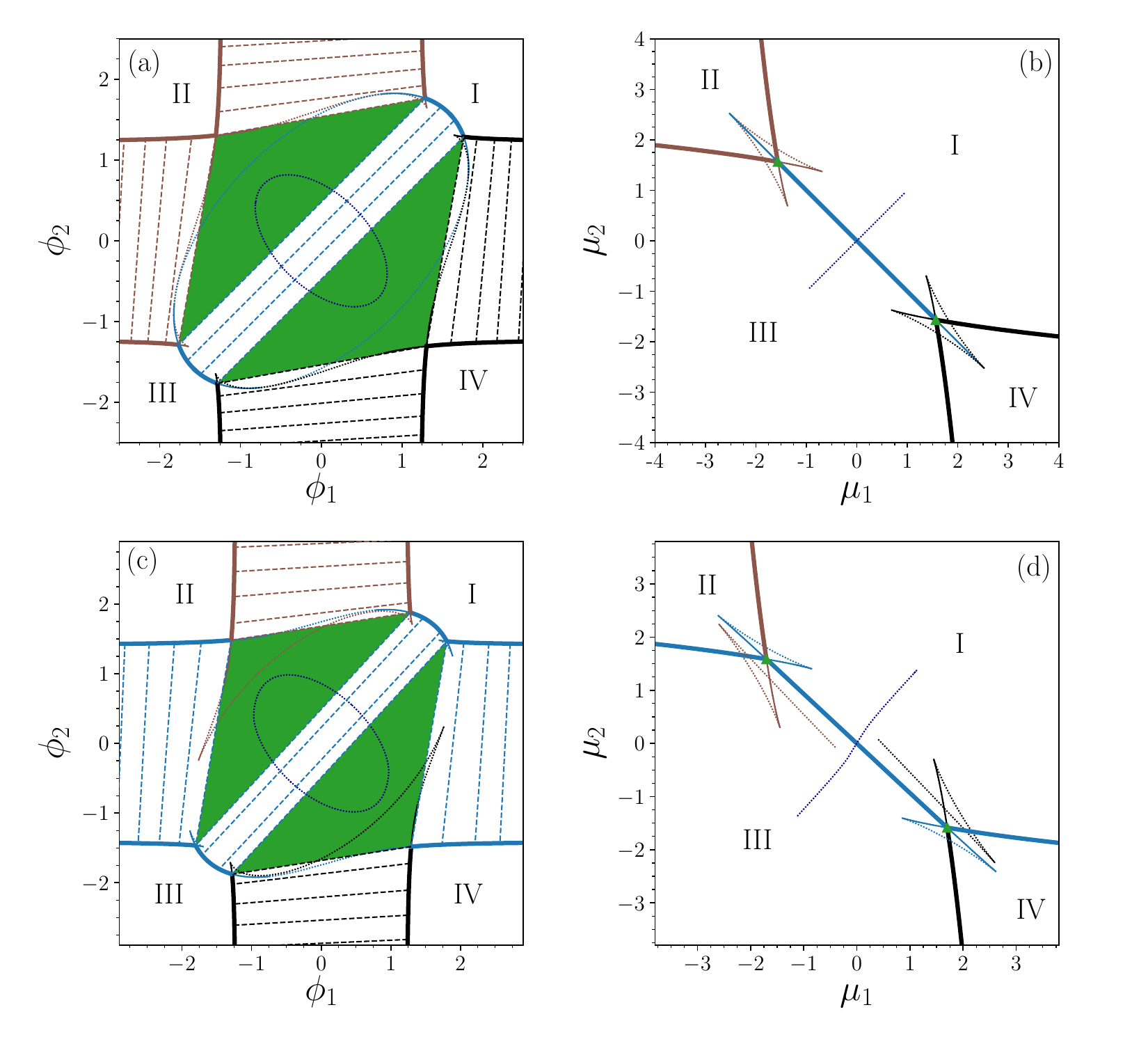}
 \caption{\small \it The phase behavior of the ternary system modeled by two variationally coupled CH equations [Eq.~(\ref{eq:model_int})] represented in planes spanned by (a, c) the mean concentrations and (b, d) the chemical potentials. Panels (a) and (b) display the fully symmetric case $a_\Delta=0$ while (c) and (d) give an asymmetric case with $a_\Delta=-0.5$. Phases I to IV are described in the main text. The heavy solid lines in (a, c) [(b, d)] represent the various binodals [phase boundaries], i.e., the states at coexistence. The thin solid and dotted lines give coexisting metastable and unstable states, respectively, while the straight dashed lines in (a, c) are tie lines connecting particular coexisting stable states. The triangular green shaded regions in (a, c) indicate three-phase coexistence and correspond to the triple points (green triangle symbols) in (b, d). The remaining parameters are $a=-1.5$ and $\rho=1$. 
} \label{fig:phase_diagram}
\end{figure}

To calculate uniform steady states in the thermodynamic limit, all spatial derivatives in Eqs.~(\ref{eq:steady_stateEq}) are set to zero to obtain
\begin{align}\label{eq:model_int}
\begin{split}
0=& f_1'(\phi_1) - \rho \phi_2 - \mu_1 \\
0=& f_2'(\phi_2)  - \rho \phi_1 -\mu_2\, .
\end{split}
\end{align}
Next, we consider two uniform states in different boxes ``A'' and ``B'' with concentrations $\phi^\text{A}_1$, $\phi^\text{A}_2$ and $\phi^\text{B}_1$, $\phi^\text{B}_2$, respectively. At coexistence, the two boxes are at equal temperature (by definition for our isothermal system), at identical chemical potentials $\mu_1$ and $\mu_2$ and at identical pressure $p$, i.e., equal grand potential density
\begin{equation}
  \omega = -p = f_1 + f_2 - \rho\phi_1\phi_2 - \mu_1\phi_1-\mu_2\phi_2\, .
  \label{eq:omega}
\end{equation}
As a result we have the three conditions
\begin{eqnarray}
\mu^{\mathrm{A}}_1&=&\mu^{\mathrm{B}}_1\, ,\nonumber\\
\mu^{\mathrm{A}}_2&=&\mu^{\mathrm{B}}_2\, ,\label{eq:coex}\\
p^{\mathrm{A}}&=&p^{\mathrm{B}}
  \nonumber
\end{eqnarray}
to determine the four unknown concentrations at coexistence leaving one of them a free parameter. Fig.~\ref{fig:phase_diagram} gives resulting phase diagrams in planes spanned by the mean concentrations and the chemical potentials, respectively. For details on the employed continuation procedure see Ref.~\cite{HoAT2021jpcm}.

 \begin{figure}
 \includegraphics[width=0.7\textwidth]{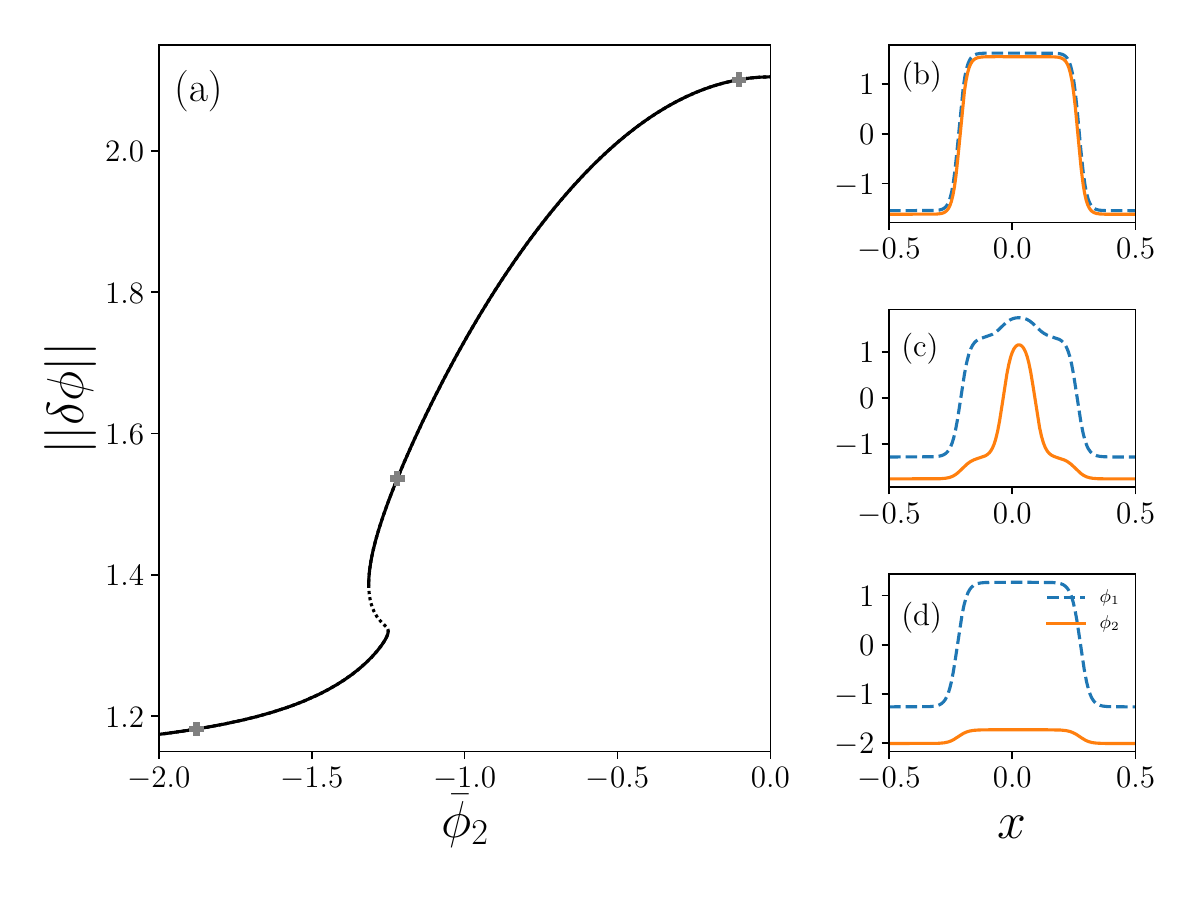}
  \caption{\small \it (a) Bifurcation diagram with control parameter $\bar\phi_2$ at fixed $\bar\phi_1=0$ for $a_\Delta=0$ [i.e., straight vertical central cut through Fig.~\ref{fig:phase_diagram}~(a)] and finite domain size $\ell=10\pi$ (and $\kappa=1$). Panels (b)-(d) give examples of concentration profiles at points marked by plus symbols in (a). The remaining parameters are as in Fig.~\ref{fig:phase_diagram}.} \label{fig:phases_finitedomain}
 \end{figure}

For $a_\Delta = 0$ the steady equations \eqref{eq:model_int} and the pressure \eqref{eq:omega} are symmetric w.r.t.~an exchange of the two fields and chemical potentials $(\phi_1,\mu_1,\phi_2,\mu_2) \to (\phi_2,\mu_2,\phi_1,\mu_1)$ and also w.r.t.\ the inversion $(\phi_1,\mu_1,\phi_2,\mu_2) \to (-\phi_1,-\mu_1,-\phi_2,-\mu_2)$. These symmetries are inherited by the phase diagrams in Figs.~\ref{fig:phase_diagram}~(a) and (b), namely, they are reflection symmetric w.r.t.\ both diagonals (exchange symmetry) and w.r.t.\ the origin (inversion symmetry). In contrast, for $a_\Delta\neq 0$ as in Figs.~\ref{fig:phase_diagram}~(c) and (d), the reflection symmetry w.r.t.\ both diagonals is broken.

 In the four corners of panel (a) and (c) one finds the four phases I to IV with various extended coexistence regions in between. The four phases may be called (I) high-$\phi_1$, high-$\phi_2$ phase, (II) low-$\phi_1$, high-$\phi_2$ phase, (III) low-$\phi_1$, low-$\phi_2$ phase, and (IV) high-$\phi_1$, low-$\phi_2$ phase.

For the present $\rho>0$ case, all phases with the exception of II and IV can pairwise coexist (for an $\rho<0$ the excluded combination will be I-III). This is best seen in the $(\mu_2,\mu_1)$-plane [Figs.~\ref{fig:phase_diagram}~(b) and (d)]. There heavy solid lines [green triangle symbols] indicate phase boundaries [triple points] where two [three] phases coexist. In the $(\phi_1,\phi_2)$-plane [Fig.~\ref{fig:phase_diagram}~(a) and (c)] two coexisting states lie on binodal lines (heavy solid lines) and are connected by tie lines (thin dashed lines) that represent the Maxwell construction in the ternary system. States between binodals are unstable w.r.t.\ phase decomposition and decompose along the tie lines. In Figs.~\ref{fig:phase_diagram}~(a) and (c), triple points become extended (green shaded) regions. States within such a region decompose into the three coexisting states at the corners of the triangle.

Note that for large $|\phi_1|$ or $|\phi_2|$ the two fields practically decouple. For instance, for $|\phi_1|\gg 1$, to leading order $\phi_1$ is uniform, and $\phi_2$ separates into states $\approx 1.25$ and $\approx-1.25$. This is already well visible in Fig.~\ref{fig:phase_diagram}~(a), even at $\phi_1=\pm 2.5$ and can also be seen in the concentration profiles discussed below. Actually, in the slightly artificial limit $\rho\to0$ the two fields entirely decouple and the $(\phi_1,\phi_2)$-phase diagram converges to vertical and horizontal binodal lines at $\pm\sqrt{-a}$ \footnote{Note, that for $a_\Delta \neq 0$ the binodals converge to horizontal lines at $\pm\sqrt{-a-a_\Delta}$ and vertical lines at $\pm\sqrt{-a}$ which reflects the broken exchange symmetry, see e.g.~Fig.~\ref{fig:phase_diagram}~(c).}. Their crossing points define a square that contains a ``four-phase coexistence'' region. In the $(\mu_2,\mu_1)$-plane vertical and horizontal lines at zero chemical potential cross at the origin that corresponds to a quadruple point of four-phase coexistence.
% \tobias{Discussion of the asymptotics? I.e. for $|\phi_2|\rightarrow \infty$ for both boxes (with the same sign, i.e., $\phi^{\text{A}}_2=\phi^{\text{B}}_2=\phi_2$  ) we make an educated guess that $\phi^\text{A,B}_1=\pm 1$. Then Eqs.~\eqref{eq:model_int} and \eqref{eq:omega} give
% \begin{align*}
% \mu^{\text{A,B}}_1 = -\rho \phi_2~\\
% \mu^{\text{A,B}}_2 = a \phi_2 + \phi_2^3 -\rho \phi^{\text{A,B}}_1 \approx  a \phi_2 + \phi_2^3 ~\\
% p^{\text{A,B}} = -f1 - f2 -\rho \phi^{\text{A,B}}_1 \phi_2 + \mu^{\text{A,B}}_1 \phi^{\text{A,B}}_1 - \mu^{\text{A,B}}_2 \phi^{\text{A,B}}_2 \approx  -f1 - f2 -\mu_2 \phi_2
% \end{align*} 
% I.e. all three conditions for coexistence are fulfilled and we used the approximation $|\phi_2|^3>>|\phi_1|$. \textit{Explanation okay? At least it fits with the results of the continuation}
% %
% }

%In sum there are five pairs of associated binodals representing different transitions.
%However, we only use three colors representing these binodals, since some transitions are connected with each other.

Now we come back to the case of $\rho>0$ in Fig.~\ref{fig:phase_diagram} to discuss the remaining features. The concept of coexistence can be extended beyond the thermodynamic limit of a globally stable coexistence: Two coexisting phases may still exist even if the resulting state is only metastable (thin solid lines in Fig.~\ref{fig:phase_diagram}). Such ``metastable binodals'' exists for a small parameter range after the binodals cross a triple point. Mathematically, one or both of the ``coexisting'' states can even be linearly unstable (dotted lines) - such ``unstable binodals'' are important for the understanding of the topology of the phase diagrams: In the exchange symmetric case [Fig.~\ref{fig:phase_diagram}~(a)] each unstable binodal (except the purple line) connects the ends of two metastable binodals and represents a threshold state that has to be overcome when going from a metastable coexistence to a stable one. The purple dotted ellipse in (a) and the corresponding purple dotted line in (b) represent unstable coexistences of phases II and IV. These states are remnants of the discussed four-phase coexistence in the limiting case $\rho\to0$. When the sign of $\rho$ changes, the blue and purple ellipses exchange their roles.

The phase diagrams in Figs.~\ref{fig:phase_diagram}~(c) and~(d) present the more generic case of a broken field exchange symmetry (at $a_\Delta=-0.5$). Although the symmetry w.r.t.~the diagonals is broken, qualitatively the stable two- and three-phase coexistences are unchanged. However, a qualitative difference is observed for metastable and unstable coexistence lines: They connect differently, e.g., the blue ellipse in (a) is broken in (c). Also, in (a) and (b) metastable and unstable binodals connect stable phase coexistences I-IV [II-III] and III-IV [I-II], in (c) and (d) they connect the stable phase coexistence I-IV via I-III to II-III (visible through the different respective line colors). Knowledge of such metastable and linearly unstable states is particularly important when the dynamics of phase transitions is considered, e.g., when considering the motion of fronts \cite{TaNi1985prl,VTPS2008pre}. 
For more extensive analyses of the phase behavior of ternary mixtures including the dependency on a third parameter, e.g. the temperature, see Refs.~\cite{KoWK2006jcp,PTPR2003pre}.

%In the fully decoupled limit, metastable coexistence is limited \tobiasbf{delete: "by an ``inner square'' at $|\phi_i|=\sqrt{-a/3}$", because its only a square if $a_\Delta=0$} within the 4-phase coexistence region.

% E.g. phase II is characterized by low $\phi_1$ and high $\phi_2$ values and thus sits in the upper left corner in Fig.~\ref{fig:phase_diagram}~(a). Since the chemical potentials in the pure phases are proportional to the concentration values to the third, when we neglect the smaller linear contributions, the different phases sit in the same corners in the chemical potential plane (see Fig.~\ref{fig:phase_diagram}~(b)).

% E.g. starting with low $\phi_1$ concentration, the gray pair of binodals represent the II-III-phase transition and is thermodynamic stable.
% Where the lower gray binodal crosses the blue binodal, i.e., at the triple point in Fig.~\ref{fig:phase_diagram}~(b) it becomes metastable (thin solid line) and then unstable (dotted line). 
% There it also changes its direction in the concentration plane and roughly goes straight up.
% Then it becomes stable again and it regains its thermodynamic stability at the triple point.
% However from there on together with its associated other binodal it represents the I-II-phase transition.
% An analogous connection is observed for the I-IV and III-IV-phase transitions (black lines) resulting from the field exchange symmetry.

Next, we consider phase coexistence in finite systems, where energies of interfaces between coexisting phases become important. Then, transitions between phases can be described by bifurcation diagrams giving a property of states as a function of a control parameter. %Fig.~\ref{fig:phases_finitedomain}~(a) with example profiles given in panel~(b)-(d).
Increasing the system size, one can systematically study how coexistence in the thermodynamic limit emerges from the bifurcation diagrams. See Ref.~\cite{TFEK2019njp} for the case of a one-field CH equation.

%Here, the bifurcation diagrams provide us with the passive reference case for the active system investigated in Section~\ref{sec:nonlinear-nonvari}. 
In Fig.~\ref{fig:phases_finitedomain}~(a) we use the mean concentration $\bar\phi_2$ as control parameter, keep $\bar{\phi}_1=0$ fixed, and employ a suitable norm as solution measure [see Eq.~\eqref{eq:norm}]. That is, we consider a straight vertical cut through the phase diagram in Fig.~\ref{fig:phase_diagram}~(a). At small $\bar\phi_2$, we start in the coexistence region of phases~III and IV, i.e., the low-$\phi_1$, low-$\phi_2$ phase and the high-$\phi_1$, low-$\phi_2$ phase coexist, see panel (d) for an example profile. %There, the $\phi_1(x)$ and $\phi_2(x)$ profiles are given as blue dotted and orange solid lines, respectively.
%Close inspection shows that the concentration values of the plateaus are not exactly the binodal values in the phase diagram. This is due to the finite size of the system.
%Furthermore we use a system size of $\ell=10\pi$ and set $\kappa=1$ which together determine the steepness of the phase boundaries.

Increasing $\bar{\phi}_2$, the triple point region in Fig.~\ref{fig:phase_diagram}~(a) is crossed, before reaching the I-III coexistence region. In the bifurcation diagram, the branch undergoes two saddle-node bifurcations where the states loose and regain linear stability, respectively. This is related to the nucleation of a third phase within the profile, namely, the high-$\phi_1$, high-$\phi_2$ phase (i.e., phase-I). It appears at the center of the phase-IV plateau [see panel (c)]. Phase~IV is still visible as two shoulder-like plateaus between the expanding phase~I and phase~III. Further increasing $\bar{\phi}_2$, the plateaus of phase~IV shrink and are replaced by phase~I [see panel (d)]. Beyond the maximum at $\bar{\phi}_2=0$, the two concentration fields exactly reverse roles due to the inversion symmetry and phase~III is replaced by phase~II in a similar sequence of events (not shown).
% Beyond the r.h.s.\ pair of saddle-node bifurcations, i.e., at $\bar \phi_2 \gtrsim 1.2$, we end in the coexistence region of phases~I and II, i.e., the high-$\phi_1$, high-$\phi_2$ phase and the low-$\phi_1$, high-$\phi_2$ phase (not shown).

%If $a$ decreases, the binodal lines in the phase diagram move apart and the three state coexistence region becomes larger. Concentration profiles show steeper interfaces between phases rendering plateaus more pronounced. 

 \section{Weakly nonlinear analysis}
 \label{sec:app-wna}
 To better characterize primary pitchfork bifurcations and to identify parameter values where subcritical behavior occurs we use a weakly nonlinear analysis to derive amplitude equations.
%that describes any stationary solution in the vicinity of its primary bifurcation.
Although our main interest is in the onset of subcritical behavior in the case of zero mean concentrations, the theory is developed for the general case employing the ansatz
\begin{align}\label{eq:ansatz}
\vecg{\phi}&= \vecg{\bar\phi} + \sqrt{|\varepsilon|} \vecg{ A} e^{i q_n \ell x}  + |\varepsilon| \vecg{C} e^{2 i q_n \ell x} + \textrm{c.c.}~\\
{\rm with } \, \, \,  \vecg{A} &=  \vecg{v} A_0 +  |\varepsilon| \vecg{A}_1 \, .
\end{align}
The smallness parameter $\varepsilon$ gives the distance to the primary bifurcation at $\varepsilon=\varepsilon_\textrm{c}=0$ and $q_n=\frac{2 n \pi}{\ell}$ is the discretized rescaled wavenumber. All amplitudes are constants, i.e., we only consider homogeneous steady patterns. Then, as the mean concentrations are fixed at $\vecg{\bar\phi}$, the perturbation does not contain a mean or neutral mode $\sim e^0$. The bifurcations of interest are simple codimension-1-points, i.e., they are represented by lines in the $(f''_2,f''_1)$-plane (see Fig.~\ref{fig:stab-diagram}) where $f_{1}''$ and $f_{2}''$ directly depend on $\bar\phi_1$ and $\bar\phi_2$, respectively. However, to keep results general, we use $f''_1$ and $f''_2$ as control parameters. %Hence, one of the two parameters can be chosen arbitrarily in some range, then the other one is adjusted to the corresponding line of the bifurcation at $(f''_2,f''_1)=(f''_{2,\mathrm{c}},f''_{1,\mathrm{c}})$.
In Fig.~\ref{fig:stab-diagram} the colored lines represent the onset of linear instability, i.e., the first mode to become unstable. If one considers the onset of linear instability in a Turing bifurcation for arbitrary $f''_2$, then $f''_{1,\mathrm{c}}={f''_1}^T$ [Eq.~\eqref{eq:a11tu2}] and $q_n= k_\mathrm{c}/\ell$ [Eq.~\eqref{eq:kcTuring}].

However, our calculations hold for any stationary primary bifurcation at fixed wavenumber $q_n$ even far away from the onset of linear instability.
Hence, the critical parameters depend on $q_n$, i.e.~$f''_{i,\mathrm{c}}=f''_{i,\mathrm{c}}(q_n)$ -- this is not explicitly indicated in the following. Starting from a bifurcation at a particular $(f''_{2,\mathrm{c}},f''_{1,\mathrm{c}})$ we consider an arbitrary direction in the $(f''_2,f''_1)$-plane characterized by an angle $\vartheta \in \left[0, \pi \right[$. %(and $\varepsilon>0$ or $\varepsilon<0$) and 
The bifurcation is then crossed on a line defined by
\begin{equation}
f''_1=f''_{1,\mathrm{c}} + \varepsilon  \sin \vartheta \, , \quad f''_2= f''_{2,\mathrm{c}} + \varepsilon  \cos \vartheta \,.
\end{equation}
At the bifurcation, the eigenvector $\vecg{v}$ of the critical mode solves the linear equation
\begin{equation}
\tens{B}\Big|_{q_n,\varepsilon_\textrm{c}} \vecg{v}=
\left( \begin{array}{c c}
q_n^2 + f_{1,\mathrm{c}}'' & -\left(\rho + \alpha \right)\\
-Q\left(\rho - \alpha\right) & Q\left(\kappa q_n^2 + f_{2,\mathrm{c}}''\right)
\end{array}\right)  
\vecg{v}
=0\,.
\end{equation}
%Excluding the special case $\rho=\alpha$, we scale the second component of the eigenvector to one and obtain
The resulting eigenvector $\vecg{v}$ and adjoint eigenvector $\vecg{v}^\dagger$ are
\begin{align} \label{eq:eigenvectors}
\vecg{v} &= \left( \begin{array}{c} 
\rho + \alpha
~\\
q_n^2 + f''_{1,\mathrm{c}}
\end{array} \right)
\, , \quad ~
\vecg{v}^\dagger = \left( \begin{array}{c} 
Q \left( \rho - \alpha \right)
~\\
q_n^2 + f''_{1,\mathrm{c}}
\end{array} \right)\, .
\end{align}
The latter is in the kernel of the adjoint linear operator $\tens{B}^\dagger\Big|_{q_n,\varepsilon_\textrm{c}}$ and is needed when applying the Fredholm alternative.
We insert ansatz \eqref{eq:ansatz} into the model Eqs.~\eqref{eq:nondim_bG_final}, compare Fourier coefficients and sort in orders in $|\varepsilon|$. At $\mathcal{O}(|\varepsilon|)$ from the coefficients of $e^{2 i q_n \ell x}$ we obtain
\begin{align} 
\tens{B}\Big|_{2 q_n,\varepsilon_{\mathrm{c}}} \vecg{C} =& - \frac{1}{2!}\left( \begin{array}{c} 
f'''_1 \, v_1^2
~\\
Q f'''_2 \, v_2^2
\end{array} \right) A_0^2 
\end{align}
\begin{alignat}{1}
\begin{aligned}
\Rightarrow C_1 =& - \frac{\left(4\kappa q_n^2 + f_{2,\mathrm{c}}''\right)f'''_1 \, v_1^2 +\left(\rho + \alpha \right) f'''_2 \, v_2^2}{ \frac2Q \det \tens{B}\big|_{2 q_n,\varepsilon_{\mathrm{c}}}} \, A_0^2 \equiv \gamma_1  A_0^2  \, ,\\
 C_2 =& - \frac{\left(4 q_n^2 + f_{1,\mathrm{c}}''\right)f'''_2 \, v_2^2 +\left(\rho - \alpha \right) f'''_1 \, v_1^2}{ \frac2Q  \det \tens{B}\big|_{2 q_n,\varepsilon_{\mathrm{c}}}}\, A_0^2 \equiv \gamma_2 A_0^2\, ,
\end{aligned}
\label{eq:CasA0}
\end{alignat}
% \tobias{we could also write 
% \begin{equation}
% C_i =\frac{L_{ij} a_j - L_{jj} a_i}{L_{ij} L_{ji} - L_{ii} L_{jj}}\Big|_{2 q_n,\varepsilon_{\mathrm{c}}} A_0^2 \, , (i,j)\in [(1,2),(2,1)]  
% \end{equation}}
%
with $C_1$ [$C_2$] and $v_1$ [$v_2$] being the first [second] component of $\vecg{C}$ and $\vecg{v}$, respectively. Furthermore,  $\det \tens{B}\big|_{2 q_n,\varepsilon_{\mathrm{c}}}$ is the determinant of $\tens{B}$ taken at $q=2q_n$ and $\varepsilon=\varepsilon_\text{c}$.
Both components $C_i$ of the higher harmonic mode are determined by $A_0$ (slaving principle) via Eqs.~\eqref{eq:CasA0}. They are independent of the mobility ratio $Q$ (since $\det \tens{B}$ carries a factor $Q$). For brevity we define the proportionality constants as $\gamma_i$.

At order $\mathcal{O}(|\varepsilon|^{3/2})$ the coefficients of $e^{i q_n \ell x}$ give
\begin{align}
\tens{B}\Big|_{q_n,\varepsilon_\mathrm{c}} \vecg{A}_1 + \frac{\partial}{\partial \varepsilon} \tens{B}\Big|_{q_n,\varepsilon_{\textrm{c}}} \vecg{v}
\,  A_0 +  \left( \begin{array}{c} 
f'''_1 \, v_1 C_1
~\\
Q f'''_2 \, v_2 C_2
\end{array} \right) A^*_0
~~\nonumber \\
+\frac{1}{2} \left( \begin{array}{c} 
f''''_1 v_1^3
~\\
Q f''''_2 v_2^3
\end{array} \right) |A_0|^2 A_0 =0  \, .\mylab{eq:Order3}
\end{align}
%~
To apply the Fredholm alternative we multiply Eq.~\eqref{eq:Order3} from the left by $\vecg{v}^\dagger$. Then, the first term vanishes. We insert the expressions for amplitudes $C_1$ and $C_2$ [Eqs.~\eqref{eq:CasA0}], we re-incorporate the smallness parameter into the amplitude $A_0$ and finally with $\frac{\partial}{\partial \varepsilon} \tens{B}\Big|_{q_n,\varepsilon_{\textrm{c}}} \vecg{v}= \left( \begin{array}{c} 
v_1 \sin \vartheta ~\nonumber
~\\
Q v_2 \cos \vartheta
\end{array} \right)$ we obtain the stationary amplitude equation 
\begin{align}
\varepsilon \left(v_1 v_1^\dagger \sin \vartheta + Q v_2 v_2^\dagger \cos \vartheta \right)A_0 +
 \left(f'''_1 \, v_1 v_1^\dagger \gamma_1 + Q f'''_2 \, v_2 v_2^\dagger \gamma_2\right) |A_0|^2 A_0
~\nonumber 
~\\
+\frac{1}{2} \left( f''''_1 v_1^3 v_1^\dagger +  Q f''''_2 v_2^3 v_2^\dagger \right)  |A_0|^2 A_0.
=0 \mylab{eq:AEgeneral}
\end{align}
Next, it is used to characterize the branches emerging at primary bifurcations. As $v_1^\dagger\sim Q$ the amplitude equation is independent of the mobility ratio $Q$ (as expected for steady states).

First, we reproduce the result for the one-field CH equation as obtained in the decoupled limit ($\rho=\alpha=0$): One finds the simple expressions  $v_1=v_1^\dagger=\gamma_1=0$, $v_2=v_2^\dagger=1$ and $\gamma_2= - \frac{f'''_2}{2\left(4 \kappa q_n^2 + f''_{2,\textrm{c}}\right)}$, here in the case of an instability related to the second CH equation. %and the other option of an instability related to the first equation is equally possible.}
As the second control parameter, here $f''_1$, does not enter, the only possible direction in the $(f''_2,f''_1)$-plane is $\vartheta=0$ and the linear stability threshold is $f''_{2,\textrm{c}}= - \kappa q_n^2$. The amplitude equation [Eq.~\eqref{eq:AEgeneral}] reduces to
\begin{equation}
 \varepsilon A_0 + \frac12 \left[f''''_2 - \frac{{f'''_2}^2}{ 3 \kappa q_n^2} \right] |A_0|^2 A_0 = 0\,.
\end{equation}
%\ttuwe{ expressions for the decoupled CH's will come out as limiting cases, see (33) of \cite{TALT2020n} with $D=0$ and $Q=1$.}
For the particular energy $f_2 = \frac{a}{2}\phi_2^2 + \frac{1}{4}\phi_2^4$, this further reduces to
\begin{equation}
\varepsilon A_0 + \left(3 - \frac{6 \bar{\phi}_2^2}{\kappa q_n^2}\right) |A_0|^2 A_0 = 0,
\end{equation}
i.e., the transition from supercritical to subcritical pitchfork bifurcation occurs at $\bar{\phi}_2^2=\frac{\kappa q_n^2}{2}$ as expected \cite{Novi1985jsp} \footnote{Note that in Ref.~\cite{Novi1985jsp} a different scaling is used. There, the transition for a critical wavelength $k=1$ occurs at $B^2= 4.5$. In our scaling $B^2= 9 \bar{\phi}_2^2$ and $\kappa=q_n=1$. }.

% which describes supercritical pitchfork bifurcations, since the homogeneous state ($A_0=0$) looses stability for decreasing control paramter $\varepsilon<0$.}

Second, we discuss the case of coupled fields with zero mean concentrations, i.e.~$\vecg{\bar\phi}=(0,0)$. With the standard double-well potentials $f_i \sim \phi_i^2 + \phi_i^4$ the third derivatives taken at $\vecg{\bar\phi}$ vanish and the mode $\sim e^{2i q_n \ell x}$ is not excited, i.e.~$\gamma_{1,2}=0$ and
the amplitude equation \eqref{eq:AEgeneral} reduces to
\begin{align}
\varepsilon \left(v_1 v_1^\dagger \sin \vartheta +Q  v_2 v_2^\dagger \cos \vartheta \right)A_0 
+\frac{1}{2} \left( f''''_1 v_1^3 v_1^\dagger + Q f''''_2 v_2^3 v_2^\dagger \right)  |A_0|^2 A_0
=0 \,. \mylab{eq:AE_zero}
\end{align}
Inserting the components of the eigenvectors [Eqs.~\eqref{eq:eigenvectors}] we obtain
\begin{align}
\varepsilon \left(- \Delta \sin \vartheta + \left(q_n^2 + f''_{1,\textrm{c}}\right)^2 \cos \vartheta \right)A_0 ~\nonumber ~\\
+\frac{1}{2} \left( - f''''_1 \Delta \Sigma + f''''_2 \left(q_n^2 +f''_{1,\textrm{c}}\right)^4 \right)  |A_0|^2 A_0
  =0\, ,
  \label{eq:appD5}
\end{align}
with $\Delta = \alpha^2 - \rho^2$ and $\Sigma=\left(\alpha + \rho\right)^2$.
We can choose the amplitude $A_0$ to be real and positive (since we use periodic boundary conditions). Solving \eqref{eq:appD5} yields
\begin{equation}\label{eq:Asol}
A_0=0 \quad\mathrm{and}\quad A_0= \sqrt{2 \varepsilon \, \frac{\Delta \sin \vartheta  - \xi_n \cos \vartheta}{ -f''''_1  \Delta \Sigma +f''''_2 \xi_n^2}}\,,
\end{equation}
with 
\begin{align}
\xi_n = \left(q_n^2 + f''_{1,\textrm{c}}\right)^2\, .
\end{align}
%\tobias{general discussion difficult, since it depends on where we are in the stability plane whether $A_0=0$ becomes unstable for $\varepsilon<0$ or $>0$}
Our bifurcation diagrams use $a$ as control parameter, i.e., for $\bar \phi_i=0$ correspond to diagonal cuts through the $(f_1'',f_2'')$- plane, i.e.~
\begin{align}
 \vartheta = \pi/4 \, \, \, \text{and} \, \,\,  f''_1 = a\, , \quad f''_2= a+ a_\Delta \, ,\quad f''''_1 = f''''_2=6 \\
 \Rightarrow 
 A_0= \sqrt{\frac{\varepsilon}{3 \sqrt{2}} \, \frac{\Delta -  \xi_n }{ - \Delta \Sigma + \xi_n^2}} \,.
\end{align}
Then the critical parameter at given wavenumber $q_n$ is given by $f''_{1,\textrm{c}}=a_{\pm}(q_n)$ [Eq.~\eqref{eq:azero} with $\bar \phi_i=0$] where $+$ [$-$] refers to the eigenvalue $\lambda_+$ [$\lambda_-$].
Furthermore
\begin{equation}\label{eq:xi}
\xi_{n,\pm}= \left(q_n^2 + a_{\pm}(q_n)\right)^2 = \left(\frac{M_n}{2} \pm \sqrt{\frac{M_n^2}{4} - \Delta}\right)^2 \quad \text{with}\,\, M_n= q^2_n\left(1-\kappa\right)- a_\Delta\,.
\end{equation}
%\tobias{$M$ is the difference between the decoupled systems,i.e.~$M= q^2 + f''_1 -\left(\kappa q^2 + f''_2\right)$. Isn't that interesting? I suggest that it has a similar form for two coupled (c)SH Eqs}
The trivial state ($A_0=0$) looses stability for decreasing $a$, i.e.~for $\varepsilon<0$. Then the corresponding bifurcation is subcritical if 
\begin{align}
 \frac{\Delta -  \xi_{n,\pm} }{ - \Delta \Sigma + \xi_{n,\pm}^2}&>0~\nonumber \\
% \Rightarrow \left( \rho^2 - \alpha^2 < -\xi^2  \wedge \rho^2 - \alpha^2  > -\frac{\xi^4}{(\rho + \alpha)^2} \right) &  \vee  \left( \rho^2 - \alpha^2 > -\xi^2  \wedge \rho^2 - \alpha^2  < -\frac{\xi^4}{(\rho + \alpha)^2} \right) ~\\
\Rightarrow \text{min}\left(\xi_{n,\pm}, \frac{\xi_{n,\pm}^2}{\Sigma}\right) < \Delta &<  \text{max}\left(\xi_{n,\pm}, \frac{\xi_{n,\pm}^2}{\Sigma}\right)\,.\label{eq:sub_criterion}
\end{align}
This can only occur if $\Delta>0$, i.e., $|\alpha|>|\rho|$. As $\xi_{n,+}(M_n)=\xi_{n,-}(-M_n)$ [see Eq.~\eqref{eq:xi}], branches of the same periodicity $n$ related to $\lambda_+$ and $\lambda_-$ exchange sub- and supercritical behavior when $M_n$ switches sign. 
In particular, for $\kappa=1$ the parameter $\xi_{n,\pm}=\xi_{\pm}=\left(-\frac{a_\Delta}{2} \pm \sqrt{\frac{a_\Delta^2}{4} - \Delta}\right)^2$ is independent of $n$. Then, the inequality~\eqref{eq:sub_criterion} gives the same threshold for subcriticality \textit{for all} stationary primary bifurcations [cf.~Fig.~\ref{fig:linsupp_bif_sub}~(d)].

There exist two nongeneric cases where subcritical behavior can not occur for any stationary primary bifurcation, namely, for identical subsystems ($\kappa=1$ and $a_\Delta=0$, i.e. $M=0$ and $\xi_{n,\pm}= |\Delta|$), and for vanishing variational coupling ($\rho=0$, i.e. $\Sigma=\Delta$).
The criterion~\eqref{eq:sub_criterion} is illustrated in Fig.~\ref{fig:sub_ranges} of the main text. Its impact on the onset of time-periodic behavior is discussed in Sec.~\ref{sec:OnsetLargeOscill}.

\section{Time-periodic behavior arbitrarily close to equilibrium}
\label{sec:app:small}
In Sec.~\ref{sec:OnsetLargeOscill} two generic scenarios are discussed for the emergence of time-periodic behavior of the fully phase separated state. Here, we highlight the particular case of purely nonvariational coupling ($\rho=0$, $\alpha \neq 0$). In this special situation the necessary condition for time-periodic behavior $|\alpha|>|\rho|$ is fulfilled at arbitrarily small $\alpha$, and oscillatory behavior can be expected arbitrarily close to a classical gradient dynamics describing systems evolving towards thermodynamic equilibrium. In the nongeneric case of identical decoupled subsystems, i.e., for $f_1''=f_2''$ and $\kappa=1$, all primary pitchfork bifurcations become Hopf bifurcations for any $\alpha\neq0$. Time-periodic states then only exist with small amplitude (not shown).
%in the vicinity of the uniform state. Hence, it can not be expected that this behavior plays an important role in the nonlinear regime.

\begin{figure}[tbh]
\includegraphics[width=0.7\textwidth]{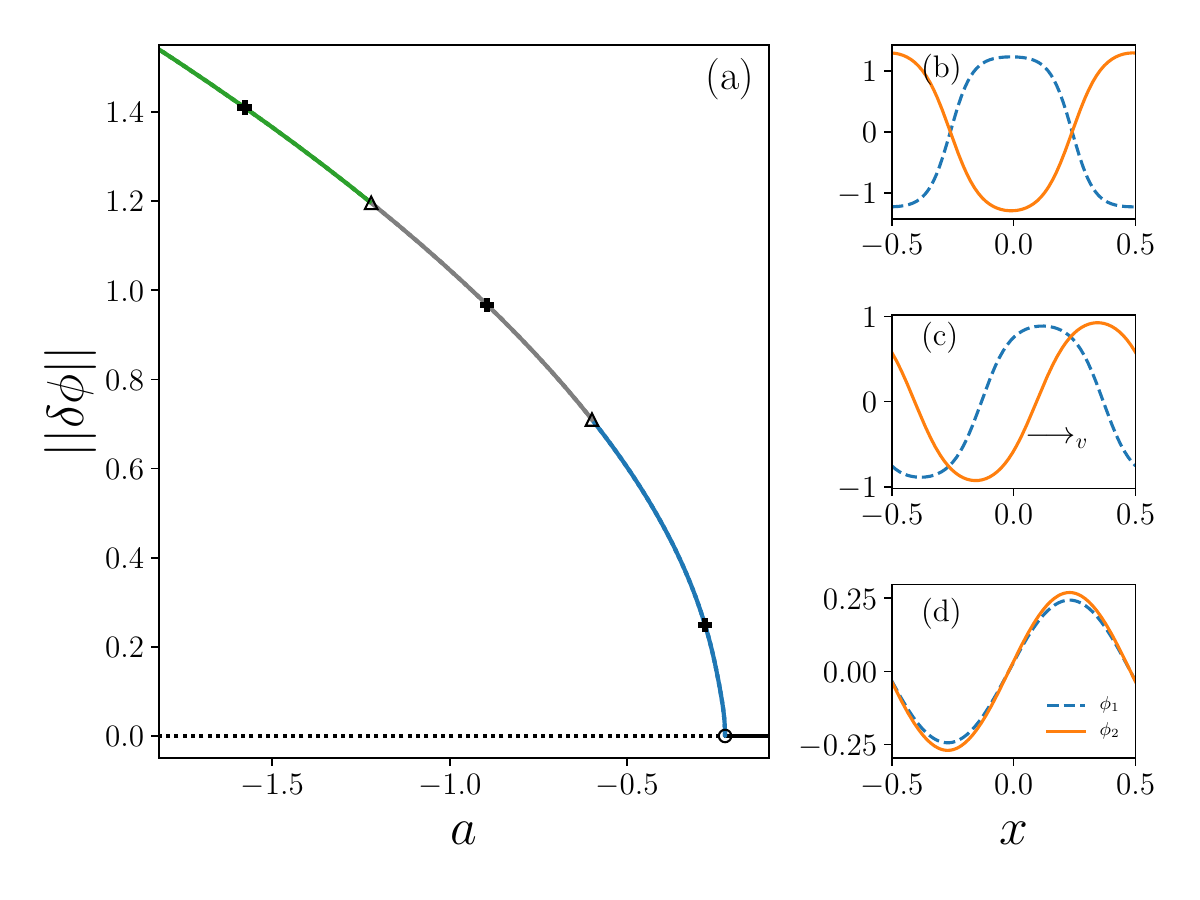}
\caption{\small \it Emergence of drifting states close to equilibrium for $\alpha=0.01$. Panel (a) shows a bifurcation diagram corresponding to a CH instability of the uniform state (black line). The loci of the selected profiles in (b)-(d) are marked by bold ``+'' symbols in (a). Stable branches of steady $n=1$ states exist with in-phase [blue line in (a), profile in (d)] and anti-phase [green line in (a), profile in (b)] fields. They are connected by a branch of stable drifting states [gray line in (a), profile in (c)]. The remaining parameters are $\rho=0\,,\,\,a_\Delta=-0.38\,,\,\, \kappa=2.4\,,\,\, \bar{\phi}_1=0\,,\,\, \bar{\phi}_2=0\,,\,\,\ell=4\pi$ and $Q=1$.}\label{fig:alpha0}
\end{figure}

Significantly more relevant is the generic case of unequal subsystems. A corresponding bifurcation diagram is given in Fig.~\ref{fig:alpha0}~(a) for a small nonvariational coupling $\alpha=0.01$. It shows the uniform state and the linearly stable parts of three different phase-separated states. They almost lie on top of each other and can not be distinguished by eye. The unstable parts are omitted. The uniform state (black horizontal line) looses stability to a CH instability where the stationary $n=1$ branch (blue line) emerges supercritically, and hence, stable. Panel~(d) illustrates the emerging state and shows that the two fields are in-phase, i.e., near onset the nonvariational coupling acts attractively.
  
However, far in the nonlinear regime, an anti-phase arrangement is favored, e.g., for $a\lesssim -1.22 $ the state of panel~(b) is stable (green line).
The stable in-phase and anti-phase stationary $n=1$ states are connected by a branch of stable drifting states (gray line) that ends at two drift-pitchfork bifurcations (triangles). The example profile in panel (c) indicates an intermediate phase shift that allows one to move from in-phase to anti-phase along the branch. For any phase between $0$ and $\pi$ the states drift with constant velocity in the direction indicated by an arrow in panel (c). Note that both drift-pitchfork bifurcations emerge together in a codimension-2 point at $\alpha=0$. %for increasing and decreasing $\alpha$.
For instance, at $\alpha=-0.01$ the branch of drifting states connects stable anti-phase states emerging in the primary bifurcation with stable in-phase states far in the nonlinear regime (not shown).
% One can give further details on the conditions: $f_2''>f_1''$ [$f_2''<f_1''$ and $\kappa <1$ [$\kappa>1$]. 

In addition to the unique property of time-dependent behavior arbitrarily close to equilibrium, the purely nonvariational coupling also represents a special case regarding model classification: For $\rho=0$, we can write Eq.~\eqref{eq:nondim_bG_final} in a gradient dynamics form, namely,
\begin{align}
\partial_t\phi_i=&\partial_x \left(\frac{Q_i}{\ell^2}\, \partial_x \frac{\delta  \widetilde{\mathcal{F}}}{ \delta \phi_i}\right)\, ,\quad i=1,2~\\
\text{with}\, \, \widetilde{\mathcal{F}} =& \int \left[-\frac{1}{2\ell^2}|\partial_x \phi_1|^2 - f_1(\phi_1) + \frac{\kappa}{2\ell^2}\, |\partial_x \phi_2|^2  + f_2(\phi_2) + \alpha \phi_1 \phi_2 \right] \mathrm{d} x~\\
\text{and}\, \,  Q_1=&-1, \,\, Q_2= Q\,.
\end{align}
However, the ``energy'' $\widetilde{\mathcal{F}}$ has now destabilizing and stabilizing gradient-square terms, and is not bounded from below. However, the active character is encoded in the negative mobility constant $Q_1$, implying that $\widetilde{\mathcal{F}}$ does not necessarily decrease in time in contrast to its variational pendant.
%\begin{figure}[tbh]
%\includegraphics[width=0.49\textwidth]{./kappa_3deltaT-1,9phasediagram.pdf}
%\includegraphics[width=0.49\textwidth]{./kappa_0,14deltaT-1,9phasediagram.pdf}
%\includegraphics[width=0.49\textwidth]{./q3_kappa_1,5deltaT-1,9phasediagram.pdf}
%\includegraphics[width=0.49\textwidth]{./q1_kappa_1,5deltaT1,9phasediagram.pdf}
%\caption{\small \it blue region: $n^+$ subcritical, orange region: $n^-$ subcritical}\label{fig:example}
%\end{figure}

\clearpage

\bibliography{cCH}
\end{document}